\documentclass[titlepage,11pt,amstex]{article}
\usepackage{eurosym}
\usepackage[utf8]{inputenc}
\usepackage{amssymb}
\usepackage{amsfonts}
\usepackage{amsmath}
\usepackage{mathptmx}
\usepackage{comment}
\usepackage[dvips]{graphicx}
\usepackage{supertabular}
\usepackage{longtable}
\usepackage{tabularx}
\usepackage{dsfont}
\usepackage{psfont}
\usepackage{rotating}

\usepackage{xcomment} 
\usepackage{ifthen} 
\usepackage{xifthen} 

\usepackage{calc}  
\usepackage{time}  
\usepackage{clock}  

\usepackage{harvardD}
\usepackage{bibunits}

\usepackage{section}
\usepackage{theorem}
\usepackage{verbatim}
\providecommand{\bTentative}{}

\usepackage{eurosym} 
\usepackage{fix-cm}  
\usepackage{textcomp} 
\usepackage{pifont}  
\usepackage[T1]{fontenc}
    
\providecommand{\SpecialFonts}{
    \usepackage{calligra} 
    \usepackage[varumlaut]{yfonts}  
    }

\usepackage{pdfpages} 

\usepackage{gloss}
\usepackage{makeidx} 

\usepackage{enumerate}

\usepackage{amssymb}  
\usepackage{amsfonts}  
\usepackage{amsmath}  

\usepackage{array}
\usepackage{bbold} 
\usepackage{dsfont}  
\usepackage{fixmath} 
\usepackage{mathtools}  
\usepackage{mathrsfs} 

\usepackage{endnotes}

\usepackage{afterpage} 
\usepackage{layout} 
\usepackage{lscape} 

\usepackage{setspace} 

\usepackage{url}  

\usepackage{longtable} 

\providecommand{\paraNumbering}{\theoremstyle{change}}

\SpecialFonts

\providecommand{\sectitlesize}{}

\newcommand{\mcitepYp}{\citeyear}
\newcommand{\mciteA}{\citename}
\newcommand{\mciteY}{\citeyear*}

\newcommand{\mciteAYY}[3][]{\mciteA{#2} (\mciteY{#2}, \mciteY{#3}#1)}

\newcommand{\mciteAYYYY}[5][]{\mciteA{#2} (\mciteY{#2}, \mciteY{#3}, \mciteY{#4}, \mciteY{#5}#1)}

\newcommand{\mcite}[3][]{\mciteA{#2} \mcitepYp{#3}#1}

\newcommand{\cbu}{,\xspace}
\newcommand{\dbu}{.\xspace}
\newcommand{\bbibuniteconometrica}{\begin{bibunit}[econometrica] 
    \renewcommand{\cite}{\nocite*}
    \renewcommand\refname{} \renewcommand{\cbu}{} 
    \renewcommand{\dbu}{} \renewcommand{\newline}{}
    \vspace{-3.5\baselineskip}}
\newcommand{\bbibunitagsm}{\begin{bibunit}[agsm] \renewcommand{\cite}{\nocite*}
    \renewcommand\refname{} \renewcommand{\cbu}{} \renewcommand{\dbu}{} 
    \renewcommand{\newline}{}
    \vspace{-3.5\baselineskip}}
\newcommand{\ebibunit}{\putbib[refer1] \end{bibunit}}

\newcommand{\bbibunitunsrt}{\begin{bibunit}[unsrt] \renewcommand{\cite}{\nocite*}
    \renewcommand\refname{} \renewcommand{\cbu}{} \renewcommand{\dbu}{} \renewcommand{\newline}{}
    \vspace{-3.5\baselineskip}}
\newcommand{\ebibunitunsrt}{\putbib[refer1] \end{bibunit}}

\makeindex  
\makeglossary  
\makegloss

\newcommand{\NumericNumberedLists}{
\def\labelenumi{\arabic{enumi}.}
\def\theenumi{\arabic{enumi}}
\def\labelenumii{\arabic{enumii}.}
\def\theenumii{\arabic{enumii}}
\def\p@enumii{\theenumi.}
\def\labelenumiii{\arabic{enumiii}.}
\def\theenumiii{\arabic{enumiii}}
\def\p@enumiii{\theenumi.\theenumii.}
\def\labelenumiv{\arabic{enumiv}.}
\def\theenumiv{\arabic{enumiv}}
\def\p@enumiv{\p@enumiii.\theenumiii}
}

\allowdisplaybreaks

\providecommand{\sectitlesize}{}

\newlength{\totalhormargin}
\newlength{\totalvermargin}

\newcommand{\December}{December }

\def\todayMY{\ifcase\month\or
  January\or February\or March\or April\or May\or June\or
  July\or August\or September\or October\or November\or
  December\fi\ \number\year}

\providecommand{\vartitleadjust}{}
\providecommand{\vartitle}{}

\providecommand{\varAuthors}{Jean-Marie Dufour \thanks{\ \ \DufourAddress} \\
  McGill University}

\providecommand{\vardate}{\today, \texttime}

\providecommand{\standardthanksIFM}{}
\renewcommand{\standardthanksIFM}{This work was supported by the William Dow Chair in Political
  Economy (McGill University), the Bank of Canada (Research Fellowship), 
  the Toulouse School of Economics (Pierre-de-Fermat Chair of excellence), 
  the Universitad Carlos III de Madrid (Banco Santander de Madrid Chair of excellence), 
  a Guggenheim  Fellowship,
  a Konrad-Adenauer  Fellowship (Alexander-von-Humboldt Foundation, Germany),
  the Canadian Network of Centres of Excellence
  [program on \emph{Mathematics of Information Technology and
  Complex Systems} (MITACS)], the Natural Sciences and Engineering Research
  Council of Canada, the Social Sciences and Humanities Research
  Council of Canada, and the Fonds de recherche sur la soci\'{e}t\'{e}
  et la culture (Qu\'{e}bec).}

\providecommand{\varthanks}{\ \ \standardthanksIFM}
\providecommand{\thanksvar}{\thanks{\varthanks}}

\providecommand{\DufourAddress}{William Dow Professor of Economics, McGill University,
  Centre interuniversitaire de recherche en analyse des
  organisations (CIRANO), and Centre interuniversitaire de recherche en
  \'{e}conomie quantitative (CIREQ). Mailing address:
  Department of Economics, McGill University, Leacock Building, Room 414,
  855 Sherbrooke Street West, Montr\'{e}al, Qu\'{e}bec H3A 2T7, Canada.
  TEL: (1) 514 398 4400 ext. 09156; FAX: (1) 514 398 4800; e-mail: 
  \protect\url=jean-marie.dufour@mcgill.ca= . Web page:
  \protect\url{http://www.jeanmariedufour.com} }

\providecommand{\DokoAddress}{School of Economics, The University of Adelaide, 
  10 Pulteney Street, Adelaide, SA 5005. Tel: +618 8313 1174;  
  e-mail: \protect\url=firmin.dokotchatoka@adelaide.edu.au=. 
  Homepage: \protect\url{http://www.adelaide.edu.au/directory/firmin.dokotchatoka}}

\newcommand{\sptha}{\hspace{-0.01em}}
\newcommand{\spthb}{\hspace{-0.01em}}

\newcommand{\sppr}{}
\newcommand{\theoremname}{Theorem}
\newcommand{\acknowledgementname}{Acknowledgement}
\newcommand{\algorithmname}{Algorithm}
\newcommand{\assumptionname}{Assumption}
\newcommand{\axiomname}{Axiom}
\newcommand{\casename}{Case}
\newcommand{\claimname}{Claim}
\newcommand{\conclusionname}{Conclusion}
\newcommand{\conditionname}{Condition}
\newcommand{\conjecturename}{Conjecture}
\newcommand{\corollaryname}{Corollary}
\newcommand{\criterionname}{Criterion}
\newcommand{\definitionname}{Definition}
\newcommand{\examplename}{Example}
\newcommand{\exercisename}{Exercise}
\newcommand{\lemmaname}{Lemma}
\newcommand{\notationname}{Notation}
\newcommand{\problemname}{Problem}
\newcommand{\proofname}{\sppr Proof}

\newcommand{\propertyname}{\sppr Property}
\newcommand{\propositionname}{Proposition}
\newcommand{\reflistname}{References}
\newcommand{\remarkname}{Remark}
\newcommand{\resultname}{Result}
\newcommand{\solutionname}{Solution}

\newcommand{\summaryname}{Summary}

\providecommand{\proofof}{Proof of }
\renewcommand{\proofof}{Proof of }

\newenvironment{proof}[1][\sptha \proofname]{\par
  \normalfont
  \trivlist
  \item[\hskip\labelsep\scshape
    #1{.}]\ignorespaces}
    {\qed\endtrivlist\vspace{\baselineskip}}

\newenvironment{proofnoname}[1][\sptha \proofname]{\par
  \noindent
  \normalfont}%
  {\qed\vspace{\baselineskip}}

\newenvironment{proofflexc}[1][\spthb \proofname]{\par
  \noindent
  \normalfont}
  {\qed\vspace{\baselineskip}}

\newenvironment{npar}{\noindent \bf{\thesection.\thenpar}}{}
\newenvironment{subnpar}{\noindent \bf{\thesubsection.\thesubnpar}}{}

\newcounter{paran}

\newcounter{npar}[section]
\newcounter{nparr}[section]

\newcounter{subnpar}[subsection]
\newcounter{subnparr}[subsection]

\DeclareRobustCommand{\qed}{%
  \ifmmode 
  \else \leavevmode\unskip\penalty9999 \hbox{}\nobreak\hfill
  \fi
  \quad\hbox{\qedsymbol}}
\newcommand{\openbox}{\leavevmode
  \hbox to.77778em{%
  \hfil\vrule
  \vbox to.675em{\hrule width.6em\vfil\hrule}%
  \vrule\hfil}}
\DeclareRobustCommand{\qeddirect}{%
  \ifmmode 
  \else \leavevmode\unskip\penalty9999 \hbox{}\nobreak\hfill
  \fi
  \quad\hbox{\qedsymbol}}
\providecommand{\qedsymbol}{\openbox} 

\newcommand{\bproofin}{\begin{proof}}
\newcommand{\eproofin}{\end{proof}}
\newcommand{\bproofend}{\begin{proofnoname}}
\newcommand{\eproofend}{\end{proofnoname}}
\newcommand{\bproofth}{\begin{proofth}}
\newcommand{\eproofth}{\end{proofth}}

\providecommand{\proofof}{Proof of }
\renewcommand{\proofof}{Proof of }

\newcommand{\thsection}{\thesection}

\newcommand{\thsectioneq}{\thesection}

\newcommand{\thsec}{\thsection}
\newcommand{\thseceq}{\thsectioneq}
\providecommand{\thetheorem}{{\bf \thsec.\arabic{theorem}}}
\renewcommand{\thetheorem}{{\bf \thsec.\arabic{theorem}}}
\renewcommand{\theequation}{{\rm \thseceq.\arabic{equation}}}

\makeatletter
   \long
\def\@makecaption#1#2{\vskip 0\p@
   \setbox\@tempboxa\hbox{#1 #2}}
\makeatother

\makeatletter

\providecommand{\l@theorem}{\@dottedtocline{1}{0em}{5em}}
\renewcommand{\l@theorem}{\@dottedtocline{1}{0em}{5em}}

\providecommand{\listttheoremnameb}{\sectitlesize List of Definitions, Assumptions, Propositions and Theorems}

\newcommand\listoftheorems{
 \section*{\listttheoremnameb
           \@mkboth{\MakeUppercase\listttheoremnameb}
           {\MakeUppercase\listttheoremnameb}}
           \@starttoc{lth}
           }
\newcommand{\listoftheoremscont}[1]{
 \providecommand{\listttheoremnameb}{#1}
 \renewcommand{\listttheoremnameb}{#1}
 \section*{\listttheoremnameb
           \@mkboth{\MakeUppercase\listttheoremnameb}
           {\MakeUppercase\listttheoremnameb}}
           \@starttoc{lth}
                      \addcontentsline{toc}{section}{\listttheoremnameb}
           }

\makeatother
\providecommand{\contentshift}{\hspace{-3.5em}}
\renewcommand{\contentshift}{\hspace{-3.5em}}
\providecommand{\contentshiftS}{\hspace{0em}}
\renewcommand{\contentshiftS}{\hspace{0em}}

\newcommand{\sppra}{\hspace{-0.2em}}
\newcommand{\indentprc}{}

\newcommand{\captionproofflex}[2]{}

\newcommand{\captionproofflexc}[2]{
 \sppra\indentprc\textsc{Proof of #1 \protect\ref{#2} }\quad
 \addcontentsline{lth}{theorem}{\protect\numberline{}
  {\contentshift \proofof #1 \protect\ref{#2} }}}

\newcommand{\bAssumptionA}{\begin{assumption}}
\newcommand{\eAssumptionA}{\end{assumption}}

\providecommand{\paraNumbering}{}
\renewcommand{\paraNumbering}{}
\paraNumbering

\newtheorem{theorem}{\sptha \theoremname}[section]

{\theorembodyfont{\normalfont}%
 }

\newtheorem{assumption}{\sptha \assumptionname}[section]

\newtheorem{corollary}[theorem]{\sptha \corollaryname}

{\theorembodyfont{\normalfont}%
 }
{\theorembodyfont{\normalfont}
 }

\newtheorem{lemma}[theorem]{\sptha \lemmaname}

{\theorembodyfont{\normalfont} 
 \newtheorem{proofth}{\sptha \proofname}[section]}

\newtheorem{proposition}[theorem]{\sptha \propositionname}

{\theorembodyfont{\normalfont} 
 }
{\theorembodyfont{\normalfont}%
 }

{\theorembodyfont{\normalfont} 
 }

\providecommand{\resetcountersSection}{\renewcommand{\thsec}{\thsection} 
  \renewcommand{\thseceq}{\thsectioneq} 
  \setcounter{theorem}{0} \setcounter{definition}{0} \setcounter{equation}{0}}
  
\newcommand{\captiontheorem}[2]{\textsc{ #2}.
 \addcontentsline{lth}{theorem}{{\bf #1 \contentshiftS  \protect\numberline{\thetheorem}}
 {\contentshift   : \contentshiftS #2}}}

\newcommand{\captionproofemptynocontent}[2]{}

\newcommand{\captionproofin}[2]{}

\providecommand{\vartitleadjust}{}
\renewcommand{\vartitleadjust}{}

\providecommand{\varthanks}{\ \ \standardthanksIFM}
\renewcommand{\varthanks}{\ \ \standardthanksIFM}
\providecommand{\thanksvar}{\thanks{\varthanks}}
\renewcommand{\thanksvar}{\thinspace \thanks{\ \ The authors thank Nazmul Ahsan, Marine Carrasco, Atsushi Inoue, Jan Kiviet, 
 Vinh Nguyen, Benoit Perron, Pascale Val\'ery, and Hui Jun Zhang for several useful comments. \varthanks}}

\providecommand{\vartitle}{\vartitleadjust Finite-sample distributional theory for exogeneity tests: 
  \\ weak identification, incomplete models \\ and non-Gaussian distributions}
\renewcommand{\vartitle}{\vartitleadjust Exogeneity tests, incomplete models, weak identification \\ and non-Gaussian distributions: 
  \\ invariance and finite-sample distributional theory}

\providecommand{\DokoAddress}{School of Economics, The University of Adelaide, 
  10 Pulteney Street, Adelaide, SA 5005. Tel: +618 8313 1174. 
  e-mail: firmin.dokotchatoka@adelaide.edu.au. 
  Homepage: http://www.adelaide.edu.au/directory/firmin.dokotchatoka}
\renewcommand{\DokoAddress}{School of Economics, The University of Adelaide, 
  10 Pulteney Street, Adelaide, SA 5005. Tel: +618 8313 1174; 
  e-mail: firmin.dokotchatoka@adelaide.edu.au. 
  Homepage: http://www.adelaide.edu.au/directory/firmin.dokotchatoka}
\providecommand{\varAuthors}{Firmin Doko Tchatoka\thanks{\ \ \DokoAddress} \\
  The University of Adelaide \and Jean-Marie Dufour \thanks{\ \ \DufourAddress} \\
  McGill University}
\renewcommand{\varAuthors}{Firmin Doko Tchatoka\thanks{\ \ \DokoAddress} \\
  The University of Adelaide \and Jean-Marie Dufour \thanks{\ \ \DufourAddress} \\
  McGill University}

\providecommand{\vardate}{\firstversion{\May 2007} \\
  \revised{\October 2007, \November 2008, \December 2009, \March 2010, \January 2012, \July 2013, \September 2015, \December 2015, \May 2016, \December 2016} 
  \\ \thisversion{\December 2016} \\ \compiled{\today, \texttime}}
\renewcommand{\vardate}{\firstversion{\May 2007} \\
  \revised{\October 2007, \November 2008, \December 2009, \March 2010, \January 2012, \July 2013, \September 2015, \December 2015, \May 2016, \December 2016} 
  \\ \thisversion{\December 2016} \\ \compiled{\today, \texttime}}
\renewcommand{\thefootnote}{\alph{footnote}}

\providecommand{\vardate}{\December 2016}
\renewcommand{\vardate}{\December 2016}

\begin{document}

\title{%
\vartitle%
\thanksvar }
\author{%
\varAuthors%
}
\date{%
\vardate%
}
\maketitle

\newpage 

\pagestyle{plain}\pagenumbering{roman}\setcounter{page}{1}

\begin{center}
ABSTRACT

\quad
\end{center}

\noindent We study the distribution of Durbin-Wu-Hausman (DWH) and
Revankar-Hartley (RH) tests for exogeneity from a finite-sample viewpoint,
under the null and alternative hypotheses. We consider linear structural
models with possibly non-Gaussian errors, where structural parameters may
not be identified and where reduced forms can be incompletely specified (or
nonparametric). On level control, we characterize the null distributions of
all the test statistics. Through conditioning and invariance arguments, we
show that these distributions do not involve nuisance parameters. In
particular, this applies to several test statistics for which no
finite-sample distributional theory is yet available, such as the standard
statistic proposed by Hausman (1978). The distributions of the test
statistics may be non-standard -- so corrections to usual asymptotic
critical values are needed -- but the characterizations are sufficiently
explicit to yield finite-sample (Monte-Carlo) tests of the exogeneity
hypothesis. The procedures so obtained are robust to weak identification,
missing instruments or misspecified reduced forms, and can easily be adapted
to allow for parametric non-Gaussian error distributions. We give a general
invariance result (\emph{block triangular invariance}) for exogeneity test
statistics. This property yields a convenient \emph{exogeneity canonical form%
} and a parsimonious reduction of the parameters on which power depends. In
the extreme case where no structural parameter is identified, the
distributions under the alternative hypothesis and the null hypothesis are
identical, so the power function is flat, for all the exogeneity statistics.
However, as soon as identification does not fail completely, this phenomenon
typically disappears. We present simulation evidence which confirms the
finite-sample theory. The theoretical results are illustrated with two
empirical examples: the relation between trade and economic growth, and the
widely studied problem of the return of education to earnings.

\quad


\quad

\noindent JEL \thinspace\ classification: C3; C12; C15; C52.

\newpage 

\tableofcontents%

\newpage

\listoftables%

\listoftheorems
\addcontentsline{toc}{section}{\listttheoremnameb}%

\newpage

\pagenumbering{arabic} \setcounter{section}{0} \setcounter{page}{1} 
\renewcommand{\thefootnote}{\arabic{footnote}}%

\section{ \sectitlesize Introduction \label{Sec: Introduction}}

\resetcountersSection

The literature on weak instruments is now considerable and has often focused
on inference for the coefficients of endogenous variables in so-called
\textquotedblleft instrumental-variable regressions\textquotedblright\ (or
\textquotedblleft IV regressions\textquotedblright ); see the reviews of 
\cite{Stock-Wright-Yogo(2002)}, \cite{Dufour(2003)}, \cite%
{Andrews-Stock(2007b)}, and \cite{Poskitt-Skeels(2012)}. Although research
on tests for exogeneity in IV regressions is considerable, most of these
studies either deal with cases where instrumental variables are strong (thus
leaving out issues related to weak instruments), or focus on the asymptotic
properties of exogeneity tests.\footnote{%
See, for example, \cite{Durbin(1954)}, 
\mciteAYYYY{Wu(1973)}{Wu(1974)}{Wu(1983b)}{Wu(1983)}%
, \cite{Revankar-Hartley(1973)}, \cite{Farebrother(1976)}, \cite%
{Hausman(1978)}, \cite{Revankar(1978)}, 
\mciteAYY{Dufour(1979)}{Dufour(1987)}%
, 
\mcite{Hwang(1980)}{Hwang(1980),Hwang(1985)}%
, \cite{Kariya-Hodoshima(1980)}, \cite{Hausman-Taylor(1981)}, \cite%
{Spencer-Berk(1981)}, 
\mcite{Nakamura-Nakamura(1981)}{Nakamura-Nakamura(1981),Nakamura-Nakamura(1985)}%
, \cite{Engle(1982b)}, 
\mcite{Holly(1982)}{Holly(1982),Holly(1983),Holly(1983b)}%
, \cite{Holly-Monfort(1986)}, \cite{Reynolds(1982)}, 
\mcite{Smith(1983)}{Smith(1983),Smith(1984),Smith(1985),Smith(1994)}%
, \cite{Thurman(1986)}, \cite{Rivers-Vuong(1988)}, \cite{Smith-Pesaran(1990)}%
, 
\mcite{Ruud(1984)}{Ruud(1984),Ruud(2000)}%
, 
\mcite{Newey(1985a)}{Newey(1985a),Newey(1985)}%
, 
\mcite{Davidson-Mackinnon(1985)}{Davidson-Mackinnon(1985),Davidson-Mackinnon(1985b),Davidson-Mackinnon(1989),Davidson-Mackinnon(1990),Davidson-Mackinnon(1993)}%
, \cite{Meepagala(1992)}, 
\mcite{Wong(1996)}{Wong(1996),Wong(1997)}%
,, \cite{Ahn(1997)}, \cite{Staiger-Stock(1997)}, \cite{Hahn-Hausman(2002)}, 
\cite{Baum-Schaffer-Stillman(2003)}, 
\mcite{Kiviet-Niemczyk(2006)}{Kiviet-Niemczyk(2006),Kiviet-Niemczyk(2007)}%
, \cite{Blundell-Horowitz(2007)}, \cite{Guggenberger(2008)}, \cite%
{Hahn-Ham-Moon(2010)}, \cite{Jeong-Yoon(2010)}, \cite{Chemelarova-Hill(2010)}%
, \cite{Kiviet-Pleus(2012)}, \cite{Lee-Okui(2012)}, \cite{Kiviet(2013)}, 
\mcite{Wooldridge(2014)}{Wooldridge(2014),Wooldridge(2015)}%
, \cite{Caetano(2015)}, \cite{DokoTchatoka(2015)}, \cite%
{Kabaila-Mainzer-Farchione(2015)}, and \cite{Lochner-Moretti(2015)}.} To the
best of our knowledge, there is no study on the finite-sample performance of
exogeneity tests when IVs can be arbitrary weak, when the errors may follow
a non-Gaussian distribution, or when the reduced form is incompletely
specified. The latter feature is especially important to avoid losing the
validity of the test procedure when important instruments are
\textquotedblleft left-out\textquotedblright\ when applying an exogeneity
test, as happens easily for some common \textquotedblleft
identification-robust\textquotedblright\ tests on model structural
coefficients [see \cite{Dufour-Taamouti(2007)}].

In this paper, we investigate the finite-sample properties (size and power)
of exogeneity tests of the type proposed by \cite{Durbin(1954)}, \cite%
{Wu(1973)}, \cite{Hausman(1978)}, and \cite{Revankar-Hartley(1973)},
henceforth DWH and RH tests, allowing for: (a) the possibility of
identification failure (weak instruments); (b) model errors with
non-Gaussian distributions, including heavy-tailed distributions which may
lack moments (such as the Cauchy distribution); and (c) incomplete reduced
forms (\emph{e.g.}, situations where important instruments are missing or
left out) and arbitrary heterogeneity in the reduced forms of potentially
endogenous explanatory variables.

As pointed out early by \cite{Wu(1973)}, a number of economic hypotheses can
be formulated in terms of independence (or \textquotedblleft
exogeneity\textquotedblright ) between stochastic explanatory variables and
the disturbance term in an equation. These include, for example, the
permanent income hypothesis, expected profit maximization, and recursiveness
hypotheses in simultaneous equations. Exogeneity (or \textquotedblleft
predetermination\textquotedblright ) assumptions can also affect the
\textquotedblleft causal interpretation\textquotedblright\ of model
coefficients [see \cite{Simon(1953)}, \cite{Engle-Hendry-Richard(1983)}, 
\cite{Angrist-Pischke(2009)}, \cite{Pearl(2009)}], and eventually the choice
of estimation method.

To achieve the above goals, we consider a general setup which allows for
non-Gaussian distributions and arbitrary heterogeneity in reduced-form
errors. Under the assumption that the distribution of the structural errors
(given IVs) is specified up to an unknown factor (which may depend on IVs),
we show that exact exogeneity tests can be obtained from all DWH and RH
statistics [including \cite{Hausman(1978)} statistic] through the Monte
Carlo test (MCT) method [see \cite{Dufour(2006)}]. The null distributions of
the test statistics typically depend on specific instrument values, so
\textquotedblleft critical values\textquotedblright\ should also depend on
the latter. Despite this, the MCT procedure automatically controls the level
irrespective of this complication, and thus \emph{avoids} the need to
compute \emph{critical values}. Of course, as usual, the null hypothesis is
interpreted here as the conjunction of all model assumptions (including
\textquotedblleft distributional\textquotedblright\ ones) with the
exogeneity restriction.

The finite-sample tests built in this way are also robust to weak
instruments, in the sense that they never over-reject the null hypothesis of
exogeneity even when IVs are weak. This entails that size control is
feasible in finite samples for all DWH and RH tests [including the \cite%
{Hausman(1978)} test]. All exogeneity tests considered can also be described
as identification-robust in finite samples. These conclusions stand in
contrast with ones reached by \cite[Section D]{Staiger-Stock(1997)} who
argue -- following a local asymptotic theory -- that size adjustment may not
be feasible due to the presence of nuisance parameters in the asymptotic
distribution. Of course, this underscores the fundamental difference between
a finite-sample theory and an asymptotic approximation, even when the latter
is \textquotedblleft improved\textquotedblright .

More importantly, we show that the proposed Monte Carlo test procedure
remains valid even if the right-hand-side (possibly) endogenous regressors
are heterogenous and the reduced-form model is incompletely specified
(missing instruments). Because of the latter property, we say that the DWH
and RH tests are \emph{robust to incomplete reduced forms}. For example,
robustness to incomplete reduced forms is relevant in macroeconomic models
with structural breaks in the reduced form: this shows that exogeneity tests
remain applicable without knowledge of break dates. In such contexts,
inference on the structural form may be more reliable than inference on the
reduced form. This is of great practical interest, for example, in inference
based on IV regressions and DSGE models. For further discussion of this
issue, see \cite{Dufour-Taamouti(2007)}, \cite{Dufour-Khalaf-Kichian(2013)}
and \cite{DokoTchatoka(2015b)}.

We study analytically the power of the tests and identify the crucial
parameters of the power function. In order to do this, we first prove a
general invariance property (\emph{block triangular invariance}) for
exogeneity test statistics -- a result of separate interest, \emph{e.g.} to
study how nuisance parameters may affect the distributions of exogeneity
test statistics. This property yields a convenient \emph{exogeneity
canonical form} and a parsimonious reduction of the parameters on which
power depends. In particular, we give conditions under which exogeneity
tests have no power, and conditions under which they have power. We show
formally that the tests have little power when instruments are weak. In
particular, the power of the tests cannot exceed the nominal level if all
structural parameters are completely unidentified. Nevertheless, power may
exist as soon as one instrument is strong (partial identification).

We present a Monte Carlo experiment which confirms our theoretical findings.
In particular, simulation results confirm that the MCT versions of all
exogeneity statistics considered allow one to control test size perfectly,
while usual critical values (under a Gaussian error assumption) are either
exact or conservative. The conservative property is visible in particular
when the two-stage-least-squares (2SLS) estimator of the structural error
variance is used in covariance matrices. In such cases, the MCT version of
the tests allows sizable power gains.

The results are also illustrated through two empirical examples: the
relation between trade and economic growth, and the widely studied problem
of the return of education to earnings.

The paper is organized as follows. Section \ref{Sec: Model} formulates the
model studied, and Section \ref{Sec: Exogeneity tests} describes the
exogeneity test statistics, including a number of alternative formulations (%
\emph{e.g.}, linear-regression-based interpretations) which may have
different analytical and numerical features. In Section \ref{Sec: Incomplete
models and pivotal properties}, we give general characterizations of the
finite-sample distributions of the test statistics and show how they can be
implemented as Monte Carlo tests, with either Gaussian or non-Gaussian
errors. In Section \ref{Sec: Block-triangular invariance}, we give the
general block-triangular invariance result and describe the associated
exogeneity canonical representation. Power is discussed in Section \ref{Sec:
Power}. The simulation experiment is presented in Section \ref{Sec:
Simulation experiment}, and the empirical illustration in Section \ref{Sec:
Empirical illustrations}. We conclude in Section \ref{sec: Conclusion}.
Additional details on the formulation of the different test statistics and
the proofs are supplied in Appendix.

Throughout the paper, $I_{m}$ stands for the identity matrix of order $m.\,$
For any full-column-rank $T\times m$ matrix $A,\,$ $\mathrm{\bar{P}}%
[A]=A(A^{\prime }A)^{-1}A^{\prime }$ is the projection matrix on the space
spanned by the columns of $A$, and $\mathrm{\bar{M}}[A]=I_{T}-\mathrm{\bar{P}%
}[A].$ For arbitrary $m\times m$ matrices $A$ and $B,$ the notation $A>0$
means that $A$ is positive definite (p.d.), $A\geq 0$ means $A$ is positive
semidefinite (p.s.d.), and $A\leq B$ means $B-A\geq 0$. Finally, $\Vert
A\Vert $ is the Euclidian norm of a vector or matrix, \emph{i.e.}, $\Vert
A\Vert =[\mathrm{tr}(A^{\prime }A)]^{\frac{1}{2}}.\,$

\section{\sectitlesize Framework \label{Sec: Model}}

\resetcountersSection

We consider a structural model of the form: 
\begin{equation}
y=Y\beta +X_{1}\gamma +u\,,\,  \label{eq: y StrucEqch2}
\end{equation}%
\begin{equation}
Y=g(X_{1},\,X_{2},\,X_{3},\,V,\,\bar{\Pi})\,,
\label{eq: Y non-linearRedFch2b}
\end{equation}%
where (\ref{eq: y StrucEqch2}) is a linear structural equation, $y\in 
\mathbb{R}^{T}$ is a vector of observations on a dependent variable, $Y\in 
\mathbb{R}^{T\times G}$ is a matrix of observations on (possibly) endogenous
explanatory variables which are determined by equation (\ref{eq: Y
non-linearRedFch2b}), $X_{1}\in \mathbb{R}^{T\times k_{1}}$ is a matrix of
observations on exogenous variables included in the structural equation (\ref%
{eq: y StrucEqch2}), $X_{2}\in \mathbb{R}^{T\times k_{2}}$ and $X_{3}\in 
\mathbb{R}^{T\times k_{3}}$ are matrices of observations on exogenous
variables excluded from the structural equation, $u=(u_{1},\,\ldots
\,,\,u_{T})^{\prime }\in \mathbb{R}^{T}$ is a vector of structural
disturbances, $V=\left[ V_{1},\,\ldots \,,\,V_{T}\right] ^{\prime }\in 
\mathbb{R}^{T\times G}$ is a matrix of random disturbances, $\beta \in 
\mathbb{R}^{G}$ and $\gamma \in \mathbb{R}^{k_{1}}$ are vectors of unknown
fixed structural coefficients, and $\bar{\Pi}$ is a matrix of fixed
(typically unknown) coefficients. We suppose $G\geq 1$, $k_{1}\geq 0$, $%
k_{2}\geq 0$, $k_{3}\geq 0,$ and denote:%
\begin{gather}
X=[X_{1},\,X_{2}]=[x_{1},\ldots ,\,x_{T}]^{\prime }\,,\quad \bar{X}%
=[X_{1},\,X_{2},\,X_{3}]=[\bar{x}_{1},\ldots ,\,\bar{x}_{T}]^{\prime }\,,
\label{eq: X Xbar} \\
\bar{Y}=[Y,\,X_{1}]\,,\quad Z=[Y,\,X_{1},\,X_{2}]=[z_{1},\ldots
,\,z_{T}]^{\prime }\,,\quad \bar{Z}=[Y,\,X_{1},\,X_{2},\,X_{3}]=[\bar{z}%
_{1},\ldots ,\,\bar{z}_{T}]^{\prime }\,,  \label{eq: Z Zbar} \\
U=[u,\,V]=[U_{1},\,\ldots \,,\,U_{T}]^{\prime }\ .  \label{eq: U = [u, V]}
\end{gather}%
Equation (\ref{eq: Y non-linearRedFch2b}) usually represents a reduced-form
equation for $Y$. The form of the function $g(\cdot )$ may be nonlinear or
unspecified, so model (\ref{eq: Y non-linearRedFch2b}) can be viewed as
\textquotedblleft nonparametric\textquotedblright\ or \textquotedblleft
semiparametric\textquotedblright . The inclusion of $X_{3}$ in this setup
allows for $Y$ to depend on exogenous variables not used by the exogeneity
tests. This assumption is crucial, because it characterizes the fact that we
consider here \textquotedblleft incomplete models\textquotedblright\ where
the reduced form for $Y$ may not be specified and involves unknown exogenous
variables. It is well known that several \textquotedblleft
identification-robust\textquotedblright\ tests for $\beta $ [such as those
proposed by \cite{Kleibergen(2002)} and \cite{Moreira(2003)}] are not robust
to allowing a general reduced form for $Y$ such as the one in (\ref{eq: Y
non-linearRedFch2b}); see \cite{Dufour-Taamouti(2007)} and \cite%
{DokoTchatoka(2015b)}.

We also make the following rank assumption on the matrices $[Y,\,X]\;$and $%
\big[\mathrm{\bar{P}}[X]Y,\,X_{1}\big]$:%
\begin{equation}
\lbrack Y,\,X]\;\text{and\ }\big[\,\mathrm{\bar{P}}[X]Y,\,X_{1}\big]\text{
have full-column rank with probability one }(\text{conditional on }X).
\label{eq: Rank conditions}
\end{equation}%
This (fairly standard) condition ensures that the matrices $X$, $\mathrm{%
\bar{M}}[X_{1}]Y$ and $\mathrm{\bar{M}}[X]Y$ have full column rank, hence
the unicity of the least-squares (LS) estimates when each column of $Y$ is
regressed on $X$, as well as the existence of a unique
two-stage-least-squares (2SLS) estimate for $\beta $ and $\gamma $ based on $%
X$ as the instrument matrix. Clearly, (\ref{eq: Rank conditions}) holds when 
$X$ has full column rank and the conditional distribution of $Y$ given $X$
is absolutely continuous (with respect to the Lebesgue measure).

A common additional maintained hypothesis in this context consists in
assuming that $g(\cdot )$ is a linear equation of the form%
\begin{equation}
Y=X_{1}\Pi _{1}+X_{2}\Pi _{2}+V=X\Pi +V  \label{eq: Linear reduced form}
\end{equation}%
where $\Pi _{1}\in \mathbb{R}^{k_{1}\times G}$ and $\Pi _{2}\in \mathbb{R}%
^{k_{2}\times G}$ are matrices of unknown reduced-form coefficients. In this
case, the reduced form for $y$ is 
\begin{equation}
y=X_{1}\pi _{1}+X_{2}\pi _{2}+v  \label{eq: [y,Y] Linear reduced form}
\end{equation}%
where $\pi _{1}=\gamma +\Pi _{1}\,\beta ,$ $\pi _{2}=\Pi _{2}$\thinspace $%
\beta ,$ and $v=u+V\beta $. When the errors $u$ and $V$ have mean zero
(though this assumption may also be replaced by another \textquotedblleft
location assumption\textquotedblright , such as zero medians), the usual
necessary and sufficient condition for identification of this model is 
\begin{equation}
\mathrm{rank}(\Pi _{2})=G\,.  \label{eq: rank(Pi2)=G}
\end{equation}%
If $\Pi _{2}=0,\,$ the instruments $X_{2}$ are irrelevant, and $\beta $ is
completely unidentified. If $1\leq \mathrm{rank}(\Pi _{2})<G,$ $\beta $ is
not identifiable, but some linear combinations of the elements of $\beta $
are identifiable$\,$ [see \cite{Dufour-Hsiao(2008)} and \cite%
{DokoTchatoka(2015b)}]. If $\Pi _{2}$ is close not to have full column rank [%
\emph{e.g.}, if some eigenvalues of $\Pi _{2}^{\prime }\Pi _{2}$ are close
to zero], some linear combinations of $\beta $ are ill-determined by the
data, a situation often called \textquotedblleft weak
identification\textquotedblright\ in this type of setup [see \cite%
{Dufour(2003)}, \cite{Andrews-Stock(2007b)}].

We study here, from a finite-sample viewpoint, the size and power properties
of the exogeneity tests of the type proposed by \cite{Durbin(1954)}, \cite%
{Wu(1973)}, \cite{Hausman(1978)}, and \cite{Revankar-Hartley(1973)} for
assessing the exogeneity of $Y$ in (\ref{eq: y StrucEqch2})\thinspace
-\thinspace (\ref{eq: Linear reduced form}) when: (a) instruments may be
weak; (b) $[u,\,V]$ may not follow a Gaussian distribution [\emph{e.g.},
heavy-tailed distributions which may lack moments (such as the Cauchy
distribution) are allowed]; and (c) the usual reduced-form specification (%
\ref{eq: Linear reduced form}) is misspecified, and $Y$ follows the more
general model (\ref{eq: Y non-linearRedFch2b}) which allows for omitted
instruments, an unspecified nonlinear form and heterogeneity. To achieve
this, we consider the following distributional assumptions on model
disturbances (where $\mathbb{P}[\cdot ]$ refers to the relevant probability
measure).

\begin{assumption}
\label{Ass: Reduced-form heterogeneity} 
\captiontheorem{\assumptionname}{Conditional scale model for the structural error distribution}
For some fixed vector $a$ in $\mathbb{R}^{G},$ we have: 
\begin{equation}
u=Va+e\,,  \label{eq: Linear structural error decomposition}
\end{equation}%
\begin{equation}
e=(e_{1},\,\ldots \,,\,e_{T})^{\prime }=\sigma _{1}(\bar{X})\,\varepsilon \,%
\text{,}  \label{eq: Conditional homoskedasticity}
\end{equation}%
where $\sigma _{1}(\bar{X})$ is a $($possibly random$)$ function of $\bar{X}$
such that $\mathbb{P}[\sigma _{1}(\bar{X})\neq 0\,|\,\bar{X}]=1,$ and the
conditional distribution of $\varepsilon $ given $\bar{X}$ is completely
specified.
\end{assumption}

\begin{assumption}
\label{Ass: Conditional mutual independence of e and V} 
\captiontheorem{\assumptionname}{Conditional mutual independence of  $e$ and $V$}
$V$ and $\varepsilon $ are independent, conditional on $\bar{X}$.
\end{assumption}

In the above assumptions, possible dependence between $u$ and $V$ is
parameterized by $a$, while $\varepsilon $ is independent of $V$
(conditional on $\bar{X})$, and $\sigma _{1}(\bar{X})$ is an arbitrary
(possibly random) \emph{scale parameter} which may depend on $\bar{X}$
(except for the non-degeneracy condition $\mathbb{P}[\sigma _{1}(\bar{X}%
)\neq 0\,|\,\bar{X}]=1$). So we call $a$ the \textquotedblleft endogeneity
parameter\textquotedblright\ of the model. Assumption \ref{Ass: Reduced-form
heterogeneity} is quite general and allows for heterogeneity in the
distributions of the reduced-form disturbances $V_{t},\;t=1,\ldots ,\,T.$ In
particular, the rows of $V$ need not be identically distributed or
independent. Further, non-Gaussian distributions are covered, including
heavy-tailed distributions which may lack second moments (such as the Cauchy
distribution). In such cases, $\sigma _{1}(\bar{X})^{2}$ \emph{does not
represent a variance}. Since the scale factor may be random, we can have $%
\sigma _{1}(\bar{X})=\bar{\sigma}(\bar{X},\,V,\,e).$ Of course, these
conditions hold when $u=\sigma \,\varepsilon ,$ where $\sigma $ is an
unknown positive constant and $\varepsilon $ is independent of $X$ with a
completely specified distribution. In this context, the standard Gaussian
assumption is obtained by taking: $\varepsilon \thicksim \mathrm{N}%
[0,\,I_{T}]\,.$ The distributions of $\varepsilon $ and $\sigma _{1}$ may
also depend on a subset of $\bar{X}$, such as $X=[X_{1},\,X_{2}]$. Note also
the parameter $a$ is not presumed to be identifiable, and $e$ may not be
independent of $V$ -- though this would be a reasonable additional
assumption to consider in the present context.

In this context, we consider the hypothesis that $Y$ can be treated as
independent of $u$ in (\ref{eq: y StrucEqch2}), deemed the (strict) \emph{%
exogeneity} of $Y$ with respect to $u$, so no simultaneity bias would show
up if (\ref{eq: y StrucEqch2}) is estimated by least squares. Under the
Assumptions \ref{Ass: Reduced-form heterogeneity} and \ref{Ass: Conditional
mutual independence of e and V}, $a=0$ is clearly a sufficient condition for 
$u$ and $e$ to be independent. Further, as soon as $V$ has full column rank
with probability one, $a=0$ is also necessary for the latter independence
property. This leads one to test: 
\begin{equation}
H_{0}:\,a=0\,.  \label{eq: H0: a=0}
\end{equation}%
We stress here that \textquotedblleft exogeneity\textquotedblright\ may
depend on a set of conditioning variables ($\bar{X}$), though of course we
can have cases where it does not depend on $\bar{X}$ or holds
unconditionally. The setup we consider in this paper allows for both
possibilities.

Before we move to describe tests of exogeneity, it will be useful to study
how $H_{0}$ can be reinterpreted in the more familiar language of covariance
hypotheses, provided standard second-moment assumptions are made.

\begin{assumption}
\label{Ass: Homoskedasticity} 
\captiontheorem{\assumptionname}{Homoskedasticity}
The vectors $U_{t}=[u_{t},\,V_{t}^{\prime }]^{\prime },$ $t=1,\,\ldots
\,,\,T,$ have zero means and the same $($finite$)$ nonsingular covariance
matrix:%
\begin{equation}
\mathbb{E}[U_{t}U_{t}^{^{\prime }}\,|\,\bar{X}]=\Sigma =\left[ 
\begin{array}{cc}
\sigma _{u}^{2} & \sigma _{Vu}^{^{\prime }} \\ 
\sigma _{Vu} & \Sigma _{V}%
\end{array}%
\right] >0\,,\text{\quad }t=1,\,\ldots \,,\,T.  \label{eq: covarutilde}
\end{equation}%
where $\sigma _{u}^{2}$, $\sigma _{Vu}$ and $\Sigma _{V}$ may depend on $%
\bar{X}$.
\end{assumption}

\begin{assumption}
\label{Ass: Orthogonality between e and V} 
\captiontheorem{\assumptionname}{Orthogonality between $e$ and $V$}
$\mathbb{E}[V_{t}\,e_{t}\,|\,\bar{X}]=0$, $\mathbb{E}[e_{t}\,|\,\bar{X}]=0$
and $\mathbb{E}[e_{t}^{2}\,|\,\bar{X}]=\sigma _{e}^{2}$, for $t=1,\ldots
,\,T $.
\end{assumption}

Under the above assumptions, the reduced-form disturbances 
\begin{equation}
W_{t}=[v_{t},\,V_{t}^{\prime }]^{\prime }=[u_{t}+V_{t}^{\prime }\beta
,\,V_{t}^{\prime }]^{\prime },\quad t=1,\,\ldots \,,\,T,  \label{eq: W_t}
\end{equation}%
also have a nonsingular covariance matrix (conditional on $\bar{X}$), 
\begin{equation}
\Omega =\left[ 
\begin{array}{cc}
\sigma _{u}^{2}+\beta ^{\prime }\Sigma _{V}\beta +2\beta ^{\prime }\sigma
_{Vu} & \beta ^{\prime }\Sigma _{V}+\sigma _{Vu}^{\prime } \\ 
\Sigma _{V}\beta +\sigma _{Vu} & \Sigma _{V}%
\end{array}%
\right] \,.  \label{eq: covareduerrors}
\end{equation}%
In this context, the exogeneity hypothesis of $Y$ can be formulated as 
\begin{equation}
H_{0}:\sigma _{Vu}=0\,.  \label{eq: H0delta=0}
\end{equation}%
Further, 
\begin{equation}
\sigma _{Vu}=\Sigma _{V}a\,,\quad \sigma _{u}^{2}=\sigma _{e}^{2}+a^{\prime
}\Sigma _{V}a=\sigma _{e}^{2}+\sigma _{Vu}^{\prime }\Sigma _{V}^{-1}\sigma
_{Vu}\,,  \label{eq: delta-vs-a}
\end{equation}%
so $\sigma _{Vu}=0\,\Leftrightarrow \,a=0$, and the exogeneity of $Y$ can be
assessed by testing whether $a=0.$ Note, however, that Assumptions \ref{Ass:
Homoskedasticity} and \ref{Ass: Orthogonality between e and V} will not be
needed for the results presented in this paper.

In order to study the power of exogeneity tests, it will be useful to
consider the following separability assumptions.

\begin{assumption}
\label{Ass: Endogeneity-parameter distributional separability} 
\captiontheorem{\assumptionname}{Endogeneity-parameter distributional  separability}
$\bar{\Pi}$ is not restricted by $a$, and the conditional distribution of $%
[V,$\thinspace $e]$ given $\bar{X}$ does not depend on the parameter $a.$
\end{assumption}

\begin{assumption}
\label{Ass: Reduced-form error separability for Y} 
\captiontheorem{\assumptionname}{Reduced-form linear separability for $Y$}
$Y$ satisfies the equation%
\begin{equation}
Y=g(X_{1},\,X_{2},\,X_{3},\,\bar{\Pi})+V\,.  \label{eq: Y = g + V}
\end{equation}
\end{assumption}

Assumption \ref{Ass: Endogeneity-parameter distributional separability}
means that the distributions of $V$ and $e$ do not depend on the endogeneity
parameter $a$, while Assumption \ref{Ass: Reduced-form error separability
for Y} means that $V$ can be linearly separated from $g(X_{1},\,X_{2},%
\,X_{3},\,\bar{\Pi})$ in (\ref{eq: Y non-linearRedFch2b}).

\section{\sectitlesize Exogeneity tests \label{Sec: Exogeneity tests}}

\resetcountersSection

We consider the four statistics proposed by \cite{Wu(1973)} $[\mathcal{T}%
_{l},\,l=1,2,3,4],$ the statistic proposed by \cite{Hausman(1978)} $[%
\mathcal{H}_{1}]$ as well as some variants $[\mathcal{H}_{2},\,\mathcal{H}%
_{3}]$ occasionally considered in the literature [see, for example, \cite%
{Hahn-Ham-Moon(2010)}], and the test suggested by \cite[RH]%
{Revankar-Hartley(1973)} $[\mathcal{R]}$. These statistics can be formulated
in two alternative ways: (1) as Wald-type statistics for the difference
between the two-stage least squares (2SLS) and the ordinary least squares
(OLS) estimators of $\beta $ in equation (\ref{eq: y StrucEqch2}), where
different statistics are obtained by changing the covariance matrix; or (2)
a $F$-type significance test on the coefficients of an \textquotedblleft
extended\textquotedblright\ version of (\ref{eq: y StrucEqch2}), so the
different statistics can be written in terms of the difference between
restricted and unrestricted residual sum of squares.

\subsection{\sectitlesize Test statistics \label{Sec: Test statistics}}

We now give a unified presentation of different available DWH-type
statistics. The test statistics considered can be written as follows:%
\begin{gather}
\mathcal{T}_{i}=\kappa _{i}(\tilde{\beta}-\hat{\beta})^{\prime }\tilde{\Sigma%
}_{i}^{-1}(\tilde{\beta}-\hat{\beta})\,,\text{\quad }i=1,\,2,\,3,\,4,\,
\label{eq: statT1} \\
\mathcal{H}_{j}=T(\tilde{\beta}-\hat{\beta})^{\prime }\hat{\Sigma}_{j}^{-1}(%
\tilde{\beta}-\hat{\beta})\,,\text{\quad }j=1,\,2,\,3,
\label{eq: statHausm1} \\
\mathcal{R}=\kappa _{R}\,\big(y^{\prime }\,\Psi _{R}\,y\,/\,\hat{\sigma}%
_{R}^{2}\big)\,,\,  \label{eq: statRH}
\end{gather}%
where $\hat{\beta}$ and $\tilde{\beta}$ are the ordinary least squares (OLS)
estimator and two-stage least squares (2SLS) estimators of $\beta $, \emph{%
i.e.} 
\begin{equation}
\hat{\beta}=(Y^{\prime }M_{1}Y)^{-1}Y^{\prime }M_{1}\,y\,,  \label{eq: b_OLS}
\end{equation}%
\begin{equation}
\tilde{\beta}=[(PY)^{\prime }M_{1}(PY)]^{-1}(PY)^{\prime
}M_{1}\,y=(Y^{\prime }N_{1}Y)^{-1}Y^{\prime }N_{1}\,y\,,  \label{eq: b_2SLS}
\end{equation}%
while we denote $\hat{\gamma}$ and $\tilde{\gamma}$ the corresponding OLS
and 2SLS estimators of $\gamma $, and%
\begin{gather}
M_{1}=\mathrm{\bar{M}}[X_{1}]\,,\quad P=\mathrm{\bar{P}}[X]\,,\quad M=%
\mathrm{\bar{M}}[X]=I_{T}-\mathrm{\bar{P}}[X]\,,\quad N_{1}=M_{1}P\,,
\label{eq: M_1} \\
\tilde{\Sigma}_{1}=\tilde{\sigma}_{1}^{2}\hat{\Delta}\,,\text{\quad }\tilde{%
\Sigma}_{2}=\tilde{\sigma}_{2}^{2}\hat{\Delta}\,,\text{\quad }\tilde{\Sigma}%
_{3}=\tilde{\sigma}^{2}\hat{\Delta}\,,\text{\quad }\tilde{\Sigma}_{4}=\hat{%
\sigma}^{2}\hat{\Delta}\,,  \label{eq: SigmaTilde} \\
\hat{\Sigma}_{1}=\tilde{\sigma}^{2}\hat{\Omega}_{IV}^{-1}-\hat{\sigma}^{2}%
\hat{\Omega}_{LS}^{-1}\,,\text{\quad }\hat{\Sigma}_{2}=\tilde{\sigma}^{2}%
\hat{\Delta}\,,\text{\quad }\hat{\Sigma}_{3}=\hat{\sigma}^{2}\hat{\Delta}%
\,,\,  \label{eq: SigmaHat} \\
\hat{\Delta}=\hat{\Omega}_{IV}^{-1}-\hat{\Omega}_{LS}^{-1}\,,\quad \hat{%
\Omega}_{IV}=\frac{1}{T}Y^{\prime }N_{1}Y\,,\text{\quad }\hat{\Omega}_{LS}=%
\frac{1}{T}Y^{\prime }M_{1}Y,  \label{eq: SigmaRhat} \\
\hat{u}=y-Y\hat{\beta}-X_{1}\hat{\gamma}=M_{1}(y-Y\hat{\beta})\,,\quad 
\tilde{u}=y-Y\tilde{\beta}-X_{1}\tilde{\gamma}=M_{1}(y-Y\tilde{\beta})\,,
\label{eq: u^hat u^tilde} \\
\hat{\sigma}^{2}=\frac{1}{T}\hat{u}^{\prime }\hat{u}=\frac{1}{T}(y-Y\hat{%
\beta})^{\prime }M_{1}(y-Y\hat{\beta})\,,\quad \tilde{\sigma}^{2}=\frac{1}{T}%
\tilde{u}^{\prime }\tilde{u}=\frac{1}{T}(y-Y\tilde{\beta})^{\prime }M_{1}(y-Y%
\tilde{\beta})\,,  \label{eq: sigma hat} \\
\tilde{\sigma}_{1}^{2}=\frac{1}{T}(y-Y\tilde{\beta})^{\prime }N_{1}(y-Y%
\tilde{\beta})=\tilde{\sigma}^{2}-\tilde{\sigma}_{e}^{2}\,,\quad \tilde{%
\sigma}_{e}^{2}=\frac{1}{T}(y-Y\tilde{\beta})^{\prime }M(y-Y\tilde{\beta})\,,
\label{eq: sigma1hat} \\
\tilde{\sigma}_{2}^{2}=\hat{\sigma}^{2}-(\tilde{\beta}-\hat{\beta})^{\prime }%
\hat{\Delta}^{-1}(\tilde{\beta}-\hat{\beta})\,,  \label{eq: sigma2hat} \\
\Psi _{R}=\frac{1}{T}\{\mathrm{\bar{M}}[\bar{Y}]-\mathrm{\bar{M}}%
[Z]\}\,,\quad \hat{\sigma}_{R}^{2}=y^{\prime }\Lambda _{_{R}}\,y\,,\quad
\Lambda _{_{R}}=\frac{1}{T}\mathrm{\bar{M}}[Z]\,,  \label{eq:  Psi_R}
\end{gather}%
$\kappa _{1}=(k_{2}-G)/G,\,$ $\,\kappa _{2}=(T-k_{1}-2G)/G,\,$ $\,\kappa
_{3}=\kappa _{4}=T-k_{1}-G,\,$ and $\kappa _{R}=(T-k_{1}-k_{2}-G)/k_{2}.$
Here, $\hat{u}$ is the vector of OLS residuals from equation (\ref{eq: y
StrucEqch2}) and $\hat{\sigma}^{2}$ is the corresponding OLS-based estimator
of $\sigma _{u}^{2}$ (without correction for degrees of freedom), while $%
\tilde{u}$ is the vector of the 2SLS residuals and $\tilde{\sigma}^{2}$ the
usual 2SLS-based estimator of $\sigma _{u}^{2}$; $\tilde{\sigma}_{1}^{2},\,$ 
$\tilde{\sigma}_{2}^{2},\,$ $\tilde{\sigma}_{e}^{2}\,$ and $\hat{\sigma}%
_{R}^{2}$ may be interpreted as alternative IV-based scaling factors. Note
also that $P_{1}\,P=P\,P_{1}=P_{1}$, $M_{1}\,M=M\,M_{1}=M$, and%
\begin{eqnarray}
N_{1}
&=&M_{1}P=P\,M_{1}=P\,M_{1}P=M_{1}P\,M_{1}=N_{1}M_{1}=M_{1}N_{1}=N_{1}N_{1} 
\notag \\
&=&M_{1}-M=P-P_{1}=\mathrm{\bar{P}}[X]-\mathrm{\bar{P}}[X_{1}]=\mathrm{\bar{P%
}}[M_{1}X_{2}].  \label{eq: N_1 forms}
\end{eqnarray}%
Each one of the corresponding tests rejects $H_{0}$ when the statistic is
\textquotedblleft large\textquotedblright . We also set 
\begin{equation}
\hat{V}=:MY\,,\quad \hat{\Sigma}_{V}=:\frac{1}{T}\hat{V}^{\prime }\hat{V}\,,
\end{equation}%
\emph{i.e.} $\hat{\Sigma}_{V}$ is the usual sample covariance matrix of the
LS residuals $(\hat{V})$ from the reduced-form linear model (\ref{eq: Linear
reduced form}).

The tests differ through the use of different \textquotedblleft covariance
matrix\textquotedblright\ estimators. $\mathcal{H}_{1}$ uses two different
estimators of $\sigma _{u}^{2}$, while the others resort to a single scaling
factor (or estimator of $\sigma _{u}^{2}$). We think the expressions given
here for $\mathcal{T}_{l},\,l=1,\,2,\,3,\,4,\,$ in (\ref{eq: statT1}) are
easier to interpret than those of \cite{Wu(1973)}, and show more clearly the
relation with Hausman-type tests. The statistic $\mathcal{H}_{1}$ can be
interpreted as the statistic proposed by \cite{Hausman(1978)}, while $%
\mathcal{H}_{2}$ and $\mathcal{H}_{3}$ are sometimes interpreted as variants
of $\mathcal{H}_{1}$ [see \cite{Staiger-Stock(1997)} and \cite%
{Hahn-Ham-Moon(2010)}]. We use the above notations to better see the
relation between Hausman-type tests and Wu-type tests. In particular, $%
\tilde{\Sigma}_{3}=\hat{\Sigma}_{2}$ and $\tilde{\Sigma}_{4}=\hat{\Sigma}%
_{3},$ so $\mathcal{T}_{3}=(\kappa _{3}/T)\mathcal{H}_{2}$ and $\mathcal{T}%
_{4}=(\kappa _{4}/T)\mathcal{H}_{3}.$ Further, $\mathcal{T}_{4}$ is a
nonlinear monotonic transformation of $\mathcal{T}_{2}$:%
\begin{equation}
\mathcal{T}_{4}=\frac{\kappa _{4}\,\mathcal{T}_{2}}{\mathcal{T}_{2}+\kappa
_{2}}=\frac{\kappa _{4}}{(\kappa _{2}/\mathcal{T}_{2})+1}\,.
\label{eq: link T_4 T_2}
\end{equation}%
Despite these relations, the tests based on $\mathcal{T}_{3}$ and $\mathcal{H%
}_{2}$ are equivalent only if exact critical values are used, and similarly
for the pairs $(\mathcal{T}_{4},$\thinspace $\mathcal{H}_{3})$ and $(%
\mathcal{T}_{2},$\thinspace $\mathcal{T}_{4}$). We are not aware of a simple
equivalence between $\mathcal{H}_{1}$ and $\mathcal{T}_{i},\,i=1,\,2,\,3,%
\,4,\,$ and similarly between $\mathcal{T}_{1}$ and $\mathcal{H}%
_{j},\,j=1,\,2,\,3.\,$

The link between the formulation of \cite{Wu(1973)} and the one above is
discussed in Appendix \ref{sec: Wu and Hausman test statistics}.\footnote{%
When the errors $U_{1},\ldots ,\,U_{T}$ are i.i.d. Gaussian [in which case
Assumptions \ref{Ass: Homoskedasticity} and \ref{Ass: Orthogonality between
e and V} hold], the $\mathcal{T}_{2}$ test of \cite{Wu(1973)} can also be
interpreted as the LM test of $a=0$; see \cite{Smith(1983)} and \cite%
{Engle(1982b)}.} Condition (\ref{eq: Rank conditions}) entails that $\hat{%
\Omega}_{IV}$, $\hat{\Omega}_{LS}$ and $\hat{\Sigma}_{V}$ are (almost
surely) nonsingular, which in turn implies that $\hat{\Delta}$ is
invertible; see Lemma \ref{Th: Difference of matrix inverses} in Appendix.
In particular, it is of interest to observe that%
\begin{eqnarray}
\hat{\Delta}^{-1} &=&\hat{\Omega}_{IV}+\hat{\Omega}_{IV}(\hat{\Omega}_{LS}-%
\hat{\Omega}_{IV})^{-1}\hat{\Omega}_{IV}=\hat{\Omega}_{IV}+\hat{\Omega}_{IV}%
\hat{\Sigma}_{V}^{-1}\hat{\Omega}_{IV}=\hat{\Omega}_{LS}\,\hat{\Sigma}%
_{V}^{-1}\,\hat{\Omega}_{LS}-\hat{\Omega}_{LS}  \notag \\
&=&\frac{1}{T}Y^{\prime }N_{1}\left[ I_{T}+Y(Y^{\prime }MY)^{-1}Y^{\prime }%
\right] N_{1}Y=\frac{1}{T}Y^{\prime }M_{1}[Y(Y^{\prime }MY)^{-1}Y^{\prime
}-I_{T}]M_{1}Y\,.  \label{eq: Inv(Deltahat)}
\end{eqnarray}%
from which we see easily that $\hat{\Delta}^{-1}$ is positive definite.
Further, $\hat{\Delta}^{-1}$ only depends on the least-squares residuals $%
M_{1}Y$ and $MY$ from the regressions of $Y$ on $X_{1}$ and $X$
respectively, and $\hat{\Delta}^{-1}$ can be bounded as follows: 
\begin{equation}
\hat{\Omega}_{IV}\leq \hat{\Delta}^{-1}\leq \hat{\Omega}_{LS}\,\hat{\Sigma}%
_{V}^{-1}\,\hat{\Omega}_{LS}
\end{equation}%
so that%
\begin{equation}
(\tilde{\beta}-\hat{\beta})^{\prime }\,\hat{\Omega}_{IV}\,(\tilde{\beta}-%
\hat{\beta})\leq (\tilde{\beta}-\hat{\beta})^{\prime }\,\hat{\Delta}^{-1}\,(%
\tilde{\beta}-\hat{\beta})\leq (\tilde{\beta}-\hat{\beta})^{\prime }\,\hat{%
\Omega}_{LS}\,\hat{\Sigma}_{V}^{-1}\,\hat{\Omega}_{LS}\,(\tilde{\beta}-\hat{%
\beta})\,.
\end{equation}%
To the best of our knowledge, the additive expressions in (\ref{eq:
Inv(Deltahat)}) are not available elsewhere.

Finite-sample distributional results are available for $\mathcal{T}_{1},$ $%
\mathcal{T}_{2}$ and $\mathcal{R}$ when the disturbances $u_{t}$ are i.i.d.
Gaussian. If $u\sim N[0,\,\sigma ^{2}I_{T}]$ and $X$ is independent of $u$,
we have: 
\begin{equation}
\mathcal{T}_{1}\,{\sim }\,F(G,\,k_{2}-G)\,,\text{\quad }\mathcal{T}_{2}\,{%
\sim }\,F(G,\,T-k_{1}-2G)\,,\text{\quad }\mathcal{R}\,{\sim }%
\,F(k_{2},\,T-k_{1}-k_{2}-G)\,,  \label{eq: T_1 T_2 Null Gaussian}
\end{equation}%
under the null hypothesis of exogeneity. Furthermore, for large samples, we
have under the null hypothesis (along with standard asymptotic regularity
conditions): 
\begin{equation*}
\mathcal{H}_{i}\overset{L}{\rightarrow }\chi ^{2}(G)\,,\,i=1,\,2,\,3\,\;%
\text{and\ }\mathcal{T}_{l}\overset{L}{\rightarrow }\chi ^{2}(G)\,,\,l=3,\,4,
\end{equation*}%
when $\mathrm{rank}(\Pi _{2})=G$.

Finite-sample distributional results are not available in the literature for 
$\mathcal{H}_{i}$, $\,i=1,\,2,\,3\,$ and $\mathcal{T}_{l},$ $\,l=3,\,4$,
even when errors are Gaussian and usual full identification assumptions are
made. Of course, the same remark applies when usual conditions for
identification fail [$\text{rank}(\Pi _{2})<G$] or get close to do so --%
\emph{\ e.g.}, some eigenvalues of $\Pi _{2}^{\prime }\Pi _{2}$ are close to
zero (weak identification) -- and disturbances may not be Gaussian. This
paper provides a formal characterization of the size and power of the tests
when IVs may be arbitrary weak, with and without Gaussian errors.

\subsection{\sectitlesize Regression-based formulations of exogeneity
statistics \label{Sec: Regression-based formulations of exogeneity
statistics}}

We now show that all the above test statistics can be computed from
relatively simple linear regressions, which may be analytically revealing
and computationally convenient. We consider again the regression of $u$ on $%
V $ in (\ref{eq: Linear structural error decomposition}): 
\begin{equation}
u=Va+e\,  \label{eq: StructPojVhat}
\end{equation}%
for some constant vector $a\in \mathbb{R}^{G}$, where $e$ has mean zero and
variance $\sigma _{e}^{2},$ and is uncorrelated with $V$ and $X$. We can
write the structural equation (\ref{eq: y StrucEqch2}) in three different
ways as follows: 
\begin{eqnarray}
y &=&Y\beta +X_{1}\gamma +\hat{V}a+e_{\ast }=\hat{Z}\theta +e_{\ast }\,,
\label{eq: StructVhat} \\
y &=&\hat{Y}\beta +X_{1}\gamma +\hat{V}b+e_{\ast }=Z_{\ast }\theta _{\ast
}+e_{\ast }\,,  \label{eq: Struct(Yhat-Vhat)} \\
y &=&Yb+X_{1}\bar{\gamma}+X_{2}\bar{a}+e=\bar{Z}_{\ast }\bar{\theta}+e\,,
\label{eq: StructYZ}
\end{eqnarray}%
where%
\begin{equation}
\hat{Z}=[Y,\,X_{1},\,\hat{V}]\,,\;\theta =(\beta ^{\,\prime },\,\gamma
^{\,\prime },\,a^{\,\prime })^{\prime },\;Z_{\ast }=[\hat{Y},\,X_{1},\,\hat{V%
}]\,,\;\theta _{\ast }=(\beta ^{\,\prime },\,\gamma ^{\,\prime
},\,b^{\,\prime })^{\prime },\;\bar{Z}_{\ast }=[Y,\,X_{1},\,X_{2}]\,,
\end{equation}%
\begin{equation}
\bar{\theta}=(b^{\,\prime },\,\bar{\gamma}^{\,\,\prime },\,\bar{a}%
^{\,\,\prime })^{\prime },\;b=\beta +a,\;\bar{\gamma}=\gamma -\Pi
_{1}\,a,\quad \bar{a}=-\Pi _{2}\,\,a\,,
\end{equation}%
\begin{equation}
\hat{Y}=\mathrm{\bar{P}}[X]\,Y,\quad \hat{V}=\mathrm{\bar{M}}[X]\,Y\,,\quad
e_{\ast }=\mathrm{\bar{P}}[X]\,Va+e\,.
\end{equation}%
Clearly, $\beta =b\,$ if and only if $a=0.$ Equations (\ref{eq:
StructPojVhat})\thinspace -\thinspace (\ref{eq: StructYZ}) show that the
endogeneity of $Y$ in (\ref{eq: y StrucEqch2})\thinspace -\thinspace (\ref%
{eq: Linear reduced form}) can be interpreted as an omitted-variable problem
[for further discussion of this view, see 
\mcite{Dufour(1979)}{Dufour(1979),Dufour(1987)}
and \cite{Doko-Dufour(2014)}]. The inclusion of $\hat{V}$ in equations (\ref%
{eq: StructVhat})\thinspace -\thinspace (\ref{eq: Struct(Yhat-Vhat)}) may
also be interpreted as an application of control function methods [see \cite%
{Wooldridge(2015)}]. We also consider the intermediate regression:%
\begin{equation}
y-Y\tilde{\beta}=X_{1}\bar{\gamma}+X_{2}\bar{a}+e_{\ast \ast }=X\theta
_{\ast \ast }+e_{\ast \ast }\,  \label{eq: y - Xbtilde}
\end{equation}%
where $\tilde{\beta}$ is the 2SLS estimator of $\beta $.

Let $\hat{\theta}\,$ be the OLS estimator of $\theta $ and $\hat{\theta}%
^{0}\,$ the restricted OLS estimator of $\theta $ under the constraint $%
H_{0}:a=0\,$ [in (\ref{eq: StructVhat})], $\,\hat{\theta}_{\ast }\,$ the OLS
estimator of $\theta _{\ast }$ and $\hat{\theta}_{\ast }^{0}\,$ the
restricted OLS estimate of $\theta _{\ast }$ under $H_{0}^{\ast }:\beta =b\,$
[in (\ref{eq: Struct(Yhat-Vhat)})], $\check{\theta}$ the OLS estimate of $%
\bar{\theta}$ and $\,\check{\theta}^{0}$ the restricted OLS estimate of $%
\bar{\theta}$ under $\,\bar{H}_{0}:\bar{a}=0\,$[in (\ref{eq: StructYZ})].
Similarly, the OLS estimate of $\theta _{\ast \ast }$ based on (\ref{eq: y -
Xbtilde}) is denoted $\hat{\theta}_{\ast \ast },$ while $\hat{\theta}_{\ast
\ast }^{0}$ represents the corresponding restricted estimate under $\,\bar{H}%
_{0}:\bar{a}=0$. The sum of squared error functions associated with (\ref%
{eq: StructVhat})\thinspace -\thinspace (\ref{eq: StructYZ}) are denoted: 
\begin{gather}
S(\theta )=\Vert y-\hat{Z}\theta \Vert ^{2},\quad S_{\ast }(\theta _{\ast
})=\Vert y-Z_{\ast }\theta _{\ast }\Vert ^{2},\quad \bar{S}(\bar{\theta}%
)=\Vert y-\bar{Z}_{\ast }\bar{\theta}\Vert ^{2}\,,
\label{eq: SumsquareResiduals} \\
\tilde{S}(\theta _{\ast \ast })=\Vert y-Y\tilde{\beta}-X\theta _{\ast \ast
}\Vert ^{2}\,.
\end{gather}%
Using $Y=\hat{Y}+\hat{V}$, we see that:%
\begin{equation}
S(\hat{\theta})=S_{\ast }(\hat{\theta}_{\ast })=\bar{S}(\check{\theta}%
^{0})\,,\quad S(\hat{\theta}^{0})=S_{\ast }(\hat{\theta}_{\ast }^{0})=\tilde{%
S}(\hat{\theta}_{\ast \ast }^{0})\,,
\end{equation}%
\begin{equation}
S(\hat{\theta})=T\,\tilde{\sigma}_{2}^{2}\,\,,\quad S(\hat{\theta}^{0})=T\,%
\hat{\sigma}^{2}\,,\quad S_{\ast }(\hat{\theta}_{\ast }^{0})=T\,\tilde{\sigma%
}^{2}\,,\quad \tilde{S}(\hat{\theta}_{\ast \ast })=T\,\tilde{\sigma}%
_{e}^{2}\,.
\end{equation}%
We then get the following expressions for the statistics in (\ref{eq: statT1}%
)\thinspace -\thinspace (\ref{eq: statRH}): 
\begin{gather}
\mathcal{T}_{1}=\kappa _{1}\,\left( \frac{S(\hat{\theta}^{0})-S(\hat{\theta})%
}{S_{\ast }(\hat{\theta}_{\ast }^{0})-\tilde{S}(\hat{\theta}_{\ast \ast })}%
\right) =\kappa _{1}\,\left( \frac{S(\hat{\theta}^{0})-S(\hat{\theta})}{%
\tilde{S}(\hat{\theta}_{\ast \ast }^{0})-\tilde{S}(\hat{\theta}_{\ast \ast })%
}\right) \,,  \label{eq: T1T2auxiliary} \\
\mathcal{T}_{2}=\kappa _{2}\,\left( \frac{S(\hat{\theta}^{0})-S(\hat{\theta})%
}{S(\hat{\theta})}\right) \,,\quad \mathcal{T}_{3}=\kappa _{3}\,\left( \frac{%
S(\hat{\theta}^{0})-S(\hat{\theta})}{S_{\ast }(\hat{\theta}_{\ast }^{0})}%
\right) \,,\quad \mathcal{T}_{4}=\kappa _{4}\,\left( \frac{S(\hat{\theta}%
^{0})-S(\hat{\theta})}{S(\hat{\theta}^{0})}\right) \,,
\label{eq: T3T4auxiliary} \\
\mathcal{H}_{2}=T\,\left( \frac{S(\hat{\theta}^{0})-S(\hat{\theta})}{S_{\ast
}(\hat{\theta}_{\ast }^{0})}\right) \,,\quad \mathcal{H}_{3}=T\,\left( \frac{%
S(\hat{\theta}^{0})-S(\hat{\theta})}{S(\hat{\theta}^{0})}\right) \,,
\label{eq: H2H3auxiliary} \\
\mathcal{R}=\kappa _{R}\,[\bar{S}(\check{\theta}^{0})-\bar{S}(\check{\theta}%
)]/\bar{S}(\check{\theta})\,.  \label{eq: statRevreg}
\end{gather}%
Details on the derivation of the above formulas are given in Appendix \ref%
{sec: Appendix : Regression Formula}.

(\ref{eq: H2H3auxiliary})\thinspace -\thinspace (\ref{eq: statRevreg})
provide simple regression formulations of the DWH and RH statistics in terms
of restricted and unrestricted sum of squared errors in linear regressions.
However, we did not find such a simple expression for the Hausman statistic $%
\mathcal{H}_{1}$. While DWH-type tests consider the null hypothesis $%
H_{0}:a=0,\,$ the RH test focuses on the null hypothesis $H_{0}^{\ast }:\,%
\bar{a}=-\Pi _{2}$\thinspace $a=0.\,$ If $\text{rank}(\Pi _{2})=G\,,$ we
have: $a=0$ if and only if $\bar{a}=0.$ However, if $\mathrm{rank}(\Pi
_{2})<G,\,$ $\bar{a}=0$ does not imply $a=0$: $H_{0}$ entails $H_{0}^{\ast }$%
, but the converse does not hold in this case.

The regression interpretation of the $\mathcal{T}_{2}$ and $\mathcal{H}_{3}$
statistics was mentioned earlier in 
\mciteAYY{Dufour(1979)}{Dufour(1987)}
and \cite{Nakamura-Nakamura(1981)}. The $\mathcal{R}$ statistic was also
derived as a standard regression test by \cite{Revankar-Hartley(1973)}. To
our knowledge, the other regression interpretations given here are not
available elsewhere.

\section{\sectitlesize Incomplete models and pivotal properties \label{Sec:
Incomplete models and pivotal properties}}

\resetcountersSection

In this section, we study the finite-sample null distributions of DWH-type
and RH exogeneity tests under Assumption \ref{Ass: Reduced-form
heterogeneity}, allowing for the possibility of identification failure (or
weak identification) and model incompleteness. The proofs of these results
rely on two lemmas of independent interest (Lemmas \ref{Th: Properties of
exogeneity statistics components} - \ref{Th: Properties of weighting
matrices in exogeneity statistics}) given in Appendix.

\subsection{\sectitlesize Distributions of test statistics under exogeneity 
\label{Sec: Distributions of test statistics under exogeneity}}

We first show that the exogeneity test statistics in (\ref{eq: statT1}%
)\thinspace -\thinspace (\ref{eq: statRH} ) can be rewritten as follows,
irrespective whether the null hypothesis holds or not.

\begin{proposition}
\label{Th: Quadratic-form representations of exogeneity statistics} 
\captiontheorem{\propositionname}{Quadratic-form representations of exogeneity statistics}
The exogeneity test statistics in $(\ref{eq: statT1})$\thinspace -\thinspace 
$(\ref{eq: statRH})$ can be expressed as follows: 
\begin{gather}
\mathcal{T}_{l}=\kappa _{l}\,\left( \frac{y^{\prime }\,\Psi _{_{0}}\,y}{%
y^{\prime }\,\Lambda _{l}\,y}\right) \,,\;\text{for\ }l=1,\,2,\,3,\,4,\,
\label{eq: T_l y} \\
\mathcal{H}_{1}=T\,(y^{\prime }\,\Psi _{_{1}}[y]\,y)\,=T\,(C_{1}y)^{\prime
}\,\big[(y^{\prime }\Lambda _{3}\,y)\,\hat{\Omega}_{IV}^{-1}-(y^{\prime
}\Lambda _{4}\,y)\,\hat{\Omega}_{LS}^{-1}\big]^{-1}\,(C_{1}y)\,,\,
\label{eq: H_1 y} \\
\mathcal{H}_{2}=T\,\left( \frac{y^{\prime }\,\Psi _{_{0}}\,y}{y^{\prime
}\,\Lambda _{3}\,y}\right) \,,\quad \mathcal{H}_{3}=T\,\left( \frac{%
y^{\prime }\,\Psi _{_{0}}\,y}{y^{\prime }\,\Lambda _{4}\,y}\right) \,,\quad 
\mathcal{R}=\kappa _{_{R}}\,\left( \frac{y^{\prime }\,\Psi _{R}\,y}{%
y^{\prime }\,\Lambda _{_{R}}\,y}\right) \,,  \label{eq:H_2 y}
\end{gather}%
where 
\begin{gather}
\Lambda _{1}=\frac{1}{T}N_{1}\,\mathrm{\bar{M}}[N_{1}Y]\,N_{1}\,,\quad
\Lambda _{2}=M_{1}\left( \frac{1}{T}\mathrm{\bar{M}}[M_{1}Y]-\Psi
_{_{0}}\right) M_{1}\,,  \label{eq: Lambda_1} \\
\Lambda _{3}=\frac{1}{T}M_{1}\,N_{2}^{\prime }N_{2}M_{1}\,,\quad \Lambda
_{4}=\frac{1}{T}\mathrm{\bar{M}}[\bar{Y}]=\frac{1}{T}M_{1}\mathrm{\bar{M}}%
[M_{1}Y]M_{1}\,,  \label{eq: Lambda_3} \\
\Psi _{_{1}}[y]=C_{1}^{\prime }\hat{\Sigma}_{1}^{-1}C_{1}=C_{1}^{\prime }%
\big[(y^{\prime }\Lambda _{3}\,y)\hat{\Omega}_{IV}^{-1}-(y^{\prime }\Lambda
_{4}\,y)\hat{\Omega}_{LS}^{-1}\big]^{-1}C_{1}\,,  \label{eq: Psi1}
\end{gather}%
and $\Psi _{_{0}}$, $B_{2}$, $C_{1}$, $\Psi _{R}$ and $\Lambda _{_{R}}$ are
defined as in Lemma \ref{Th: Properties of exogeneity statistics components}.
\end{proposition}

The following theorem characterizes the distributions of all exogeneity
statistics under the null hypothesis of exogeneity ($H_{0}:\,a=0$).

\begin{theorem}
\label{Th: Null distributions of exogeneity test statistics} 
\captiontheorem{\theoremname}{Null distributions of exogeneity statistics}
Under the model described by $(\ref{eq: y StrucEqch2})$\thinspace - $(\ref%
{eq: Rank conditions})$, suppose Assumption \ref{Ass: Reduced-form
heterogeneity} holds. If $H_{0}:\,a=0$ also holds, then the test statistics
defined in $(\ref{eq: statT1})$\thinspace -\thinspace $(\ref{eq: statRH})$
have the following representations:%
\begin{gather}
\mathcal{T}_{l}=\kappa _{l}\,\left( \frac{\varepsilon ^{\prime }\,\Psi
_{_{0}}\,\varepsilon }{\varepsilon ^{\prime }\,\Lambda _{l}\,\varepsilon }%
\right) \,,\;\text{for\ }l=1,\,2,\,3,\,4,\,  \label{eq: T_l e} \\
\mathcal{H}_{1}=T\,(\varepsilon ^{\prime }\,\Psi _{_{1}}[\varepsilon
]\,\varepsilon )=T\,(C_{1}\varepsilon )^{\prime }\,\big[(\varepsilon
^{\prime }\Lambda _{3}\,\varepsilon )\,\hat{\Omega}_{IV}^{-1}-(\varepsilon
^{\prime }\Lambda _{4}\,\varepsilon )\,\hat{\Omega}_{LS}^{-1}\big]%
^{-1}\,(C_{1}\varepsilon )\,,  \label{eq: H_i e} \\
\mathcal{H}_{2}=T\,\left( \frac{\varepsilon ^{\prime }\,\Psi
_{_{0}}\,\varepsilon }{\varepsilon ^{\prime }\,\Lambda _{3}\,\varepsilon }%
\right) \,,\quad \mathcal{H}_{3}=T\,\left( \frac{\varepsilon ^{\prime
}\,\Psi _{_{0}}\,\varepsilon }{\varepsilon ^{\prime }\,\Lambda
_{4}\,\varepsilon }\right) \,,\quad \mathcal{R}=\kappa _{_{R}}\,\left( \frac{%
\varepsilon ^{\prime }\,\Psi _{R}\,\varepsilon }{\varepsilon ^{\prime
}\,\Lambda _{_{R}}\,\varepsilon }\right) \,,
\end{gather}%
where $\Psi _{_{0}}$, $\Lambda _{1},\ldots ,\,\Lambda _{4}$, $\Psi _{_{1}}$, 
$\Psi _{R}$ and $\Lambda _{_{R}}$ are defined as in Proposition \ref{Th:
Quadratic-form representations of exogeneity statistics}. If Assumption \ref%
{Ass: Conditional mutual independence of e and V} also holds, the
distributions of the test statistics $\mathcal{T}_{1}$, $\mathcal{T}_{2}$, $%
\mathcal{T}_{3}$, $\mathcal{T}_{4}$, $\mathcal{H}_{1}$, $\mathcal{H}_{2}$, $%
\mathcal{H}_{3}$ and $\mathcal{R}$, conditional on $\bar{X}$ and $Y,$ only
depend on the conditional distribution of $\varepsilon $ given $\bar{X}$, as
specified in Assumption \ref{Ass: Reduced-form heterogeneity}, and the
values of $Y$ and $X$.
\end{theorem}

The last statement of Theorem \ref{Th: Null distributions of exogeneity test
statistics} comes from the fact that the weighting matrices defined in (\ref%
{eq: Lambda_1})\thinspace -\thinspace (\ref{eq: Psi1}) only depend on $X$, $%
Y $ and $\varepsilon $. Given $X$ and $Y,$ the null distributions of the
exogeneity test statistics only depend on the distribution of $\varepsilon $%
: provided the distribution of $\varepsilon \,|\,\bar{X}$ can be simulated,
exact tests can be obtained through the Monte Carlo test method [see Section %
\ref{Sec: Exact Monte Carlo exogeneity tests}]. Furthermore, the tests
obtained in this way are robust to weak instruments in the sense that the
level is controlled even if identification fails (or is weak). This result
holds even if the distribution of $\varepsilon \,|\,\bar{X}$ does not have
moments (the Cauchy distribution, for example). This may be useful, for
example, in financial models with fat-tailed error distributions, such as
the Student $t$ distribution. There is no further restriction on the
distribution of $\varepsilon \,|\bar{X}$. For example, the distribution of $%
\varepsilon \,|\bar{X}$ may depend on $\bar{X},\,$provided it can be
simulated.

It is interesting to observe that the distribution of $V$ plays no role
here, so the vectors $V_{1},\ldots ,\,V_{T}$ may follow arbitrary
distributions with unspecified heterogeneity (or heteroskedasticity) and
serial dependence. In addition to finite-sample validity of all the
exogeneity tests in the presence of identification failure (or weak
identification), Theorem \ref{Th: Null distributions of exogeneity test
statistics} entails robustness to \emph{incomplete reduced forms }and \emph{%
instrument exclusion} under the null hypothesis of exogeneity. No further
information is needed on the form of the reduced form for $Y$ in (\ref{eq: Y
non-linearRedFch2b}): $g(\cdot )$ can be an unspecified nonlinear function, $%
\Pi =[\Pi _{1}$\thinspace $,$\thinspace $\Pi _{2}$\thinspace $]\,$ an
unknown parameter matrix, and $V$ may follow an arbitrary distribution. This
result extends to the exogeneity tests the one given in \cite%
{Dufour-Taamouti(2007)} on Anderson-Rubin-type tests (for structural
coefficients).

As long as the distribution of $\varepsilon \,($given $\bar{X}$ and $Y$) can
be simulated, all tests remain valid under $H_{0}$, and test sizes are
controlled conditional on $\bar{X}$ and $Y$, hence also unconditionally. In
particular, Monte-Carlo test procedures remain valid even if the instrument
matrix $X_{3}$ is not used by the test statistics. A similar property is
underscored in \cite{Dufour-Taamouti(2007)} for Anderson-Rubin tests in
linear structural equation models. This observation is also useful to allow
for models with structural breaks in the reduced form: exogeneity tests
remain valid in such contexts without knowledge of the form and timing of
breaks. In such contexts, inference on the structural form may be more
reliable than inference on the reduced form, a question of great relevance
for macroeconomic models; see \cite{Dufour-Khalaf-Kichian(2013)}. However,
although the exclusion of instruments does not affect the null distributions
of exogeneity test statistics, it may lead to power losses when the missing
information is important.

\subsection{\sectitlesize Exact Monte Carlo exogeneity tests \label{Sec:
Exact Monte Carlo exogeneity tests}}

To implement the exact Monte Carlo exogeneity tests of $H_{0}$ with level $%
\alpha $ $(0<\alpha <1)$, we suggest the following methodology; for a more
general discussion, see \cite{Dufour(2006)}. Suppose that the conditional
distribution of $\varepsilon \,$(given $\bar{X}$) is continuous, so that the
conditional distribution, given $\bar{X}$, of all exogeneity statistics is
also continuous. Let $\mathcal{W}\,$ denotes any of the DWH and RH statistic
in (\ref{eq: statT1})\thinspace -\thinspace (\ref{eq: statRH}). We can then
proceed as follows:

\begin{enumerate}
\item choose $\alpha ^{\ast }$ and $N$ so that 
\begin{equation}
\alpha =\frac{I[\alpha ^{\ast }N]+1}{N+1}\,  \label{eq: alphachoice}
\end{equation}
where for any nonnegative real number $x$, $I[x]$ is the largest integer
less than or equal to $x;$

\item compute the test statistic $\mathcal{W}^{(0)}$ based on the observed
data;

\item generate $N$ \emph{i.i.d.} error vectors $\varepsilon
^{(j)}=[\varepsilon _{1}^{(j)},$\thinspace $\ldots \,,$\thinspace $%
\varepsilon _{T}^{(j)}]^{\prime },$ \thinspace\ $j=1,$\thinspace $\ldots \,,$%
\thinspace $N\,,$\thinspace\ according to the specified distribution of $%
\varepsilon \,|\bar{X}\,,$ and compute the corresponding statistics $%
\mathcal{W}^{(j)},\,j=1,\,\ldots ,\,N,\,$ following Theorem \ref{Th: Null
distributions of exogeneity test statistics}; the distribution of each
statistic does not depend on $\beta _{0}$ under the null hypothesis;

\item compute the empirical distribution function based on $\mathcal{W}%
^{(j)},\,j=1,\,\ldots ,\,N,$ 
\begin{equation}
\hat{F}_{N}(x)=\frac{\sum_{j=1}^{N}\mathds{1}\lbrack \mathcal{W}^{(j)}\leq x]%
}{N+1}\,
\end{equation}%
or, equivalently, the simulated $p$-value function 
\begin{equation}
\hat{p}_{N}[x]=\frac{1+\sum_{j=1}^{N}\mathds{1}\lbrack \mathcal{W}^{(j)}\geq
x]}{N+1}  \label{eq: algoMonteMMCpv}
\end{equation}%
where $\mathds{1}\lbrack C]=1\,$ if condition $C$ holds, and $\mathds{1}%
\lbrack C]=0\;$otherwise;

\item reject the null hypothesis of exogeneity, $H_{0}$, at level $\alpha $
when $\mathcal{W}^{(0)}\geq \hat{F}_{N}^{-1}\left( 1-\alpha ^{\ast }\right)
, $ where $\hat{F}_{N}^{-1}\left( q\right) =$ $\inf \{x:\hat{F}_{N}\left(
x\right) \geq q\}$ is the generalized inverse of $\hat{F}_{N}(\cdot ),$ or
(equivalently) when $\hat{p}_{N}[\mathcal{W}^{(0)}]\leq \alpha .$
\end{enumerate}

Under $H_{0}$, 
\begin{equation}
\mathbb{P}\big[\mathcal{W}^{(0)}\geq \hat{F}_{N}^{-1}\left( 1-\alpha ^{\ast
}\right) \big]=\mathbb{P}\big[\hat{p}_{N}[\mathcal{W}^{(0)}]\leq \alpha \big]%
=\alpha  \label{eq: alphaRegS}
\end{equation}%
so that we have a test with level $\alpha $. The property given by (\ref{eq:
alphaRegS}) is a finite-sample validity result which holds irrespective of
the sample size $T$, and no asymptotic assumption is required. If the
distributions of the statistics are not continuous, the Monte Carlo test
procedure can easily be adapted by using \textquotedblleft
tie-breaking\textquotedblright\ method described in \cite{Dufour(2006)}.%
\footnote{%
Without correction for continuity, the algorithm proposed for statistics
with continuous distributions yields a conservative test, \emph{i.e.} the
probability of rejection under the null hypothesis is not larger than the
nominal level $(\alpha )$. Further discussion of this feature is available
in \cite{Dufour(2006)}.}

It is important to note here that the distributions of the exogeneity test
statistics in Theorem \ref{Th: Null distributions of exogeneity test
statistics} generally depend on the specific \textquotedblleft instrument
matrix\textquotedblright\ $X$ used by the tests (especially when $%
\varepsilon $ is not Gaussian), so no general valid \textquotedblleft
critical value\textquotedblright\ (independent of $X$) is available. The
Monte Carlo test procedure transparently controls the level of the test
irrespective of this complication, so there is \emph{no need to compute
critical values}.

\section{\sectitlesize Block-triangular invariance and exogeneity canonical
form \label{Sec: Block-triangular invariance}}

\resetcountersSection

In this section, we establish invariance results for exogeneity tests which
will be useful to study the distributions of the test statistics under the
alternative hypothesis. This basic invariance property is given by the
following proposition.

\begin{proposition}
\label{Th: Block-triangular invariance of exogeneity tests} 
\captiontheorem{\propositionname}{Block-triangular invariance of exogeneity tests}
Let 
\begin{equation}
R=\left[ 
\begin{array}{cc}
R_{11} & 0 \\ 
R_{21} & R_{22}%
\end{array}%
\right]
\end{equation}%
be a lower block-triangular matrix such that $R_{11}\neq 0$ is a scalar and $%
R_{22}$ is a nonsingular $G\times G$ matrix. If we replace $y$ and $Y$ by $%
y^{\ast }=yR_{11}+YR_{21}$ and $Y^{\ast }=YR_{22}$ in $(\ref{eq: statT1})$%
\thinspace -\thinspace $(\ref{eq: Psi_R})$, the statistics $\mathcal{T}_{i}$ 
$(i=1,$\thinspace $2,$\thinspace $3,$\thinspace $4)$, $\mathcal{H}_{j}$ $%
(j=1,$\thinspace $2,$\thinspace $3)$ and $\mathcal{R}$ do not change.
\end{proposition}

The above result is purely algebraic, so no statistical assumption is
needed. However, when it is combined with our statistical model, it has
remarkable consequences on the properties of exogeneity tests. For example,
if the reduced-form errors $V_{1},\,\ldots \,,\,V_{T}$ for $Y$ have the same
nonsingular covariance matrix $\Sigma $, the latter can be eliminated from
the distribution of the test statistic by choosing $R_{22}$ so that $%
R_{22}^{\prime }\,\Sigma \,R_{22}=I_{G}$. This entails that the
distributions of the exogeneity statistics do not depend on $\Sigma $ under
both the null and the alternative hypotheses.

Consider now the following transformation matrix:%
\begin{equation}
R=\left[ 
\begin{array}{cc}
1 & 0 \\ 
-(\beta +a) & I_{G}%
\end{array}%
\right] \,.  \label{eq: R standard}
\end{equation}%
Then, we have $[y^{\ast },\,Y^{\ast }]=[y,\,Y]R$ with%
\begin{eqnarray}
y^{\ast } &=&y-Y(\beta +a)=Y\beta +X_{1}\gamma +Va+e-Y(\beta +a)=\mu _{y\ast
}(a)+e\,,  \label{eq: y_*} \\
Y^{\ast } &=&Y  \label{eq: Y_*}
\end{eqnarray}%
where $\mu _{y\ast }(a)$ is a $T\times 1$ vector such that 
\begin{equation}
\mu _{y^{\ast }}(a)=X_{1}\gamma +[V-g(X_{1},\,X_{2},\,X_{3},\,V,\,\bar{\Pi}%
)]a\,.  \label{eq: mu_y*}
\end{equation}

The (invertible) transformation (\ref{eq: y_*})\thinspace -\thinspace (\ref%
{eq: Y_*}) yields the following \textquotedblleft latent
reduced-form\textquotedblright\ representation:%
\begin{equation}
y^{\ast }=X_{1}\gamma +[V-g(X_{1},\,X_{2},\,X_{3},\,V,\,\bar{\Pi})]a+e\,,
\label{eq: y*}
\end{equation}%
\begin{equation}
Y=g(X_{1},\,X_{2},\,X_{3},\,V,\,\bar{\Pi})\,.  \label{eq: Y*}
\end{equation}%
We say \textquotedblleft latent\textquotedblright\ because the function $%
g(\cdot )$ and the variables $X_{3}$ are unknown or unspecified. An
important feature here is that the endogeneity parameter $a$ can be isolated
from other model parameters. This will allow us to get relatively simple
characterizations of the power of exogeneity tests. For this reason, we will
call (\ref{eq: y*})\thinspace -\thinspace (\ref{eq: Y*}), the
\textquotedblleft exogeneity canonical form\textquotedblright\ associated
with model (\ref{eq: y StrucEqch2})\thinspace -\thinspace (\ref{eq: Y
non-linearRedFch2b}) along with Assumption \ref{Ass: Reduced-form
heterogeneity}.

In the important case where reduced-form error linear separability holds
(Assumption \ref{Ass: Reduced-form error separability for Y}) in addition to
(\ref{eq: y StrucEqch2})\thinspace -\thinspace (\ref{eq: Y
non-linearRedFch2b}), we can write 
\begin{equation}
Y=g(X_{1},\,X_{2},\,X_{3},\,\bar{\Pi})+V=\mu _{Y}+V
\label{eq: Y Separability}
\end{equation}%
which, by (\ref{eq: y StrucEqch2}), entails%
\begin{equation}
y=\mu _{y}(a)+(u+V\beta )=\mu _{y}(a)+v  \label{eq: y = mu_y + v}
\end{equation}%
where $\mu _{Y}$ is a $T\times G$ matrix and $\mu _{y}$ is a $T\times 1$
vector, such that 
\begin{equation}
\mu _{Y}=g(X_{1},\,X_{2},\,X_{3},\,\bar{\Pi})\,,\quad \mu
_{y}(a)=g(X_{1},\,X_{2},\,X_{3},\,\bar{\Pi})\beta +X_{1}\gamma \,,
\label{eq: Mu_Y}
\end{equation}%
\begin{equation}
v=u+V\beta =e+V(\beta +a)\,.
\end{equation}%
Then 
\begin{equation}
\mu _{y^{\ast }}(a)=\mu _{y}(a)-\mu _{Y}(\beta +a)=X_{1}\gamma
-g(X_{1},\,X_{2},\,X_{3},\,\bar{\Pi})a\,
\end{equation}%
does not depend on $V$, and the exogeneity canonical form is:%
\begin{equation}
y^{\ast }=X_{1}\gamma -g(X_{1},\,X_{2},\,X_{3},\,\bar{\Pi})a+e\,,
\end{equation}%
\begin{equation}
Y=g(X_{1},\,X_{2},\,X_{3},\,\bar{\Pi})+V\,.
\end{equation}

\section{\sectitlesize Power \label{Sec: Power}}

\resetcountersSection

In this section, we provide characterizations of the power of exogeneity
tests. We first consider the general case where only Assumption \ref{Ass:
Reduced-form heterogeneity} is added to the basic setup (\ref{eq: y
StrucEqch2})\thinspace - (\ref{eq: Rank conditions}). To simplify the
exposition, we use the following notation: for any $T\times 1$ vector $x$
and $T\times T$ matrix $A$, we set 
\begin{equation}
S_{T}[x,\,A]=T\,x^{\prime }A\,x\,.
\end{equation}

\begin{theorem}
\label{Th: Exogeneity test distributions under the alternative hypothesis} 
\captiontheorem{\theoremname}{Exogeneity test distributions under the alternative hypothesis}
Under the model described by $(\ref{eq: y StrucEqch2})$\thinspace - $(\ref%
{eq: Rank conditions})$, suppose Assumption \ref{Ass: Reduced-form
heterogeneity} holds. Then the test statistics defined in $(\ref{eq: statT1}%
) $\thinspace -\thinspace $(\ref{eq: statRH})$ have the following
representations:%
\begin{equation}
\mathcal{T}_{l}=\kappa _{l}\,\left( \frac{S_{T}[u(\bar{a}\,),\,\Psi _{_{0}}]%
}{S_{T}[u(\bar{a}\,),\,\Lambda _{l}]}\right) \,,\quad \text{for\ }%
\,l=1,\,2,\,3,\,4,\,  \label{eq: T_l u(a)}
\end{equation}%
\begin{equation}
\mathcal{H}_{1}=T\,\{u(\bar{a}\ )^{\prime }\,\Psi _{_{1}}[u(\bar{a}\ )]\,u(%
\bar{a}\ )\}\,,\quad \mathcal{H}_{2}=T\,\left( \frac{S_{T}[u(\bar{a}\
),\,\Psi _{_{0}}]}{S_{T}[u(\bar{a}\ ),\,\Lambda _{3}]}\right) \,,\quad 
\mathcal{H}_{3}=T\,\left( \frac{S_{T}[u(\bar{a}\ ),\,\Psi _{_{0}}]}{S_{T}[u(%
\bar{a}\ ),\,\Lambda _{4}]}\right) \,,  \label{eq: H_i u(a)}
\end{equation}%
\begin{equation}
\mathcal{R}=\kappa _{_{R}}\,\left( \frac{S_{T}[u(\bar{a}\ ),\,\Psi _{_{R}}]}{%
S_{T}[u(\bar{a}\ ),\,\Lambda _{_{R}}]}\right) \,,  \label{eq: RH u(a)}
\end{equation}%
where $u(\bar{a}\ )=V\bar{a}+\varepsilon $, $\bar{a}\ =\sigma (\bar{X}%
)^{-1}a $, 
\begin{equation}
\Psi _{_{1}}[u(\bar{a}\ )]=C_{1}^{\prime }\big(S_{T}[u(\bar{a}\ ),\,\Lambda
_{3}]\,\hat{\Omega}_{IV}^{-1}-S_{T}[u(\bar{a}\ ),\,\Lambda _{4}]\,\hat{\Omega%
}_{LS}^{-1}\big)^{-1}C_{1}  \label{eq: Psi_1 Sigma_1}
\end{equation}%
and $C_{1}$, $\Psi _{_{0}}$, $\Psi _{_{1}}$, $\Psi _{R}$, $\Lambda _{_{R}}$, 
$\Lambda _{1},\ldots ,\,\Lambda _{4}$ are defined as in Theorem \ref{Th:
Null distributions of exogeneity test statistics}. If Assumption \ref{Ass:
Endogeneity-parameter distributional separability} also holds, the
distributions of the test statistics $($conditional on $\bar{X})$ depend on $%
a$ only through $\bar{a}$ in $u(\bar{a}\ )$.
\end{theorem}

By Theorem \ref{Th: Exogeneity test distributions under the alternative
hypothesis}, the distributions of all the exogeneity statistics depend on $a$%
, though possibly in a rather complex way (especially when disturbances
follow non-Gaussian distributions). If the distribution of $\varepsilon $
does not depend on $\bar{a}$ -- as would be typically the case -- power
depends on the way the distributions of the quadratic forms $S_{T}[u(\bar{a}%
\ ),\,\Psi _{_{i}}]$ and $S_{T}[u(\bar{a}\ ),\,\Lambda _{j}]$ in (\ref{eq:
T_l u(a)})\thinspace -\thinspace (\ref{eq: RH u(a)}) are modified when the
value of $\bar{a}$ changes. Both the numerator and the denominator of the
statistics in Theorem \ref{Th: Exogeneity test distributions under the
alternative hypothesis} may follow different distributions, in contrast to
what happens in standard $F$ tests in the classical linear model.

The power characterization given by Theorem \ref{Th: Exogeneity test
distributions under the alternative hypothesis} does not provide a clear
picture of the parameters which determine the power of exogeneity tests.
This can be done by exploiting the invariance result of Proposition \ref{Th:
Block-triangular invariance of exogeneity tests}, as follows.

\begin{theorem}
\label{Th: Invariance-based distributions of exogeneity test statistics} 
\captiontheorem{\theoremname}{Invariance-based distributions of exogeneity statistics}
Under the model described by $(\ref{eq: y StrucEqch2})$\thinspace - $(\ref%
{eq: Rank conditions})$, suppose Assumption \ref{Ass: Reduced-form
heterogeneity} holds. Then the test statistics defined in $(\ref{eq: statT1}%
) $\thinspace -\thinspace $(\ref{eq: statRH})$ have the following
representations:%
\begin{equation}
\mathcal{T}_{l}=\kappa _{l}\,\left( \frac{S_{T}[y_{\ast }^{\perp }(\bar{a}\
),\,\Psi _{_{0}}]}{S_{T}[y_{\ast }^{\perp }(\bar{a}\ ),\,\Lambda _{l}]}%
\right) \,,\quad \text{for\ }\,l=1,\,2,\,3,\,4,\,  \label{eq: T_l y_1*(abar)}
\end{equation}%
\begin{equation}
\mathcal{H}_{1}=S_{T}\big[y_{\ast }^{\perp }(\bar{a}\ ),\,\Psi
_{_{1}}[y_{\ast }^{\perp }(\bar{a}\ )]\big]\,,\quad \mathcal{H}%
_{2}=T\,\left( \frac{S_{T}[y_{\ast }^{\perp }(\bar{a}\ ),\,\Psi _{_{0}}]}{%
S_{T}[y_{\ast }^{\perp }(\bar{a}\ ),\,\Lambda _{3}]}\right) \,,
\label{eq: H_i y_1*(abar)}
\end{equation}%
\begin{equation}
\mathcal{H}_{3}=T\,\left( \frac{S_{T}[y_{\ast }^{\perp }(\bar{a}\ ),\,\Psi
_{_{0}}]}{S_{T}[y_{\ast }^{\perp }(\bar{a}\ ),\,\Lambda _{4}]}\right)
\,,\quad \mathcal{R}=\kappa _{_{R}}\,\left( \frac{S_{T}[y_{\ast }^{\perp }(%
\bar{a}\ ),\,\Psi _{_{R}}]}{S_{T}[y_{\ast }^{\perp }(\bar{a}\ ),\,\Lambda
_{R}]}\right) \,,  \label{eq: RH y_1*(abar)}
\end{equation}%
where%
\begin{equation}
y_{\ast }^{\perp }(\bar{a}\ )=\bar{\mu}_{y\ast }^{\perp }(\bar{a}\
)+M_{1}\varepsilon ,  \label{eq: y(abar)}
\end{equation}%
\begin{equation}
\bar{\mu}_{y\ast }^{\perp }(\bar{a}\ )=M_{1}[V-g(X_{1},\,X_{2},\,X_{3},\,V,\,%
\bar{\Pi})]\bar{a}\,,\quad \bar{a}=\sigma (\bar{X})^{-1}a\,,
\label{eq: Mu(abar)}
\end{equation}%
\begin{equation}
\Psi _{_{1}}[y_{\ast }^{\perp }(\bar{a}\ )]=C_{1}^{\prime }\big(%
S_{T}[y_{\ast }^{\perp }(\bar{a}\ ),\,\Lambda _{3}]\,\hat{\Omega}%
_{IV}^{-1}-S_{T}[y_{\ast }^{\perp }(\bar{a}\ ),\,\Lambda _{4}]\,\hat{\Omega}%
_{LS}^{-1}\big)^{-1}C_{1}\,,
\end{equation}%
and $C_{1}$, $\Psi _{_{0}}$, $\Psi _{_{1}}$, $\Psi _{R}$, $\Lambda _{_{R}}$, 
$\Lambda _{1},\ldots ,\,\Lambda _{4}$ are defined as in Theorem $\ref{Th:
Null distributions of exogeneity test statistics}$. If Assumption \ref{Ass:
Endogeneity-parameter distributional separability} also holds, the
distributions of the test statistics $($conditional on $\bar{X}$ and $V)$
depend on $a$ only through $\bar{\mu}_{y\ast }^{\perp }(\bar{a}\ )$ in $%
y_{\ast }^{\perp }(\bar{a}\ )$. If Assumption \ref{Ass: Reduced-form error
separability for Y} also holds, 
\begin{equation}
\bar{\mu}_{y\ast }^{\perp }(\bar{a}\ )=-M_{1}\mathrm{\,}g(X_{1},\,X_{2},%
\,X_{3},\,\bar{\Pi})\,\bar{a}\,.  \label{eq: Mu(abar) separable}
\end{equation}
\end{theorem}

Following Theorem \ref{Th: Invariance-based distributions of exogeneity test
statistics}, the powers of the different exogeneity tests are controlled by $%
\bar{\mu}_{y\ast }^{\perp }(\bar{a}\ )$ in (\ref{eq: Mu(abar)}). Clearly $%
\,a=0$ entails $\bar{\mu}_{y\ast }^{\perp }(\bar{a}\ )=0$, which corresponds
to the distribution under the null hypothesis [under Assumption \ref{Ass:
Endogeneity-parameter distributional separability}]. Note however, the
latter property also holds when 
\begin{equation}
M_{1}\mathrm{\,}[V-g(X_{1},\,X_{2},\,X_{3},\,V,\,\bar{\Pi})]=0
\end{equation}%
even if $a\neq 0.$

Under Assumption \ref{Ass: Reduced-form error separability for Y}, $V$ is
evacuated from $\bar{\mu}_{y\ast }^{\perp }(\bar{a}\ )$ as given by (\ref%
{eq: Mu(abar) separable}). If Assumptions \ref{Ass: Endogeneity-parameter
distributional separability} and \ref{Ass: Reduced-form error separability
for Y} hold, power is determined by this parameter. $\bar{\mu}_{y\ast
}^{\perp }(\bar{a}\ )=0$ when $a=0,$ but also when $X_{1}$ and $%
g(X_{1},\,X_{2},\,X_{3},\,\bar{\Pi})$ are orthogonal. Note also the norm of $%
\bar{\mu}_{y\ast }^{\perp }(\bar{a}\ )$ shrinks when $\sigma (\bar{X})$
increases, so power decreases when the variance of value of $\varepsilon
_{t} $ increases (as expected). Under Assumption \ref{Ass: Reduced-form
error separability for Y}, conditioning on $\ \bar{X}$ and $V$ also becomes
equivalent to conditioning on $\bar{X}$ and $Y$.

Consider the special case of a complete linear model where equations (\ref%
{eq: Linear reduced form}) and (\ref{eq: [y,Y] Linear reduced form}) hold.
We then have:%
\begin{equation}
g(X_{1},\,X_{2},\,X_{3},\,\bar{\Pi})=X_{1}\Pi _{1}+X_{2}\Pi _{2}\,,\quad \mu
_{y\ast 1}^{\perp }(\bar{a}\,)=-M_{1}X_{2}\Pi _{2}\,\bar{a}\,.
\label{eq: g(X_1,X_2,X_3) linear}
\end{equation}%
When $\Pi _{2}$ $=0$ (complete non-identification of model parameters), or $%
M_{1}X_{2}=0$ ($X_{2}$ perfectly collinear with $X_{1}$), or more generally
when $M_{1}X_{2}\Pi _{2}=0$, we have $\bar{\mu}_{y\ast }^{\perp }(\bar{a}\
)=0$. Then, under Assumption \ref{Ass: Endogeneity-parameter distributional
separability}, the distributions of the exogeneity test statistics do not
depend on $a$, and the power function is flat (with respect to $a$).

Theorem \ref{Th: Invariance-based distributions of exogeneity test
statistics} provides a conditional power characterization [given $\bar{X}$
and $V$ (or $Y)$]. Even though the level of the test does not depend on the
distribution of $V$, power typically depends on the distribution of $V$.
Unconditional power functions can be obtained by averaging over $V$, but
this requires formulating specific assumptions on the distribution of $V$.

When the disturbances $\varepsilon _{1},\ldots ,\,\varepsilon _{T}$ are
i.i.d. Gaussian, it is possible to express the power function in terms of
non-central chi-square distributions. We denote by $\chi ^{2}[n;$\thinspace $%
\delta ]$ the non-central chi-square distribution with $n$ degrees of
freedom and noncentrality parameter $\delta $, and by $F[n_{1},$\thinspace $%
n_{2};$ $\delta _{1},$\thinspace $\delta _{2}]$ the doubly noncentral $F$%
-distribution with degrees of freedom $(n_{1},$\thinspace $n_{2})$ and
noncentrality parameters $(\delta _{1},\,\delta _{2})$, \emph{i.e.} $F\sim
F[n_{1},$\thinspace $n_{2};$ $\delta _{1},$\thinspace $\delta _{2}]$ means
that $F$ can be written as $F=[Q_{1}/m_{1}]\,/\,[Q_{2}/m_{2}]$ where $%
Q_{1}\, $and $Q_{2}$ are two independent random variables such that $%
Q_{1}\sim \chi ^{2}[n_{1};$\thinspace $\delta _{1}]$ and $Q_{2}\sim \chi
^{2}[n_{2};$\thinspace $\delta _{2}]$; see \cite[Ch. 30]%
{Johnson-Kotz-Balakrishnan(1995)}. When $\delta _{2}=0,$ $F\sim F[n_{1},$%
\thinspace $n_{2};$ $\delta _{1}]$ the usual noncentral $F$-distribution.

\begin{theorem}
\label{Th: Invariance-based distributions of exogeneity statistics
components with Gaussian errors} 
\captiontheorem{\theoremname}{Invariance-based distributions of exogeneity statistics components with Gaussian errors}
Under the model described by $(\ref{eq: y StrucEqch2})$\thinspace - $(\ref%
{eq: Rank conditions})$, suppose Assumptions \ref{Ass: Reduced-form
heterogeneity} and \ref{Ass: Conditional mutual independence of e and V}
hold. If $\varepsilon \thicksim \mathrm{N}[0,\,I_{T}]$, then, conditional on 
$\bar{X}$ and $V$, we have: 
\begin{equation}
S_{T}[y_{\ast }^{\perp }(\bar{a}\ ),\,\Psi _{_{0}}]\sim \chi ^{2}[G;\,\delta
(\bar{a},\,\Psi _{_{0}})]\,,\quad S_{T}[y_{\ast }^{\perp }(\bar{a}\
),\,\Lambda _{1}]\sim \chi ^{2}[k_{2}-G\,;\,\delta (\bar{a},\,\Lambda
_{1})]\,,
\end{equation}%
\begin{equation}
S_{T}[y_{\ast }^{\perp }(\bar{a}\ ),\,\Lambda _{2}]\sim \chi
^{2}[T-k_{1}-2G\,;\,\delta (\bar{a},\,\Lambda _{2})]\,,\quad S_{T}[y_{\ast
}^{\perp }(\bar{a}\ ),\,\Lambda _{4}]\sim \chi ^{2}[T-k_{1}-G\,;\,\delta (%
\bar{a},\,\Lambda _{4})]\,,
\end{equation}%
\begin{equation}
S_{T}[y_{\ast }^{\perp }(\bar{a}\ ),\,\Psi _{R}]\sim \chi
^{2}[k_{2}\,;\,\delta (\bar{a},\,\Psi _{R})]\,,\quad S_{T}[y_{\ast }^{\perp
}(\bar{a}\ ),\,\Lambda _{_{R}}]\sim \chi ^{2}[T-k_{1}-k_{2}-G\,;\,\delta (%
\bar{a},\,\Lambda _{_{R}})]\,,
\end{equation}%
where 
\begin{equation}
\delta (\bar{a},\,\Psi _{_{0}})=S_{T}[\bar{\mu}_{y\ast }^{\perp }(\bar{a}\
),\,\Psi _{0}]\,,\quad \delta (\bar{a},\,\Lambda _{1})=S_{T}[\bar{\mu}%
_{y\ast }^{\perp }(\bar{a}\ ),\,\Lambda _{1}]\,,
\end{equation}%
\begin{equation}
\delta (\bar{a},\,\Lambda _{2})=S_{T}[\bar{\mu}_{y\ast }^{\perp }(\bar{a}\
),\,\Lambda _{2}]\,,\quad \delta (\bar{a},\,\Lambda _{4})=S_{T}[\bar{\mu}%
_{y\ast }^{\perp }(\bar{a}\ ),\,\Lambda _{4}]\,,
\end{equation}%
\begin{equation}
\delta (\bar{a},\,\Psi _{R})=S_{T}[\bar{\mu}_{y\ast }^{\perp }(\bar{a}\
),\,\Psi _{R}]\,,\quad \delta (\bar{a},\,\Lambda _{_{R}})=S_{T}[\bar{\mu}%
_{y\ast }^{\perp }(\bar{a}\ ),\,\Lambda _{_{R}}]\,,
\end{equation}%
and the other symbols are defined as in Theorem \ref{Th: Invariance-based
distributions of exogeneity test statistics}. Further, conditional on $\bar{X%
}$ and $V$, the random variable $S_{T}[y_{\ast }^{\perp }(\bar{a}\ ),\,\Psi
_{_{0}}]$ is independent of $S_{T}[y_{\ast }^{\perp }(\bar{a}\ ),\,\Lambda
_{1}]$ and $S_{T}[y_{\ast }^{\perp }(\bar{a}\ ),\,\Lambda _{2}]$, and $%
S_{T}[y_{\ast }^{\perp }(\bar{a}\ ),\,\Psi _{R}]$ is independent of $%
S_{T}[y_{\ast }^{\perp }(\bar{a}\ ),\,\Lambda _{_{R}}]$.
\end{theorem}

Note we do not have a chi-square distributional result for $S_{T}[y_{\ast
}^{\perp }(\bar{a}\ ),\,\Lambda _{3}]$ which depends on the usual 2SLS
residuals. On the other hand, $S_{T}[y_{\ast }^{\perp }(\bar{a}\ ),\,\Lambda
_{4}]$ follows a noncentral chi-square distribution, but it is not
independent of $S_{T}[y_{\ast }^{\perp }(\bar{a}\ ),\,\Psi _{_{0}}]$.

The noncentrality parameters in Theorem \ref{Th: Invariance-based
distributions of exogeneity statistics components with Gaussian errors} can
be interpreted as \emph{concentration parameters}. For example, 
\begin{eqnarray}
\delta (\bar{a},\,\Psi _{_{0}}) &=&T\,[\bar{\mu}_{y\ast }^{\perp }(\bar{a}\
)^{\prime }\Psi _{_{0}}\bar{\mu}_{y\ast }^{\perp }(\bar{a}\ )]=T\,[\bar{\mu}%
_{y\ast }^{\perp }(\bar{a}\ )^{\prime }C_{1}^{\prime }\hat{\Delta}^{-1}C_{1}%
\bar{\mu}_{y\ast }^{\perp }(\bar{a}\ )]  \notag \\
&=&\{M_{1}[V-g(X_{1},\,X_{2},\,X_{3},\,V,\,\bar{\Pi})]\bar{a}\}^{\prime
}C_{1}^{\prime }(C_{1}C_{1}^{\prime
})^{-1}C_{1}\{M_{1}[V-g(X_{1},\,X_{2},\,X_{3},\,V,\,\bar{\Pi})]\bar{a}\} 
\notag \\
&=&\{M_{1}[V-g(X_{1},\,X_{2},\,X_{3},\,V,\,\bar{\Pi})]\bar{a}\}^{\prime }%
\mathrm{\bar{P}}[C_{1}^{\prime }]\{M_{1}[V-g(X_{1},\,X_{2},\,X_{3},\,V,\,%
\bar{\Pi})]\bar{a}\}
\end{eqnarray}%
and, in the case of the simple complete linear model where (\ref{eq: Linear
reduced form}) and (\ref{eq: [y,Y] Linear reduced form}) hold,%
\begin{equation}
\delta (\bar{a},\,\Psi _{_{0}})=(M_{1}\,X_{2}\,\Pi _{2}\,\bar{a})^{\prime }%
\mathrm{\bar{P}}[C_{1}^{\prime }](M_{1}\,X_{2}\,\Pi _{2}\,\bar{a})=\bar{a}%
\,^{\prime }\Pi _{2}^{\prime }\,X_{2}^{\prime }\,M_{1}\mathrm{\bar{P}}%
[C_{1}^{\prime }]M_{1}\,X_{2}\,\Pi _{2}\,\bar{a}\,.
\end{equation}%
For $\delta (\bar{a},\,\Psi _{_{0}})$ to be different from zero, we need $%
M_{1}X_{2}\Pi _{2}\,\bar{a}\neq 0$. In particular, this requires that the
instruments $X_{2}$ not be totally weak ($\Pi _{2}\neq 0$) and linearly
independent of $X_{1}$ ($M_{1}X_{2}\neq 0$). Similar interpretations can
easily be formulated for the other centrality parameters. In particular, in
the simple complete linear model, all noncentrality parameters are zero if $%
M_{1}\,X_{2}\,\Pi _{2}\,\bar{a}=0$. Note, however, this may not hold in the
more general model described by (\ref{eq: y StrucEqch2})\thinspace -(\ref%
{eq: Rank conditions}), because of the nonlinear reduced form for $Y$ and
the presence of excluded instruments.

Theorem \ref{Th: Invariance-based distributions of exogeneity statistics
components with Gaussian errors} allows us to conclude that $\mathcal{T}_{1}$%
, $\mathcal{T}_{2}$ and $\mathcal{R}$ follow doubly noncentral $F$%
-distributions under the alternative hypothesis (conditional on $\bar{X}$
and $V$). This is spelled out in the following corollary.

\begin{corollary}
\label{Th: Doubly noncentral distributions for exogeneity statistics} 
\captiontheorem{\theoremname}{Doubly noncentral distributions for exogeneity statistics}
Under the model described by $(\ref{eq: y StrucEqch2})$\thinspace - $(\ref%
{eq: Rank conditions})$, suppose Assumptions \ref{Ass: Reduced-form
heterogeneity} and \ref{Ass: Conditional mutual independence of e and V}
hold. If $\varepsilon \thicksim \mathrm{N}[0,\,I_{T}]$, then conditional on $%
\bar{X}$ and $V$, we have:%
\begin{equation}
\mathcal{T}_{1}\sim F[G,\,k_{2}-G;\,\delta (\bar{a},\,\Psi _{_{0}}),\,\delta
(\bar{a},\,\Lambda _{1})]\,,  \label{eq: T_1 F}
\end{equation}%
\begin{equation}
\mathcal{T}_{2}\sim F[G,\,T-k_{1}-2G;\,\delta (\bar{a},\,\Psi
_{_{0}}),\,\delta (\bar{a},\,\Lambda _{2})]\,,  \label{eq: T_2 F}
\end{equation}%
\begin{equation}
\mathcal{T}_{4}=\frac{\kappa _{4}}{\kappa _{2}\mathcal{T}_{2}^{-1}+1}\leq
\left( \frac{\kappa _{4}}{\kappa _{2}}\right) \mathcal{T}_{2}\,,
\label{eq: T_4 leq F}
\end{equation}%
\begin{equation}
\mathcal{R}\sim F[k_{2},\,T-k_{1}-k_{2}-G;\,\delta (\bar{a},\,\Psi
_{R}),\,\delta (\bar{a},\,\Psi _{R})]\,,  \label{eq: R F}
\end{equation}%
where the noncentrality parameters are defined in Theorem \ref{Th:
Invariance-based distributions of exogeneity statistics components with
Gaussian errors}.
\end{corollary}

In the special case where (\ref{eq: Linear reduced form}) and (\ref{eq:
[y,Y] Linear reduced form}) hold, we have $\Lambda
_{_{R}}\,M_{1}\,g(X_{1},\,X_{2},\,X_{3},\,\bar{\Pi})=\Lambda
_{_{R}}g(X_{1},\,X_{2},\,X_{3},\,\bar{\Pi})=0$ and $\delta (\bar{a},\,\Psi
_{R})=0$, so $\mathcal{R}\sim F[k_{2},\,T-k_{1}-k_{2}-G;\,\delta (\bar{a}%
,\,\Psi _{R})]\,$the usual noncentral noncentral $F$-distribution. When $%
a=0, $ the distributions of $\mathcal{T}_{1}$, $\mathcal{T}_{2}$ and $%
\mathcal{R}$ reduce to the central chi-square in (\ref{eq: T_1 T_2 Null
Gaussian}) originally provided by \cite{Wu(1973)} and \cite%
{Revankar-Hartley(1973)}. The setup under which these are obtained here is
considerably more general than the usual linear reduced-form specification (%
\ref{eq: Linear reduced form}) considered by these authors.

Note $\mathcal{T}_{4}$ is proportional to a ratio of two noncentral
chi-square distributions, but it is not doubly-noncentral chi-square due to
the non-orthogonality of $\Psi _{_{0}}$ and $\Lambda _{4}$ [$\Psi
_{_{0}}\,\Lambda _{4}=T^{-1}\Psi _{_{0}}$, see (\ref{eq: C_1*Lambda_4})].
This observation carries to $\mathcal{H}_{3}$ through the identity $\mathcal{%
H}_{3}=(T/\kappa _{4})\mathcal{T}_{4}$. The same applies to $\mathcal{H}_{1}$
and $\mathcal{H}_{2}$, because of the presence of $S_{T}[y_{\ast }^{\perp }(%
\bar{a}\ ),\,\Lambda _{3}]$ in these statistics.

\section{\sectitlesize Simulation experiment \label{Sec: Simulation
experiment}}

\resetcountersSection

We use simulation to analyze the finite-sample performances (size and power)
of the standard and exact Monte Carlo DWH and RH tests. The DGP is described
by equations (\ref{eq: y StrucEqch2}) and (\ref{eq: Linear reduced form})
without included exogenous instruments variables $X_{1}$, $%
Y=[Y_{1}:Y_{2}]\in \mathbb{R}^{T\times 2},$ the $T\times k_{2}$ instrument
matrix $X_{2}$ is a such that $X_{2t}\overset{i.i.d.}{\sim }\mathbf{N}%
(0,\,I_{k_{2}})$ for all $\,t=1,\,\ldots ,\,T,\,$ and is fixed within
experiment. We set the true values of $\beta $ at $\beta _{0}=(2,5)^{\prime
} $ but the results are qualitatively the same for alternative choices of $%
\beta _{0}.$ The matrix $\Pi _{2}$ that describes the quality of the
instruments in the first stage regression is such that $\Pi _{2}=[\eta
_{1}\Pi _{01}:\eta _{2}\Pi _{02}]\in \mathbb{R}^{k_{2}\times 2},\,$ where $%
[\Pi _{01}:\Pi _{02}]$ is obtained by taking the first two columns of the
identity\footnote{%
We run the experiment where $[\Pi _{01}:\Pi _{02}]$ is the $k_{2}\times 2$
matrix of ones, and we found similar results as those presented here.}
matrix of order $k_{2}$. We vary both $\eta _{1}$ and $\eta _{2}$ in $%
\{0,\,0.01,\,0.5\}$, where $\eta _{1}=\eta _{2}=0$ is a design of a complete
non-identification, $\eta _{1}=\eta _{2}=0.01$ is a design of weak
identification, $\eta _{1}\in \{0,\,0.01\}\,\text{and}\,\eta _{2}=0.5$ or 
\emph{vice versa} is a design of partial identification, and finally, $\eta
_{1}=\eta _{2}=0.5$ corresponding to strong identification (strong
instruments).

The errors $u$ and $V$ are generated so that 
\begin{equation}
u=Va+e=V_{1}a_{1}+V_{2}a_{2}+e  \label{eq: (u=Va)DGPSyst}
\end{equation}%
where $a_{1}$ and $a_{2}$ are fixed scalar coefficients. In this experiment,
we set $a=(a_{1},a_{2})^{\prime }=\lambda \,a_{0},\,$ where $%
a_{0}=(0.5,0.2)^{\prime }$ and $\lambda \in \left\{ -20-5,0,1,100\right\} $\
but the results do not change qualitatively with alternative values of $%
a_{0} $ and $\lambda $. In the above setup, $\lambda $ controls the
endogeneity of $Y$: $\lambda =0$ corresponds to the exogeneity hypothesis
(level), while values of $\lambda $ different from zero represent the
alternative of endogeneity (power). We consider two specifications for the
joint distribution of $\,[e,V].\,$ In the first one, $(e_{t},V_{t}^{\prime
})^{\prime }\sim \mathbf{N}\left( 0,\,I_{3}\right) \,$ for all $t=1,\,\ldots
,\,T\,$ (Gaussian errors). In the second one, $e_{t}$ and $V_{jt},\,j=1,2,\,$
follow a $t(3)$ distribution and are uncorrelated for all $t=1,\,\ldots
,\,T. $ In both cases, $V_{1}$ and $V_{2}$ are independent. The sample size
is $T=50$, and the Monte Carlo test $p$-values are computed with $N=199$
pseudo-samples. The simulations are based on $10000$ replications. The
nominal level for both the MC critical values and the standard tests is set
at $5\%$.

\subsection{\sectitlesize Size and power with the usual critical values 
\label{sec: SizePowerDWH}}

Tables \ref{Table: Power of exogeneity tests, T=50}-\ref{Table: Power of
exogeneity tests t(3), T=50} present the empirical rejections of the
standard DWH and RH tests for both Gaussian errors (Table \ref{Table: Power
of exogeneity tests, T=50}) and $t(3)$ errors (Table \ref{Table: Power of
exogeneity tests t(3), T=50}). The first column of each table reports the
statistics, while the second column contains the values of $k_{2}$ (number
of excluded instruments). The other columns report, for each value of the
endogeneity measure ($\lambda $) and IV qualities $\eta _{1}$ and $\eta _{2}$%
, the rejection frequencies of the tests. The results confirm our
theoretical analysis.

\emph{First}, the rejection frequencies of all tests under the null
hypothesis of exogeneity ($\lambda =0$) are equal or smaller than the
nominal $5\%$ level, whether identification is weak ($\eta _{1},\eta _{2}\in
\{0,\,0.01\}$), partial ($\eta _{1}\in \{0,\,0.01\}\,\text{and}\,\eta
_{2}=0.5$ or \emph{vice versa}), or strong $(\eta _{1}=\eta _{2}=0.5)$, with
or without Gaussian errors. Thus, all DWH-type and RH tests are valid in
finite samples and robust to weak instruments (\emph{i.e.}, level is
controlled). This confirms the analysis of Section \ref{Sec: Incomplete
models and pivotal properties}. As expected, the tests $\mathcal{T}_{2}$, $%
\mathcal{T}_{4},$ $\mathcal{H}_{3},$ and $\mathcal{R}$ have rejections close
to the $5\%$ nominal level. Meanwhile, $\mathcal{T}_{3}$, $\mathcal{H}_{1}$
and $\mathcal{H}_{2}$ are highly conservative when identification is weak [$%
\eta _{1},\eta _{2}\in \{0,\,0.01\}\,$ in the tables].

\emph{Second}, all tests have power when identification is partial (columns $%
\lambda \neq 0$ and $\eta _{1}\in \{0,\,0.01\}\,\text{and}\,\eta _{2}=0.5$
or \emph{vice versa}) or strong (columns $\lambda \neq 0$ and $\eta
_{1}=\eta _{2}=0.5$), with and without Gaussian errors. Their rejection
frequencies are close to 100\% when $\lambda \neq 0$ and identification is
strong ($\eta _{1}=\eta _{2}=0.5$), despite the relatively small sample size
($T=50$). However, all tests have low power when all instruments are
irrelevant ($\lambda \neq 0$ and $\eta _{1},\eta _{2}\in \{0,\,0.01\}$). In
particular, the rejection frequencies are close to $5\%$ when $\lambda \neq
0,\,$with $\eta _{1},\eta _{2}\in \{0,\,0.01\}$, thus confirming the results
of Theorems \ref{Th: Invariance-based distributions of exogeneity test
statistics} and \ref{Th: Invariance-based distributions of exogeneity
statistics components with Gaussian errors}. The simulations also suggest
that the tests $\mathcal{T}_{2},$ $\mathcal{H}_{3},$ $\mathcal{T}_{4},$ and $%
\mathcal{R}$ have greater power than the others. However, this is not also
always the case after size correction through the exact Monte Carlo test
method, as shown in the next subsection.

\begin{sidewaystable}%

\caption{Size and power of exogeneity tests with Gaussian errors at nominal
level $5\%$ } \label{Table: Power of exogeneity tests, T=50}

\begin{center}
Table \thetable. Size and power of exogeneity tests with Gaussian errors at
nominal level $5\%$

\quad 
\scriptsize%

\begin{tabular}{|c|c|ccc|ccc|ccc|ccc|ccc|}
\hline\hline
\rule{0cm}{3ex} &  & \multicolumn{3}{c|}{$\lambda =-20$} & 
\multicolumn{3}{c|}{$\lambda =-5$} & \multicolumn{3}{c|}{$\lambda =0$} & 
\multicolumn{3}{c|}{$\lambda =1$} & \multicolumn{3}{c|}{$\lambda =100$} \\ 
\cline{3-5}\cline{6-8}\cline{9-11}\cline{12-14}\cline{15-17}
\rule{0cm}{3ex} & $k_{2}$ & $\eta _{1}=0$ & $\eta _{1}=.01$ & $\eta _{1}=.5$
& $\eta _{1}=0$ & $\eta _{1}=.01$ & $\eta _{1}=.5$ & $\eta _{1}=0$ & $\eta
_{1}=.01$ & $\eta _{1}=.5$ & $\eta _{1}=0$ & $\eta _{1}=.01$ & $\eta _{1}=.5$
& $\eta _{1}=0$ & $\eta _{1}=.01$ & $\eta _{1}=.5$ \\ 
\rule{0cm}{3ex} &  & $\eta _{2}=0$ & $\eta _{2}=0$ & $\eta _{2}=0$ & $\eta
_{2}=0$ & $\eta _{2}=0$ & $\eta _{2}=0$ & $\eta _{2}=0$ & $\eta _{2}=0$ & $%
\eta _{2}=0$ & $\eta _{2}=0$ & $\eta _{2}=0$ & $\eta _{2}=0$ & $\eta _{2}=0$
& $\eta _{2}=0$ & $\eta _{2}=0$ \\ \hline
&  &  &  &  &  &  &  &  &  &  &  &  &  &  &  &  \\ 
$\mathcal{T}_{1}$ & 5 & 5.0 & 4.8 & 74.2 & 5.3 & 4.8 & 67.7 & 4.7 & 5.0 & 5.1
& 5.1 & 4.8 & 21.1 & 5.3 & 4.4 & 74.1 \\ 
$\mathcal{T}_{2}$ & - & 4.6 & 12.4 & 100.0 & 5.1 & 5.7 & 100.0 & 4.7 & 5.2 & 
4.9 & 5.0 & 4.9 & 57.7 & 5.1 & 69.8 & 100.0 \\ 
$\mathcal{T}_{3}$ & - & 0.0 & 0.0 & 98.4 & 0.0 & 0.0 & 97.8 & 0.0 & 0.0 & 0.7
& 0.0 & 0.0 & 34.1 & 0.0 & 3.6 & 98.4 \\ 
$\mathcal{T}_{4}$ & - & 4.3 & 11.8 & 100.0 & 4.7 & 5.2 & 100.0 & 4.5 & 4.9 & 
4.6 & 4.7 & 4.5 & 56.4 & 4.8 & 69.2 & 100.0 \\ 
$\mathcal{H}_{1}$ & - & 0.0 & 0.0 & 92.4 & 0.0 & 0.0 & 90.6 & 0.0 & 0.0 & 0.3
& 0.0 & 0.0 & 20.9 & 0.0 & 2.1 & 92.1 \\ 
$\mathcal{H}_{2}$ & - & 0.0 & 0.0 & 98.5 & 0.0 & 0.0 & 98.0 & 0.0 & 0.0 & 0.8
& 0.0 & 0.1 & 36.8 & 0.0 & 4.5 & 98.5 \\ 
$\mathcal{H}_{3}$ & - & 5.0 & 12.9 & 100.0 & 5.4 & 6.0 & 100.0 & 5.0 & 5.5 & 
5.2 & 5.3 & 5.2 & 58.7 & 5.5 & 70.4 & 100.0 \\ 
$\mathcal{R}$ & - & 5.2 & 18.6 & 100.0 & 5.1 & 5.8 & 100.0 & 4.6 & 4.7 & 4.8
& 5.3 & 5.1 & 44.8 & 5.2 & 100.0 & 100.0 \\ 
&  &  &  &  &  &  &  &  &  &  &  &  &  &  &  &  \\ 
$\mathcal{T}_{1}$ & 10 & 4.9 & 3.9 & 99.5 & 5.0 & 4.7 & 98.1 & 4.7 & 5.1 & 
4.7 & 5.2 & 5.2 & 37.9 & 4.7 & 3.1 & 99.4 \\ 
$\mathcal{T}_{2}$ & - & 4.8 & 9.7 & 100.0 & 5.0 & 5.1 & 100.0 & 4.8 & 4.8 & 
5.1 & 5.1 & 5.2 & 59.1 & 4.8 & 44.6 & 100.0 \\ 
$\mathcal{T}_{3}$ & - & 0.3 & 0.7 & 100.0 & 0.4 & 0.2 & 100.0 & 0.3 & 0.3 & 
1.8 & 0.3 & 0.4 & 48.8 & 0.3 & 10.7 & 100.0 \\ 
$\mathcal{T}_{4}$ & - & 4.5 & 9.2 & 100.0 & 4.6 & 4.8 & 100.0 & 4.5 & 4.6 & 
4.8 & 4.8 & 4.9 & 57.8 & 4.5 & 43.8 & 100.0 \\ 
$\mathcal{H}_{1}$ & - & 0.2 & 0.4 & 99.1 & 0.2 & 0.1 & 98.5 & 0.2 & 0.1 & 0.8
& 0.1 & 0.1 & 32.1 & 0.1 & 7.1 & 99.2 \\ 
$\mathcal{H}_{2}$ & - & 0.4 & 0.9 & 100.0 & 0.6 & 0.3 & 100.0 & 0.5 & 0.4 & 
2.2 & 0.4 & 0.5 & 51.4 & 0.4 & 12.7 & 100.0 \\ 
$\mathcal{H}_{3}$ & - & 5.0 & 10.1 & 100.0 & 5.3 & 5.5 & 100.0 & 5.1 & 5.1 & 
5.5 & 5.4 & 5.5 & 60.0 & 5.1 & 45.6 & 100.0 \\ 
$\mathcal{R}$ & - & 5.1 & 21.5 & 100.0 & 4.8 & 5.6 & 100.0 & 5.4 & 4.9 & 5.6
& 5.3 & 5.2 & 37.8 & 5.0 & 100.0 & 100.0 \\ 
&  &  &  &  &  &  &  &  &  &  &  &  &  &  &  &  \\ 
$\mathcal{T}_{1}$ & 20 & 5.2 & 3.4 & 99.9 & 5.3 & 5.1 & 99.4 & 4.7 & 4.7 & 
5.1 & 4.9 & 5.0 & 41.7 & 4.7 & 1.5 & 99.9 \\ 
$\mathcal{T}_{2}$ & - & 5.0 & 7.0 & 100.0 & 5.2 & 5.2 & 100.0 & 4.9 & 4.6 & 
5.1 & 5.1 & 5.0 & 51.9 & 5.1 & 14.5 & 100.0 \\ 
$\mathcal{T}_{3}$ & - & 1.8 & 2.8 & 100.0 & 1.9 & 2.1 & 100.0 & 2.1 & 1.7 & 
3.3 & 2.0 & 2.0 & 47.8 & 2.0 & 7.4 & 100.0 \\ 
$\mathcal{T}_{4}$ & - & 4.6 & 6.7 & 100.0 & 4.9 & 4.9 & 100.0 & 4.5 & 4.3 & 
4.7 & 4.8 & 4.6 & 50.7 & 4.7 & 13.9 & 100.0 \\ 
$\mathcal{H}_{1}$ & - & 1.1 & 1.7 & 99.7 & 1.2 & 1.2 & 99.4 & 1.4 & 1.0 & 1.2
& 1.1 & 1.2 & 30.6 & 1.2 & 5.0 & 99.8 \\ 
$\mathcal{H}_{2}$ & - & 2.3 & 3.4 & 100.0 & 2.4 & 2.6 & 100.0 & 2.5 & 2.2 & 
3.9 & 2.5 & 2.6 & 50.3 & 2.4 & 8.5 & 100.0 \\ 
$\mathcal{H}_{3}$ & - & 5.3 & 7.4 & 100.0 & 5.6 & 5.4 & 100.0 & 5.2 & 5.0 & 
5.3 & 5.4 & 5.2 & 53.0 & 5.5 & 15.0 & 100.0 \\ 
$\mathcal{R}$ & - & 4.7 & 29.4 & 100.0 & 5.0 & 6.0 & 100.0 & 5.0 & 5.0 & 5.4
& 4.7 & 5.4 & 25.7 & 5.1 & 100.0 & 100.0 \\ 
&  &  &  &  &  &  &  &  &  &  &  &  &  &  &  &  \\ \hline\hline
\end{tabular}
\end{center}

\normalsize%

\end{sidewaystable}%

\normalsize%

\begin{sidewaystable}%


\begin{center}
Table \thetable\ (continued). Size and power of exogeneity tests with
Gaussian errors at nominal level $5\%$

\quad 
\scriptsize%

\begin{tabular}{|c|c|ccc|ccc|ccc|ccc|ccc|}
\hline\hline
\rule{0cm}{3ex} &  & \multicolumn{3}{c|}{$\lambda =-20$} & 
\multicolumn{3}{c|}{$\lambda =-5$} & \multicolumn{3}{c|}{$\lambda =0$} & 
\multicolumn{3}{c|}{$\lambda =1$} & \multicolumn{3}{c|}{$\lambda =100$} \\ 
\cline{3-5}\cline{6-8}\cline{9-11}\cline{12-14}\cline{15-17}
\rule{0cm}{3ex} & $k_{2}$ & $\eta _{1}=0$ & $\eta _{1}=.01$ & $\eta _{1}=.5$
& $\eta _{1}=0$ & $\eta _{1}=.01$ & $\eta _{1}=.5$ & $\eta _{1}=0$ & $\eta
_{1}=.01$ & $\eta _{1}=.5$ & $\eta _{1}=0$ & $\eta _{1}=.01$ & $\eta _{1}=.5$
& $\eta _{1}=0$ & $\eta _{1}=.01$ & $\eta _{1}=.5$ \\ 
\rule{0cm}{3ex} &  & $\eta _{2}=.5$ & $\eta _{2}=.5$ & $\eta _{2}=.5$ & $%
\eta _{2}=.5$ & $\eta _{2}=.5$ & $\eta _{2}=.5$ & $\eta _{2}=.5$ & $\eta
_{2}=.5$ & $\eta _{2}=.5$ & $\eta _{2}=.5$ & $\eta _{2}=.5$ & $\eta _{2}=.5$
& $\eta _{2}=.5$ & $\eta _{2}=.5$ & $\eta _{2}=.5$ \\ \hline
&  &  &  &  &  &  &  &  &  &  &  &  &  &  &  &  \\ 
$\mathcal{T}_{1}$ & 5 & 63.4 & 64.1 & 78.2 & 37.6 & 39.8 & 72.5 & 4.7 & 4.9
& 5.2 & 7.1 & 7.7 & 23.2 & 66.7 & 66.0 & 78.3 \\ 
$\mathcal{T}_{2}$ & - & 100.0 & 100.0 & 100.0 & 96.8 & 98.1 & 100.0 & 4.9 & 
5.3 & 4.9 & 11.6 & 12.3 & 61.4 & 100.0 & 100.0 & 100.0 \\ 
$\mathcal{T}_{3}$ & - & 97.3 & 97.0 & 98.4 & 81.7 & 84.0 & 98.1 & 0.6 & 0.7
& 1.1 & 3.1 & 3.1 & 39.1 & 97.2 & 97.8 & 98.6 \\ 
$\mathcal{T}_{4}$ & - & 100.0 & 100.0 & 100.0 & 96.5 & 97.9 & 100.0 & 4.5 & 
4.9 & 4.7 & 11.0 & 11.7 & 60.2 & 100.0 & 100.0 & 100.0 \\ 
$\mathcal{H}_{1}$ & - & 90.7 & 91.2 & 91.4 & 66.5 & 69.4 & 89.6 & 0.3 & 0.4
& 0.4 & 1.7 & 1.6 & 23.4 & 91.4 & 92.3 & 91.9 \\ 
$\mathcal{H}_{2}$ & - & 97.5 & 97.2 & 98.5 & 83.6 & 85.6 & 98.2 & 0.7 & 0.9
& 1.2 & 3.6 & 3.8 & 41.4 & 97.4 & 98.0 & 98.7 \\ 
$\mathcal{H}_{3}$ & - & 100.0 & 100.0 & 100.0 & 97.1 & 98.2 & 100.0 & 5.2 & 
5.6 & 5.3 & 12.2 & 12.8 & 62.5 & 100.0 & 100.0 & 100.0 \\ 
$\mathcal{R}$ & - & 100.0 & 100.0 & 100.0 & 94.7 & 96.5 & 100.0 & 5.0 & 5.3
& 5.4 & 9.3 & 9.5 & 48.4 & 100.0 & 100.0 & 100.0 \\ 
&  &  &  &  &  &  &  &  &  &  &  &  &  &  &  &  \\ 
$\mathcal{T}_{1}$ & 10 & 98.8 & 98.9 & 99.7 & 79.4 & 81.4 & 99.0 & 4.8 & 5.3
& 5.4 & 10.3 & 11.2 & 43.3 & 99.4 & 99.2 & 99.8 \\ 
$\mathcal{T}_{2}$ & - & 100.0 & 100.0 & 100.0 & 98.6 & 99.1 & 100.0 & 5.1 & 
5.3 & 5.0 & 13.1 & 14.4 & 65.6 & 100.0 & 100.0 & 100.0 \\ 
$\mathcal{T}_{3}$ & - & 100.0 & 100.0 & 100.0 & 97.3 & 98.1 & 100.0 & 1.7 & 
1.7 & 1.8 & 7.1 & 8.3 & 57.4 & 100.0 & 100.0 & 100.0 \\ 
$\mathcal{T}_{4}$ & - & 100.0 & 100.0 & 100.0 & 98.4 & 99.0 & 100.0 & 4.7 & 
5.0 & 4.7 & 12.6 & 13.6 & 64.5 & 100.0 & 100.0 & 100.0 \\ 
$\mathcal{H}_{1}$ & - & 99.2 & 99.0 & 98.1 & 87.5 & 90.6 & 97.2 & 0.7 & 0.5
& 0.4 & 3.3 & 3.9 & 33.0 & 99.1 & 99.1 & 98.4 \\ 
$\mathcal{H}_{2}$ & - & 100.0 & 100.0 & 100.0 & 97.7 & 98.4 & 100.0 & 2.1 & 
2.0 & 2.2 & 8.1 & 9.5 & 59.9 & 100.0 & 100.0 & 100.0 \\ 
$\mathcal{H}_{3}$ & - & 100.0 & 100.0 & 100.0 & 98.6 & 99.2 & 100.0 & 5.5 & 
5.6 & 5.3 & 13.9 & 15.1 & 66.5 & 100.0 & 100.0 & 100.0 \\ 
$\mathcal{R}$ & - & 100.0 & 100.0 & 100.0 & 95.5 & 97.1 & 100.0 & 5.1 & 5.1
& 5.1 & 8.4 & 9.4 & 42.8 & 100.0 & 100.0 & 100.0 \\ 
&  &  &  &  &  &  &  &  &  &  &  &  &  &  &  &  \\ 
$\mathcal{T}_{1}$ & 20 & 99.8 & 99.7 & 100.0 & 84.0 & 85.8 & 99.5 & 5.3 & 5.2
& 4.9 & 10.9 & 11.7 & 43.2 & 99.9 & 99.9 & 100.0 \\ 
$\mathcal{T}_{2}$ & - & 100.0 & 100.0 & 100.0 & 95.3 & 96.5 & 100.0 & 5.1 & 
5.0 & 5.1 & 12.1 & 12.8 & 54.6 & 100.0 & 100.0 & 100.0 \\ 
$\mathcal{T}_{3}$ & - & 100.0 & 100.0 & 100.0 & 94.5 & 95.7 & 100.0 & 3.4 & 
3.1 & 3.3 & 9.2 & 10.0 & 50.4 & 100.0 & 100.0 & 100.0 \\ 
$\mathcal{T}_{4}$ & - & 100.0 & 100.0 & 100.0 & 95.0 & 96.2 & 100.0 & 4.9 & 
4.6 & 4.7 & 11.5 & 12.2 & 53.3 & 100.0 & 100.0 & 100.0 \\ 
$\mathcal{H}_{1}$ & - & 99.7 & 99.7 & 98.9 & 85.2 & 87.2 & 97.7 & 1.1 & 1.2
& 0.8 & 4.2 & 4.4 & 26.9 & 99.8 & 99.8 & 99.0 \\ 
$\mathcal{H}_{2}$ & - & 100.0 & 100.0 & 100.0 & 95.2 & 96.4 & 100.0 & 4.0 & 
3.7 & 3.8 & 10.5 & 11.3 & 53.2 & 100.0 & 100.0 & 100.0 \\ 
$\mathcal{H}_{3}$ & - & 100.0 & 100.0 & 100.0 & 95.6 & 96.7 & 100.0 & 5.3 & 
5.4 & 5.5 & 12.6 & 13.4 & 55.6 & 100.0 & 100.0 & 100.0 \\ 
$\mathcal{R}$ & - & 100.0 & 100.0 & 100.0 & 86.9 & 90.2 & 100.0 & 5.1 & 5.3
& 4.9 & 7.5 & 7.3 & 27.4 & 100.0 & 100.0 & 100.0 \\ 
&  &  &  &  &  &  &  &  &  &  &  &  &  &  &  &  \\ \hline\hline
\end{tabular}
\end{center}

\normalsize%

\end{sidewaystable}%

\begin{sidewaystable}%

\caption{Size and Power of exogeneity tests with $t(3)$ errors at nominal
level $5\%$} \label{Table: Power of exogeneity tests t(3), T=50}

\begin{center}
Table \thetable. Size and Power of exogeneity tests with $t(3)$ errors at
nominal level $5\%$

\quad 
\scriptsize%

\begin{tabular}{|c|c|ccc|ccc|ccc|ccc|ccc|}
\hline\hline
\rule{0cm}{3ex} &  & \multicolumn{3}{c|}{$\lambda =-20$} & 
\multicolumn{3}{c|}{$\lambda =-5$} & \multicolumn{3}{c|}{$\lambda =0$} & 
\multicolumn{3}{c|}{$\lambda =1$} & \multicolumn{3}{c|}{$\lambda =100$} \\ 
\cline{3-5}\cline{6-8}\cline{9-11}\cline{12-14}\cline{15-17}
\rule{0cm}{3ex} & $k_{2}$ & $\eta _{1}=0$ & $\eta _{1}=.01$ & $\eta _{1}=.5$
& $\eta _{1}=0$ & $\eta _{1}=.01$ & $\eta _{1}=.5$ & $\eta _{1}=0$ & $\eta
_{1}=.01$ & $\eta _{1}=.5$ & $\eta _{1}=0$ & $\eta _{1}=.01$ & $\eta _{1}=.5$
& $\eta _{1}=0$ & $\eta _{1}=.01$ & $\eta _{1}=.5$ \\ 
\rule{0cm}{3ex} &  & $\eta _{2}=0$ & $\eta _{2}=0$ & $\eta _{2}=0$ & $\eta
_{2}=0$ & $\eta _{2}=0$ & $\eta _{2}=0$ & $\eta _{2}=0$ & $\eta _{2}=0$ & $%
\eta _{2}=0$ & $\eta _{2}=0$ & $\eta _{2}=0$ & $\eta _{2}=0$ & $\eta _{2}=0$
& $\eta _{2}=0$ & $\eta _{2}=0$ \\ \hline
&  &  &  &  &  &  &  &  &  &  &  &  &  &  &  &  \\ 
$\mathcal{T}_{1}$ & 5 & 4.6 & 5.0 & 50.5 & 5.3 & 5.2 & 43.9 & 5.3 & 4.9 & 5.0
& 4.9 & 4.9 & 12.9 & 5.0 & 4.4 & 50.7 \\ 
$\mathcal{T}_{2}$ & - & 4.8 & 7.8 & 99.9 & 4.9 & 5.2 & 99.5 & 5.2 & 5.0 & 4.8
& 5.1 & 5.2 & 33.7 & 5.1 & 52.6 & 99.9 \\ 
$\mathcal{T}_{3}$ & - & 0.0 & 0.0 & 91.2 & 0.0 & 0.0 & 87.6 & 0.0 & 0.0 & 0.4
& 0.0 & 0.0 & 10.6 & 0.0 & 1.5 & 91.2 \\ 
$\mathcal{T}_{4}$ & - & 4.5 & 7.3 & 99.9 & 4.6 & 4.9 & 99.4 & 4.9 & 4.7 & 4.5
& 4.7 & 4.9 & 32.6 & 4.7 & 51.7 & 99.9 \\ 
$\mathcal{H}_{1}$ & - & 0.0 & 0.0 & 85.3 & 0.0 & 0.0 & 79.4 & 0.0 & 0.0 & 0.2
& 0.0 & 0.0 & 6.4 & 0.0 & 0.8 & 84.8 \\ 
$\mathcal{H}_{2}$ & - & 0.0 & 0.0 & 91.9 & 0.0 & 0.0 & 88.6 & 0.0 & 0.0 & 0.6
& 0.0 & 0.0 & 12.3 & 0.0 & 1.8 & 91.9 \\ 
$\mathcal{H}_{3}$ & - & 5.1 & 8.1 & 99.9 & 5.3 & 5.6 & 99.5 & 5.5 & 5.3 & 5.1
& 5.4 & 5.5 & 35.0 & 5.5 & 53.2 & 99.9 \\ 
$\mathcal{R}$ & - & 4.9 & 9.8 & 100.0 & 5.0 & 5.4 & 99.6 & 5.0 & 5.2 & 4.9 & 
5.3 & 5.6 & 27.8 & 5.2 & 92.0 & 100.0 \\ 
&  &  &  &  &  &  &  &  &  &  &  &  &  &  &  &  \\ 
$\mathcal{T}_{1}$ & 10 & 5.1 & 4.6 & 86.0 & 5.0 & 4.7 & 78.6 & 4.9 & 4.9 & 
4.6 & 4.9 & 5.0 & 21.1 & 5.2 & 3.2 & 87.2 \\ 
$\mathcal{T}_{2}$ & - & 5.1 & 6.2 & 99.8 & 5.3 & 5.0 & 99.2 & 4.9 & 5.1 & 5.2
& 5.0 & 4.6 & 34.2 & 5.0 & 29.4 & 99.8 \\ 
$\mathcal{T}_{3}$ & - & 0.4 & 0.4 & 99.0 & 0.3 & 0.4 & 97.7 & 0.3 & 0.3 & 1.2
& 0.3 & 0.2 & 20.5 & 0.2 & 4.4 & 99.2 \\ 
$\mathcal{T}_{4}$ & - & 4.8 & 5.7 & 99.8 & 5.0 & 4.7 & 99.2 & 4.5 & 4.7 & 4.8
& 4.6 & 4.4 & 33.2 & 4.6 & 28.4 & 99.8 \\ 
$\mathcal{H}_{1}$ & - & 0.1 & 0.1 & 97.9 & 0.1 & 0.2 & 95.5 & 0.1 & 0.1 & 0.6
& 0.1 & 0.1 & 13.3 & 0.1 & 2.5 & 98.1 \\ 
$\mathcal{H}_{2}$ & - & 0.5 & 0.5 & 99.2 & 0.4 & 0.5 & 98.0 & 0.4 & 0.4 & 1.5
& 0.4 & 0.3 & 22.6 & 0.4 & 5.4 & 99.3 \\ 
$\mathcal{H}_{3}$ & - & 5.4 & 6.6 & 99.9 & 5.6 & 5.3 & 99.2 & 5.1 & 5.4 & 5.4
& 5.3 & 4.9 & 35.1 & 5.2 & 30.2 & 99.8 \\ 
$\mathcal{R}$ & - & 4.9 & 9.4 & 100.0 & 5.4 & 5.2 & 99.6 & 5.1 & 5.1 & 4.9 & 
5.2 & 5.1 & 23.7 & 5.2 & 93.2 & 100.0 \\ 
&  &  &  &  &  &  &  &  &  &  &  &  &  &  &  &  \\ 
$\mathcal{T}_{1}$ & 20 & 4.8 & 4.4 & 97.9 & 4.6 & 4.6 & 94.6 & 5.1 & 4.9 & 
5.4 & 4.9 & 4.9 & 29.8 & 4.9 & 1.6 & 98.4 \\ 
$\mathcal{T}_{2}$ & - & 4.9 & 5.8 & 99.8 & 4.7 & 4.6 & 99.4 & 5.2 & 5.1 & 5.5
& 4.8 & 4.8 & 38.8 & 4.6 & 12.2 & 99.9 \\ 
$\mathcal{T}_{3}$ & - & 1.8 & 2.3 & 99.8 & 1.7 & 1.9 & 99.3 & 2.1 & 2.0 & 3.3
& 1.7 & 1.9 & 33.5 & 1.7 & 5.7 & 99.8 \\ 
$\mathcal{T}_{4}$ & - & 4.5 & 5.4 & 99.8 & 4.5 & 4.2 & 99.4 & 4.9 & 4.9 & 5.1
& 4.5 & 4.5 & 37.6 & 4.3 & 11.5 & 99.9 \\ 
$\mathcal{H}_{1}$ & - & 1.1 & 1.4 & 99.6 & 0.9 & 1.0 & 98.5 & 1.2 & 1.1 & 1.6
& 1.0 & 1.1 & 24.5 & 1.0 & 3.7 & 99.7 \\ 
$\mathcal{H}_{2}$ & - & 2.3 & 2.8 & 99.8 & 2.1 & 2.2 & 99.4 & 2.5 & 2.4 & 3.8
& 2.1 & 2.3 & 35.9 & 2.2 & 6.7 & 99.8 \\ 
$\mathcal{H}_{3}$ & - & 5.2 & 6.2 & 99.9 & 5.1 & 4.7 & 99.4 & 5.5 & 5.5 & 5.7
& 5.1 & 5.1 & 39.6 & 4.8 & 12.6 & 99.9 \\ 
$\mathcal{R}$ & - & 5.2 & 11.8 & 100.0 & 4.9 & 5.4 & 99.4 & 5.1 & 4.7 & 4.7
& 5.0 & 4.9 & 23.0 & 4.4 & 98.0 & 100.0 \\ 
&  &  &  &  &  &  &  &  &  &  &  &  &  &  &  &  \\ \hline\hline
\end{tabular}
\end{center}

\normalsize%

\end{sidewaystable}%

\begin{sidewaystable}%


\begin{center}
Table \thetable\ (continued). Size and Power of exogeneity tests with $t(3)$
errors at nominal level $5\%$

\quad 
\scriptsize%

\begin{tabular}{|c|c|ccc|ccc|ccc|ccc|ccc|}
\hline\hline
\rule{0cm}{3ex} &  & \multicolumn{3}{c|}{$\lambda =-20$} & 
\multicolumn{3}{c|}{$\lambda =-5$} & \multicolumn{3}{c|}{$\lambda =0$} & 
\multicolumn{3}{c|}{$\lambda =1$} & \multicolumn{3}{c|}{$\lambda =100$} \\ 
\cline{3-5}\cline{6-8}\cline{9-11}\cline{12-14}\cline{15-17}
\rule{0cm}{3ex} & $k_{2}$ & $\eta _{1}=0$ & $\eta _{1}=.01$ & $\eta _{1}=.5$
& $\eta _{1}=0$ & $\eta _{1}=.01$ & $\eta _{1}=.5$ & $\eta _{1}=0$ & $\eta
_{1}=.01$ & $\eta _{1}=.5$ & $\eta _{1}=0$ & $\eta _{1}=.01$ & $\eta _{1}=.5$
& $\eta _{1}=0$ & $\eta _{1}=.01$ & $\eta _{1}=.5$ \\ 
\rule{0cm}{3ex} &  & $\eta _{2}=.5$ & $\eta _{2}=.5$ & $\eta _{2}=.5$ & $%
\eta _{2}=.5$ & $\eta _{2}=.5$ & $\eta _{2}=.5$ & $\eta _{2}=.5$ & $\eta
_{2}=.5$ & $\eta _{2}=.5$ & $\eta _{2}=.5$ & $\eta _{2}=.5$ & $\eta _{2}=.5$
& $\eta _{2}=.5$ & $\eta _{2}=.5$ & $\eta _{2}=.5$ \\ \hline
&  &  &  &  &  &  &  &  &  &  &  &  &  &  &  &  \\ 
$\mathcal{T}_{1}$ & 5 & 47.0 & 47.6 & 67.0 & 26.4 & 27.2 & 59.0 & 4.5 & 4.8
& 5.4 & 6.6 & 7.1 & 18.3 & 50.6 & 49.9 & 68.3 \\ 
$\mathcal{T}_{2}$ & - & 99.7 & 99.8 & 100.0 & 83.3 & 86.2 & 99.8 & 4.6 & 4.9
& 4.9 & 8.9 & 10.1 & 48.9 & 99.9 & 99.8 & 100.0 \\ 
$\mathcal{T}_{3}$ & - & 89.8 & 89.9 & 97.1 & 51.0 & 54.9 & 95.9 & 0.5 & 0.4
& 0.7 & 1.4 & 1.6 & 26.1 & 91.1 & 91.3 & 97.7 \\ 
$\mathcal{T}_{4}$ & - & 99.7 & 99.8 & 100.0 & 82.5 & 85.7 & 99.8 & 4.3 & 4.5
& 4.6 & 8.3 & 9.5 & 48.0 & 99.9 & 99.8 & 100.0 \\ 
$\mathcal{H}_{1}$ & - & 82.6 & 83.4 & 91.7 & 38.5 & 42.5 & 88.2 & 0.3 & 0.2
& 0.3 & 0.7 & 0.8 & 16.0 & 84.6 & 85.3 & 91.9 \\ 
$\mathcal{H}_{2}$ & - & 90.8 & 90.8 & 97.3 & 54.1 & 57.7 & 96.3 & 0.6 & 0.5
& 0.8 & 1.7 & 1.8 & 28.3 & 91.8 & 92.1 & 97.9 \\ 
$\mathcal{H}_{3}$ & - & 99.7 & 99.8 & 100.0 & 83.8 & 86.7 & 99.8 & 4.8 & 5.1
& 5.2 & 9.3 & 10.7 & 50.0 & 99.9 & 99.8 & 100.0 \\ 
$\mathcal{R}$ & - & 99.9 & 100.0 & 100.0 & 79.7 & 84.1 & 99.8 & 5.3 & 4.7 & 
5.0 & 7.7 & 7.9 & 38.7 & 100.0 & 100.0 & 100.0 \\ 
&  &  &  &  &  &  &  &  &  &  &  &  &  &  &  &  \\ 
$\mathcal{T}_{1}$ & 10 & 90.5 & 90.1 & 98.5 & 57.3 & 59.2 & 95.7 & 5.3 & 4.9
& 5.1 & 8.7 & 9.2 & 34.1 & 92.2 & 92.4 & 98.8 \\ 
$\mathcal{T}_{2}$ & - & 99.8 & 99.8 & 100.0 & 87.7 & 90.0 & 99.9 & 5.3 & 5.1
& 5.0 & 10.5 & 11.5 & 53.9 & 99.9 & 99.9 & 100.0 \\ 
$\mathcal{T}_{3}$ & - & 99.5 & 99.4 & 100.0 & 80.5 & 83.5 & 99.8 & 1.4 & 1.4
& 1.6 & 4.6 & 4.9 & 43.1 & 99.5 & 99.6 & 100.0 \\ 
$\mathcal{T}_{4}$ & - & 99.8 & 99.8 & 100.0 & 87.2 & 89.7 & 99.9 & 4.9 & 4.8
& 4.6 & 10.0 & 10.9 & 52.7 & 99.9 & 99.9 & 100.0 \\ 
$\mathcal{H}_{1}$ & - & 98.4 & 98.5 & 99.1 & 70.3 & 73.8 & 98.0 & 0.7 & 0.5
& 0.7 & 2.4 & 2.7 & 29.8 & 98.9 & 98.8 & 99.3 \\ 
$\mathcal{H}_{2}$ & - & 99.5 & 99.5 & 100.0 & 82.3 & 85.2 & 99.8 & 1.9 & 1.6
& 1.9 & 5.3 & 5.6 & 45.6 & 99.6 & 99.6 & 100.0 \\ 
$\mathcal{H}_{3}$ & - & 99.8 & 99.9 & 100.0 & 88.2 & 90.5 & 99.9 & 5.7 & 5.4
& 5.5 & 11.0 & 11.9 & 54.8 & 99.9 & 99.9 & 100.0 \\ 
$\mathcal{R}$ & - & 99.9 & 99.9 & 100.0 & 81.6 & 85.0 & 99.8 & 5.1 & 5.1 & 
4.8 & 7.8 & 8.1 & 36.5 & 100.0 & 100.0 & 100.0 \\ 
&  &  &  &  &  &  &  &  &  &  &  &  &  &  &  &  \\ 
$\mathcal{T}_{1}$ & 20 & 96.8 & 96.7 & 99.8 & 66.6 & 68.4 & 98.1 & 4.8 & 4.7
& 5.2 & 9.3 & 9.2 & 36.8 & 98.0 & 97.7 & 99.8 \\ 
$\mathcal{T}_{2}$ & - & 99.8 & 99.7 & 100.0 & 83.5 & 84.5 & 99.7 & 4.8 & 5.0
& 5.2 & 10.2 & 10.2 & 46.4 & 99.8 & 99.8 & 100.0 \\ 
$\mathcal{T}_{3}$ & - & 99.7 & 99.6 & 100.0 & 80.6 & 82.1 & 99.7 & 2.9 & 3.0
& 3.2 & 7.4 & 7.1 & 42.3 & 99.8 & 99.7 & 100.0 \\ 
$\mathcal{T}_{4}$ & - & 99.8 & 99.7 & 100.0 & 82.8 & 83.9 & 99.7 & 4.4 & 4.7
& 4.9 & 9.7 & 9.6 & 45.3 & 99.8 & 99.8 & 100.0 \\ 
$\mathcal{H}_{1}$ & - & 99.5 & 99.4 & 99.8 & 72.2 & 74.9 & 98.8 & 1.4 & 1.6
& 1.4 & 4.1 & 4.1 & 29.6 & 99.7 & 99.6 & 99.9 \\ 
$\mathcal{H}_{2}$ & - & 99.8 & 99.6 & 100.0 & 82.1 & 83.4 & 99.7 & 3.4 & 3.5
& 3.8 & 8.3 & 8.3 & 44.5 & 99.8 & 99.8 & 100.0 \\ 
$\mathcal{H}_{3}$ & - & 99.8 & 99.7 & 100.0 & 84.1 & 84.9 & 99.7 & 5.1 & 5.3
& 5.4 & 10.6 & 10.6 & 47.5 & 99.8 & 99.8 & 100.0 \\ 
$\mathcal{R}$ & - & 99.9 & 99.9 & 100.0 & 73.1 & 76.2 & 99.5 & 5.2 & 4.8 & 
5.1 & 7.3 & 7.5 & 25.6 & 100.0 & 100.0 & 100.0 \\ 
&  &  &  &  &  &  &  &  &  &  &  &  &  &  &  &  \\ \hline\hline
\end{tabular}
\end{center}

\normalsize%

\end{sidewaystable}%

\begin{sidewaystable}%

\caption{Size and power of exact Monte Carlo tests with Gaussian errors
at nominal level 5\%} \label{Table: Power of MCEtestsGau, T=50}

\begin{center}
Table \thetable\ . Size and power of exact Monte Carlo tests with Gaussian
errors at nominal level $5\%$

\quad 
\scriptsize%

\begin{tabular}{|c|c|ccc|ccc|ccc|ccc|ccc|}
\hline\hline
\rule{0cm}{3ex} &  & \multicolumn{3}{c|}{$\lambda =-20$} & 
\multicolumn{3}{c|}{$\lambda =-5$} & \multicolumn{3}{c|}{$\lambda =0$} & 
\multicolumn{3}{c|}{$\lambda =1$} & \multicolumn{3}{c|}{$\lambda =100$} \\ 
\cline{3-5}\cline{6-8}\cline{9-11}\cline{12-14}\cline{15-17}
\rule{0cm}{3ex} & $k_{2}$ & $\eta _{1}=0$ & $\eta _{1}=.01$ & $\eta _{1}=.5$
& $\eta _{1}=0$ & $\eta _{1}=.01$ & $\eta _{1}=.5$ & $\eta _{1}=0$ & $\eta
_{1}=.01$ & $\eta _{1}=.5$ & $\eta _{1}=0$ & $\eta _{1}=.01$ & $\eta _{1}=.5$
& $\eta _{1}=0$ & $\eta _{1}=.01$ & $\eta _{1}=.5$ \\ 
\rule{0cm}{3ex} &  & $\eta _{2}=0$ & $\eta _{2}=0$ & $\eta _{2}=0$ & $\eta
_{2}=0$ & $\eta _{2}=0$ & $\eta _{2}=0$ & $\eta _{2}=0$ & $\eta _{2}=0$ & $%
\eta _{2}=0$ & $\eta _{2}=0$ & $\eta _{2}=0$ & $\eta _{2}=0$ & $\eta _{2}=0$
& $\eta _{2}=0$ & $\eta _{2}=0$ \\ \hline
&  &  &  &  &  &  &  &  &  &  &  &  &  &  &  &  \\ 
$\mathcal{T}_{1mc}$ & 5 & 5.1 & 5.2 & 72.3 & 4.9 & 5.0 & 67.1 & 5.0 & 4.8 & 
4.9 & 5.2 & 5.1 & 21.0 & 4.8 & 4.2 & 74.9 \\ 
$\mathcal{T}_{2mc}$ & - & 5.4 & 11.2 & 100.0 & 5.3 & 5.6 & 100.0 & 5.3 & 5.4
& 5.1 & 5.3 & 5.5 & 55.7 & 5.5 & 69.4 & 100.0 \\ 
$\mathcal{T}_{3mc}$ & - & 5.2 & 9.0 & 99.3 & 5.0 & 5.4 & 99.2 & 4.9 & 5.0 & 
4.9 & 5.1 & 5.1 & 60.7 & 5.1 & 40.4 & 99.4 \\ 
$\mathcal{T}_{4mc}$ & - & 5.3 & 11.2 & 100.0 & 5.2 & 5.6 & 100.0 & 5.3 & 5.4
& 5.1 & 5.2 & 5.4 & 55.7 & 5.5 & 69.4 & 100.0 \\ 
$\mathcal{H}_{1mc}$ & - & 5.1 & 9.0 & 97.6 & 4.8 & 5.3 & 97.2 & 4.8 & 4.9 & 
4.9 & 5.0 & 5.1 & 56.5 & 5.1 & 39.9 & 97.8 \\ 
$\mathcal{H}_{2mc}$ & - & 5.2 & 9.0 & 99.3 & 5.0 & 5.4 & 99.2 & 5.0 & 5.0 & 
4.9 & 5.0 & 5.1 & 60.7 & 5.1 & 40.4 & 99.4 \\ 
$\mathcal{H}_{3mc}$ & - & 5.3 & 11.2 & 100.0 & 5.2 & 5.6 & 100.0 & 5.3 & 5.4
& 5.1 & 5.3 & 5.4 & 55.7 & 5.5 & 69.4 & 100.0 \\ 
$\mathcal{R}_{mc}$ & - & 5.5 & 16.4 & 100.0 & 5.5 & 5.7 & 100.0 & 5.4 & 5.2
& 5.3 & 5.0 & 4.9 & 43.1 & 5.8 & 100.0 & 100.0 \\ 
&  &  &  &  &  &  &  &  &  &  &  &  &  &  &  &  \\ 
$\mathcal{T}_{1mc}$ & 10 & 5.0 & 4.4 & 99.0 & 5.0 & 5.0 & 96.8 & 5.1 & 5.0 & 
5.2 & 5.1 & 5.0 & 32.9 & 4.6 & 4.0 & 98.8 \\ 
$\mathcal{T}_{2mc}$ & - & 5.2 & 8.5 & 100.0 & 5.0 & 5.3 & 100.0 & 5.2 & 5.1
& 5.0 & 5.5 & 5.6 & 54.6 & 5.7 & 40.9 & 100.0 \\ 
$\mathcal{T}_{3mc}$ & - & 5.0 & 7.8 & 100.0 & 5.0 & 5.1 & 100.0 & 4.9 & 4.7
& 4.9 & 5.0 & 5.0 & 60.9 & 5.1 & 35.1 & 100.0 \\ 
$\mathcal{T}_{4mc}$ & - & 5.1 & 8.5 & 100.0 & 5.0 & 5.3 & 100.0 & 5.2 & 5.1
& 5.0 & 5.5 & 5.6 & 54.6 & 5.7 & 40.9 & 100.0 \\ 
$\mathcal{H}_{1mc}$ & - & 5.0 & 7.7 & 99.9 & 5.0 & 5.2 & 99.9 & 4.8 & 5.0 & 
4.7 & 4.8 & 4.9 & 58.5 & 5.1 & 34.9 & 99.9 \\ 
$\mathcal{H}_{2mc}$ & - & 5.0 & 7.8 & 100.0 & 5.0 & 5.1 & 100.0 & 4.9 & 4.7
& 4.9 & 5.1 & 5.0 & 60.9 & 5.1 & 35.1 & 100.0 \\ 
$\mathcal{H}_{3mc}$ & - & 5.2 & 8.5 & 100.0 & 5.0 & 5.3 & 100.0 & 5.2 & 5.1
& 5.0 & 5.5 & 5.6 & 54.6 & 5.7 & 40.9 & 100.0 \\ 
$\mathcal{R}_{mc}$ & - & 5.6 & 16.7 & 100.0 & 5.0 & 5.6 & 100.0 & 5.1 & 5.3
& 5.4 & 5.5 & 5.8 & 35.1 & 5.0 & 100.0 & 100.0 \\ 
&  &  &  &  &  &  &  &  &  &  &  &  &  &  &  &  \\ 
$\mathcal{T}_{1mc}$ & 20 & 4.9 & 3.3 & 99.9 & 5.0 & 4.6 & 99.2 & 4.9 & 4.7 & 
4.8 & 4.8 & 5.0 & 40.7 & 4.7 & 4.3 & 99.9 \\ 
$\mathcal{T}_{2mc}$ & - & 5.1 & 6.8 & 100.0 & 5.0 & 4.8 & 100.0 & 5.1 & 4.8
& 4.9 & 5.3 & 5.7 & 51.5 & 5.6 & 14.6 & 100.0 \\ 
$\mathcal{T}_{3mc}$ & - & 4.8 & 6.6 & 100.0 & 5.0 & 4.7 & 100.0 & 5.0 & 4.6
& 4.7 & 5.0 & 5.1 & 54.3 & 5.0 & 13.9 & 100.0 \\ 
$\mathcal{T}_{4mc}$ & - & 5.0 & 6.8 & 100.0 & 5.0 & 4.8 & 100.0 & 5.1 & 4.9
& 5.0 & 5.2 & 5.7 & 51.5 & 5.6 & 14.6 & 100.0 \\ 
$\mathcal{H}_{1mc}$ & - & 4.9 & 6.6 & 100.0 & 5.0 & 4.7 & 99.9 & 5.0 & 4.6 & 
4.9 & 5.0 & 5.1 & 51.5 & 5.1 & 14.0 & 100.0 \\ 
$\mathcal{H}_{2mc}$ & - & 4.8 & 6.6 & 100.0 & 5.0 & 4.7 & 100.0 & 5.0 & 5.0
& 4.8 & 5.2 & 5.1 & 54.3 & 5.1 & 13.9 & 100.0 \\ 
$\mathcal{H}_{3mc}$ & - & 5.1 & 6.8 & 100.0 & 5.0 & 4.8 & 100.0 & 5.1 & 5.1
& 5.0 & 5.0 & 5.7 & 51.5 & 5.6 & 14.6 & 100.0 \\ 
$\mathcal{R}_{mc}$ & - & 5.8 & 30.5 & 100.0 & 5.0 & 5.9 & 100.0 & 5.2 & 5.2
& 4.9 & 5.1 & 5.9 & 26.1 & 5.5 & 100.0 & 100.0 \\ 
&  &  &  &  &  &  &  &  &  &  &  &  &  &  &  &  \\ \hline\hline
\end{tabular}
\end{center}

\normalsize%

\end{sidewaystable}%

\begin{sidewaystable}%


\begin{center}
Table \thetable\ (continued). Size and power of exact Monte Carlo tests with
Gaussian errors at nominal level $5\%$

\quad 
\scriptsize%

\begin{tabular}{|c|c|ccc|ccc|ccc|ccc|ccc|}
\hline\hline
\rule{0cm}{3ex} &  & \multicolumn{3}{c|}{$\lambda =-20$} & 
\multicolumn{3}{c|}{$\lambda =-5$} & \multicolumn{3}{c|}{$\lambda =0$} & 
\multicolumn{3}{c|}{$\lambda =1$} & \multicolumn{3}{c|}{$\lambda =100$} \\ 
\cline{3-5}\cline{6-8}\cline{9-11}\cline{12-14}\cline{15-17}
\rule{0cm}{3ex} & $k_{2}$ & $\eta _{1}=0$ & $\eta _{1}=.01$ & $\eta _{1}=.5$
& $\eta _{1}=0$ & $\eta _{1}=.01$ & $\eta _{1}=.5$ & $\eta _{1}=0$ & $\eta
_{1}=.01$ & $\eta _{1}=.5$ & $\eta _{1}=0$ & $\eta _{1}=.01$ & $\eta _{1}=.5$
& $\eta _{1}=0$ & $\eta _{1}=.01$ & $\eta _{1}=.5$ \\ 
\rule{0cm}{3ex} &  & $\eta _{2}=.5$ & $\eta _{2}=.5$ & $\eta _{2}=.5$ & $%
\eta _{2}=.5$ & $\eta _{2}=.5$ & $\eta _{2}=.5$ & $\eta _{2}=.5$ & $\eta
_{2}=.5$ & $\eta _{2}=.5$ & $\eta _{2}=.5$ & $\eta _{2}=.5$ & $\eta _{2}=.5$
& $\eta _{2}=.5$ & $\eta _{2}=.5$ & $\eta _{2}=.5$ \\ \hline
&  &  &  &  &  &  &  &  &  &  &  &  &  &  &  &  \\ 
$\mathcal{T}_{1mc}$ & 5 & 71.2 & 72.3 & 80.3 & 44.5 & 44.1 & 76.0 & 4.8 & 5.1
& 5.2 & 7.9 & 8.4 & 24.4 & 74.3 & 74.0 & 80.5 \\ 
$\mathcal{T}_{2mc}$ & - & 100.0 & 100.0 & 100.0 & 98.7 & 99.2 & 100.0 & 5.1
& 5.2 & 5.3 & 12.5 & 14.3 & 67.7 & 100.0 & 100.0 & 100.0 \\ 
$\mathcal{T}_{3mc}$ & - & 99.3 & 99.5 & 99.6 & 96.3 & 96.5 & 99.4 & 4.8 & 5.0
& 4.9 & 14.6 & 16.2 & 71.2 & 99.3 & 99.4 & 99.5 \\ 
$\mathcal{T}_{4mc}$ & - & 100.0 & 100.0 & 100.0 & 98.7 & 99.2 & 100.0 & 5.1
& 5.2 & 5.3 & 12.5 & 14.3 & 67.7 & 100.0 & 100.0 & 100.0 \\ 
$\mathcal{H}_{1mc}$ & - & 97.6 & 97.5 & 97.3 & 91.9 & 92.5 & 97.0 & 4.8 & 5.0
& 4.9 & 14.2 & 15.7 & 63.9 & 97.7 & 97.7 & 97.1 \\ 
$\mathcal{H}_{2mc}$ & - & 99.3 & 99.5 & 99.6 & 96.3 & 96.5 & 99.4 & 4.7 & 4.9
& 5.1 & 14.6 & 16.2 & 71.2 & 99.3 & 99.4 & 99.5 \\ 
$\mathcal{H}_{3mc}$ & - & 100.0 & 100.0 & 100.0 & 98.7 & 99.2 & 100.0 & 5.1
& 5.2 & 5.3 & 12.5 & 14.3 & 67.7 & 100.0 & 100.0 & 100.0 \\ 
$\mathcal{R}_{mc}$ & - & 100.0 & 100.0 & 100.0 & 97.4 & 98.6 & 100.0 & 5.0 & 
5.0 & 5.0 & 9.6 & 10.7 & 54.8 & 100.0 & 100.0 & 100.0 \\ 
&  &  &  &  &  &  &  &  &  &  &  &  &  &  &  &  \\ 
$\mathcal{T}_{1mc}$ & 10 & 98.3 & 98.3 & 99.8 & 75.6 & 79.9 & 98.5 & 4.9 & 
5.2 & 5.2 & 9.6 & 10.6 & 40.8 & 99.0 & 98.9 & 99.6 \\ 
$\mathcal{T}_{2mc}$ & - & 100.0 & 100.0 & 100.0 & 98.0 & 98.9 & 100.0 & 5.0
& 5.1 & 5.1 & 13.2 & 12.7 & 63.4 & 100.0 & 100.0 & 100.0 \\ 
$\mathcal{T}_{3mc}$ & - & 100.0 & 100.0 & 100.0 & 98.9 & 99.3 & 100.0 & 4.9
& 4.8 & 5.0 & 14.5 & 14.2 & 70.1 & 100.0 & 100.0 & 100.0 \\ 
$\mathcal{T}_{4mc}$ & - & 100.0 & 100.0 & 100.0 & 98.0 & 98.9 & 100.0 & 5.0
& 5.1 & 5.1 & 13.2 & 12.7 & 63.4 & 100.0 & 100.0 & 100.0 \\ 
$\mathcal{H}_{1mc}$ & - & 99.9 & 99.8 & 99.8 & 97.7 & 98.1 & 99.7 & 4.9 & 4.8
& 5.0 & 14.4 & 13.8 & 66.2 & 99.9 & 99.9 & 99.8 \\ 
$\mathcal{H}_{2mc}$ & - & 100.0 & 100.0 & 100.0 & 98.9 & 99.3 & 100.0 & 4.8
& 4.7 & 4.9 & 14.5 & 14.2 & 70.1 & 100.0 & 100.0 & 100.0 \\ 
$\mathcal{H}_{3mc}$ & - & 100.0 & 100.0 & 100.0 & 98.0 & 98.9 & 100.0 & 5.0
& 5.1 & 5.1 & 13.2 & 12.7 & 63.4 & 100.0 & 100.0 & 100.0 \\ 
$\mathcal{R}_{mc}$ & - & 100.0 & 100.0 & 100.0 & 94.8 & 96.6 & 100.0 & 5.2 & 
5.3 & 5.4 & 7.9 & 8.4 & 41.6 & 100.0 & 100.0 & 100.0 \\ 
&  &  &  &  &  &  &  &  &  &  &  &  &  &  &  &  \\ 
$\mathcal{T}_{1mc}$ & 20 & 99.6 & 99.5 & 99.8 & 80.5 & 82.4 & 99.3 & 5.1 & 
5.3 & 5.2 & 10.6 & 10.1 & 40.1 & 99.8 & 99.8 & 99.9 \\ 
$\mathcal{T}_{2mc}$ & - & 100.0 & 100.0 & 100.0 & 93.6 & 94.8 & 100.0 & 5.1
& 5.1 & 5.0 & 12.0 & 11.5 & 51.2 & 100.0 & 100.0 & 100.0 \\ 
$\mathcal{T}_{3mc}$ & - & 100.0 & 100.0 & 100.0 & 95.0 & 95.7 & 100.0 & 4.8
& 4.7 & 4.8 & 12.5 & 12.7 & 54.3 & 100.0 & 100.0 & 100.0 \\ 
$\mathcal{T}_{4mc}$ & - & 100.0 & 100.0 & 100.0 & 93.6 & 94.8 & 100.0 & 5.1
& 5.1 & 5.0 & 12.0 & 11.5 & 51.2 & 100.0 & 100.0 & 100.0 \\ 
$\mathcal{H}_{1mc}$ & - & 100.0 & 100.0 & 100.0 & 94.0 & 94.9 & 100.0 & 4.7
& 4.7 & 4.9 & 12.0 & 12.4 & 51.4 & 100.0 & 100.0 & 100.0 \\ 
$\mathcal{H}_{2mc}$ & - & 100.0 & 100.0 & 100.0 & 95.0 & 95.7 & 100.0 & 4.8
& 4.7 & 4.8 & 12.5 & 12.7 & 54.3 & 100.0 & 100.0 & 100.0 \\ 
$\mathcal{H}_{3mc}$ & - & 100.0 & 100.0 & 100.0 & 93.6 & 94.8 & 100.0 & 5.1
& 5.1 & 5.0 & 12.0 & 11.5 & 51.2 & 100.0 & 100.0 & 100.0 \\ 
$\mathcal{R}_{mc}$ & - & 100.0 & 100.0 & 100.0 & 84.2 & 88.2 & 100.0 & 5.3 & 
5.4 & 5.2 & 7.0 & 7.3 & 26.7 & 100.0 & 100.0 & 100.0 \\ 
&  &  &  &  &  &  &  &  &  &  &  &  &  &  &  &  \\ \hline\hline
\end{tabular}
\end{center}

\end{sidewaystable}%

\begin{sidewaystable}%

\caption{Size and power of exact Monte Carlo tests with $t(3)$ errors
at nominal level 5\%} \label{Table: Power of MCEtestsStudent, T=50}

\begin{center}
Table \thetable\ . Size and power of exact Monte Carlo tests with $t(3)$
errors at nominal level $5\%$

\quad 
\scriptsize%

\begin{tabular}{|c|c|ccc|ccc|ccc|ccc|ccc|}
\hline\hline
\rule{0cm}{3ex} &  & \multicolumn{3}{c|}{$\lambda =-20$} & 
\multicolumn{3}{c|}{$\lambda =-5$} & \multicolumn{3}{c|}{$\lambda =0$} & 
\multicolumn{3}{c|}{$\lambda =1$} & \multicolumn{3}{c|}{$\lambda =100$} \\ 
\cline{3-5}\cline{6-8}\cline{9-11}\cline{12-14}\cline{15-17}
\rule{0cm}{3ex} & $k_{2}$ & $\eta _{1}=0$ & $\eta _{1}=.01$ & $\eta _{1}=.5$
& $\eta _{1}=0$ & $\eta _{1}=.01$ & $\eta _{1}=.5$ & $\eta _{1}=0$ & $\eta
_{1}=.01$ & $\eta _{1}=.5$ & $\eta _{1}=0$ & $\eta _{1}=.01$ & $\eta _{1}=.5$
& $\eta _{1}=0$ & $\eta _{1}=.01$ & $\eta _{1}=.5$ \\ 
\rule{0cm}{3ex} &  & $\eta _{2}=0$ & $\eta _{2}=0$ & $\eta _{2}=0$ & $\eta
_{2}=0$ & $\eta _{2}=0$ & $\eta _{2}=0$ & $\eta _{2}=0$ & $\eta _{2}=0$ & $%
\eta _{2}=0$ & $\eta _{2}=0$ & $\eta _{2}=0$ & $\eta _{2}=0$ & $\eta _{2}=0$
& $\eta _{2}=0$ & $\eta _{2}=0$ \\ \hline
&  &  &  &  &  &  &  &  &  &  &  &  &  &  &  &  \\ 
$\mathcal{T}_{1mc}$ & 5 & 4.4 & 4.6 & 47.1 & 4.5 & 4.9 & 42.2 & 5.2 & 4.9 & 
4.8 & 5.1 & 5.2 & 12.9 & 4.7 & 4.6 & 49.1 \\ 
$\mathcal{T}_{2mc}$ & - & 5.3 & 7.6 & 99.9 & 5.1 & 5.1 & 99.4 & 5.3 & 5.2 & 
5.4 & 5.3 & 5.5 & 32.7 & 5.2 & 50.7 & 99.9 \\ 
$\mathcal{T}_{3mc}$ & - & 4.8 & 6.3 & 96.8 & 5.0 & 5.4 & 95.7 & 4.9 & 4.7 & 
4.9 & 4.9 & 5.1 & 35.2 & 5.1 & 29.6 & 96.8 \\ 
$\mathcal{T}_{4mc}$ & - & 5.3 & 7.6 & 99.9 & 5.1 & 5.1 & 99.4 & 5.3 & 5.2 & 
5.4 & 5.3 & 5.4 & 32.7 & 5.2 & 50.7 & 99.9 \\ 
$\mathcal{H}_{1mc}$ & - & 4.9 & 6.4 & 95.7 & 4.9 & 5.3 & 94.4 & 4.7 & 4.8 & 
4.8 & 4.8 & 5.0 & 34.5 & 5.1 & 29.1 & 95.5 \\ 
$\mathcal{H}_{2mc}$ & - & 4.8 & 6.3 & 96.8 & 5.0 & 5.4 & 95.7 & 4.9 & 4.7 & 
4.9 & 4.9 & 5.1 & 35.2 & 5.1 & 29.6 & 96.8 \\ 
$\mathcal{H}_{3mc}$ & - & 5.3 & 7.6 & 99.9 & 5.0 & 5.1 & 99.4 & 5.3 & 5.2 & 
5.4 & 5.2 & 5.4 & 32.7 & 5.2 & 50.7 & 99.9 \\ 
$\mathcal{R}_{mc}$ & - & 5.4 & 9.4 & 100.0 & 5.1 & 5.1 & 99.5 & 5.1 & 5.0 & 
5.2 & 5.4 & 5.6 & 27.9 & 5.4 & 91.0 & 100.0 \\ 
&  &  &  &  &  &  &  &  &  &  &  &  &  &  &  &  \\ 
$\mathcal{T}_{1mc}$ & 10 & 4.5 & 4.7 & 91.1 & 4.7 & 4.9 & 82.8 & 5.1 & 4.9 & 
5.1 & 5.0 & 5.2 & 23.2 & 5.1 & 4.4 & 90.5 \\ 
$\mathcal{T}_{2mc}$ & - & 5.2 & 6.9 & 99.9 & 5.4 & 5.3 & 99.5 & 5.1 & 5.2 & 
5.3 & 5.3 & 5.2 & 39.2 & 5.4 & 31.9 & 99.9 \\ 
$\mathcal{T}_{3mc}$ & - & 5.0 & 6.4 & 99.8 & 5.1 & 5.1 & 99.4 & 4.8 & 4.9 & 
4.9 & 5.1 & 5.1 & 43.3 & 5.1 & 26.7 & 99.7 \\ 
$\mathcal{T}_{4mc}$ & - & 5.2 & 6.9 & 99.9 & 5.4 & 5.3 & 99.5 & 5.1 & 5.2 & 
5.3 & 5.3 & 5.2 & 39.2 & 5.4 & 31.9 & 99.9 \\ 
$\mathcal{H}_{1mc}$ & - & 4.9 & 6.4 & 99.7 & 5.0 & 5.1 & 99.2 & 4.8 & 4.8 & 
4.7 & 5.0 & 5.1 & 42.4 & 4.9 & 26.5 & 99.7 \\ 
$\mathcal{H}_{2mc}$ & - & 5.0 & 6.4 & 99.8 & 5.1 & 5.1 & 99.4 & 4.8 & 4.9 & 
4.9 & 5.1 & 5.1 & 43.3 & 5.1 & 26.7 & 99.7 \\ 
$\mathcal{H}_{3mc}$ & - & 5.2 & 6.9 & 99.9 & 5.4 & 5.3 & 99.5 & 5.1 & 5.2 & 
5.3 & 5.3 & 5.2 & 39.2 & 5.4 & 31.9 & 99.9 \\ 
$\mathcal{R}_{mc}$ & - & 5.5 & 10.6 & 100.0 & 5.5 & 5.4 & 99.7 & 5.1 & 5.1 & 
5.2 & 5.3 & 5.5 & 27.7 & 5.7 & 95.5 & 100.0 \\ 
&  &  &  &  &  &  &  &  &  &  &  &  &  &  &  &  \\ 
$\mathcal{T}_{1mc}$ & 20 & 4.8 & 4.2 & 98.0 & 5.0 & 4.8 & 95.0 & 4.9 & 4.8 & 
4.8 & 5.0 & 5.1 & 28.7 & 5.2 & 4.8 & 98.0 \\ 
$\mathcal{T}_{2mc}$ & - & 5.4 & 5.9 & 99.9 & 5.3 & 5.1 & 99.4 & 5.1 & 5.0 & 
5.1 & 5.2 & 5.1 & 38.2 & 5.3 & 12.0 & 99.9 \\ 
$\mathcal{T}_{3mc}$ & - & 5.1 & 5.8 & 99.9 & 5.1 & 5.1 & 99.5 & 4.8 & 5.0 & 
4.7 & 4.8 & 4.9 & 40.7 & 5.1 & 11.2 & 99.8 \\ 
$\mathcal{T}_{4mc}$ & - & 5.4 & 5.9 & 99.9 & 5.3 & 5.1 & 99.4 & 5.1 & 5.0 & 
5.1 & 5.2 & 5.1 & 38.2 & 5.3 & 12.0 & 99.9 \\ 
$\mathcal{H}_{1mc}$ & - & 5.1 & 5.8 & 99.9 & 5.1 & 5.2 & 99.4 & 4.9 & 4.9 & 
4.8 & 4.8 & 4.8 & 40.3 & 5.1 & 11.3 & 99.9 \\ 
$\mathcal{H}_{2mc}$ & - & 5.1 & 5.8 & 99.9 & 5.1 & 5.1 & 99.5 & 4.8 & 5.0 & 
4.7 & 4.8 & 4.9 & 40.7 & 5.1 & 11.2 & 99.8 \\ 
$\mathcal{H}_{3mc}$ & - & 5.4 & 5.9 & 99.9 & 5.3 & 5.1 & 99.4 & 5.1 & 5.0 & 
5.1 & 5.2 & 5.1 & 38.2 & 5.3 & 12.0 & 99.9 \\ 
$\mathcal{R}_{mc}$ & - & 5.7 & 12.3 & 100.0 & 5.2 & 5.6 & 99.3 & 5.2 & 5.2 & 
5.3 & 5.3 & 5.4 & 22.9 & 5.9 & 98.3 & 100.0 \\ 
&  &  &  &  &  &  &  &  &  &  &  &  &  &  &  &  \\ \hline\hline
\end{tabular}
\end{center}

\end{sidewaystable}%

\begin{sidewaystable}%


\begin{center}
Table \thetable\ (Continued). Size and power of exact Monte Carlo tests with 
$t(3)$ errors at nominal level $5\%$

\quad 
\scriptsize%

\begin{tabular}{|c|c|ccc|ccc|ccc|ccc|ccc|}
\hline\hline
\rule{0cm}{3ex} &  & \multicolumn{3}{c|}{$\lambda =-20$} & 
\multicolumn{3}{c|}{$\lambda =-5$} & \multicolumn{3}{c|}{$\lambda =0$} & 
\multicolumn{3}{c|}{$\lambda =1$} & \multicolumn{3}{c|}{$\lambda =100$} \\ 
\cline{3-5}\cline{6-8}\cline{9-11}\cline{12-14}\cline{15-17}
\rule{0cm}{3ex} & $k_{2}$ & $\eta _{1}=0$ & $\eta _{1}=.01$ & $\eta _{1}=.5$
& $\eta _{1}=0$ & $\eta _{1}=.01$ & $\eta _{1}=.5$ & $\eta _{1}=0$ & $\eta
_{1}=.01$ & $\eta _{1}=.5$ & $\eta _{1}=0$ & $\eta _{1}=.01$ & $\eta _{1}=.5$
& $\eta _{1}=0$ & $\eta _{1}=.01$ & $\eta _{1}=.5$ \\ 
\rule{0cm}{3ex} &  & $\eta _{2}=.5$ & $\eta _{2}=.5$ & $\eta _{2}=.5$ & $%
\eta _{2}=.5$ & $\eta _{2}=.5$ & $\eta _{2}=.5$ & $\eta _{2}=.5$ & $\eta
_{2}=.5$ & $\eta _{2}=.5$ & $\eta _{2}=.5$ & $\eta _{2}=.5$ & $\eta _{2}=.5$
& $\eta _{2}=.5$ & $\eta _{2}=.5$ & $\eta _{2}=.5$ \\ \hline
&  &  &  &  &  &  &  &  &  &  &  &  &  &  &  &  \\ 
$\mathcal{T}_{1mc}$ & 5 & 46.7 & 46.9 & 67.0 & 25.6 & 27.3 & 58.7 & 4.7 & 4.9
& 5.0 & 6.3 & 6.5 & 18.4 & 50.3 & 51.8 & 68.9 \\ 
$\mathcal{T}_{2mc}$ & - & 99.9 & 99.8 & 100.0 & 83.3 & 85.7 & 99.9 & 5.2 & 
5.1 & 5.4 & 9.1 & 9.4 & 48.9 & 99.9 & 99.9 & 100.0 \\ 
$\mathcal{T}_{3mc}$ & - & 96.7 & 96.9 & 99.2 & 79.9 & 82.4 & 98.7 & 4.9 & 4.8
& 4.9 & 10.1 & 10.1 & 52.6 & 96.8 & 97.2 & 99.1 \\ 
$\mathcal{T}_{4mc}$ & - & 99.9 & 99.8 & 100.0 & 83.3 & 85.7 & 99.9 & 5.2 & 
5.1 & 5.4 & 9.1 & 9.4 & 48.9 & 99.9 & 99.9 & 100.0 \\ 
$\mathcal{H}_{1mc}$ & - & 95.2 & 95.6 & 97.5 & 77.5 & 79.7 & 96.6 & 4.6 & 4.7
& 4.9 & 9.9 & 10.1 & 50.3 & 95.6 & 96.0 & 97.7 \\ 
$\mathcal{H}_{2mc}$ & - & 96.7 & 96.9 & 99.2 & 79.9 & 82.4 & 98.7 & 4.9 & 4.8
& 4.9 & 10.1 & 10.1 & 52.6 & 96.8 & 97.2 & 99.1 \\ 
$\mathcal{H}_{3mc}$ & - & 99.9 & 99.8 & 100.0 & 83.3 & 85.7 & 99.9 & 5.2 & 
5.1 & 5.4 & 9.1 & 9.4 & 48.9 & 99.9 & 99.9 & 100.0 \\ 
$\mathcal{R}_{mc}$ & - & 100.0 & 99.9 & 100.0 & 79.6 & 82.9 & 99.8 & 5.3 & 
5.2 & 5.1 & 7.3 & 7.7 & 40.2 & 100.0 & 100.0 & 100.0 \\ 
&  &  &  &  &  &  &  &  &  &  &  &  &  &  &  &  \\ 
$\mathcal{T}_{1mc}$ & 10 & 89.6 & 89.8 & 98.6 & 56.3 & 56.9 & 95.7 & 5.1 & 
5.3 & 5.2 & 8.6 & 8.8 & 34.6 & 91.2 & 91.5 & 98.6 \\ 
$\mathcal{T}_{2mc}$ & - & 99.7 & 99.9 & 100.0 & 87.5 & 89.1 & 99.9 & 5.4 & 
5.2 & 5.2 & 10.9 & 11.2 & 53.0 & 99.8 & 99.9 & 100.0 \\ 
$\mathcal{T}_{3mc}$ & - & 99.6 & 99.7 & 100.0 & 89.7 & 91.5 & 99.9 & 5.0 & 
4.9+ & 5.1 & 11.6 & 12.4 & 56.9 & 99.6 & 99.8 & 100.0 \\ 
$\mathcal{T}_{4mc}$ & - & 99.7 & 99.9 & 100.0 & 87.5 & 89.1 & 99.9 & 5.4 & 
5.2 & 5.2 & 10.9 & 11.2 & 53.0 & 99.8 & 99.9 & 100.0 \\ 
$\mathcal{H}_{1mc}$ & - & 99.5 & 99.7 & 99.9 & 88.7 & 90.2 & 99.6 & 4.9 & 5.1
& 4.8 & 11.5 & 12.1 & 55.1 & 99.6 & 99.8 & 99.9 \\ 
$\mathcal{H}_{2mc}$ & - & 99.6 & 99.7 & 100.0 & 89.7 & 91.5 & 99.9 & 5.0 & 
4.9 & 5.1 & 11.6 & 12.4 & 56.9 & 99.6 & 99.8 & 100.0 \\ 
$\mathcal{H}_{3mc}$ & - & 99.7 & 99.9 & 100.0 & 87.5 & 89.1 & 99.9 & 5.4 & 
5.2 & 5.2 & 10.9 & 11.2 & 53.0 & 99.8 & 99.9 & 100.0 \\ 
$\mathcal{R}_{mc}$ & - & 99.9 & 100.0 & 100.0 & 82.6 & 83.9 & 99.8 & 5.5 & 
5.3 & 5.1 & 8.0 & 7.8 & 35.0 & 100.0 & 100.0 & 100.0 \\ 
&  &  &  &  &  &  &  &  &  &  &  &  &  &  &  &  \\ 
$\mathcal{T}_{1mc}$ & 20 & 97.3 & 97.6 & 99.8 & 69.8 & 71.5 & 98.2 & 4.8 & 
4.8 & 5.1 & 9.5 & 10.4 & 38.8 & 98.4 & 98.8 & 99.9 \\ 
$\mathcal{T}_{2mc}$ & - & 99.7 & 99.7 & 100.0 & 84.9 & 86.7 & 99.7 & 5.1 & 
5.0 & 5.3 & 10.9 & 10.8 & 48.3 & 99.9 & 99.9 & 100.0 \\ 
$\mathcal{T}_{3mc}$ & - & 99.8 & 99.7 & 100.0 & 87.1 & 88.4 & 99.7 & 4.9 & 
4.8 & 5.0 & 11.4 & 11.9 & 50.8 & 99.9 & 99.9 & 100.0 \\ 
$\mathcal{T}_{4mc}$ & - & 99.7 & 99.7 & 100.0 & 84.9 & 86.7 & 99.7 & 5.1 & 
5.0 & 5.3 & 10.9 & 10.8 & 48.3 & 99.9 & 99.9 & 100.0 \\ 
$\mathcal{H}_{1mc}$ & - & 99.7 & 99.7 & 100.0 & 86.3 & 87.7 & 99.6 & 4.7 & 
4.6 & 5.1 & 11.5 & 11.6 & 49.0 & 99.9 & 99.9 & 100.0 \\ 
$\mathcal{H}_{2mc}$ & - & 99.8 & 99.7 & 100.0 & 87.1 & 88.4 & 99.7 & 4.9 & 
4.8 & 5.0 & 11.4 & 11.9 & 50.8 & 99.9 & 99.9 & 100.0 \\ 
$\mathcal{H}_{3mc}$ & - & 99.7 & 99.7 & 100.0 & 84.9 & 86.7 & 99.7 & 5.1 & 
5.0 & 5.3 & 10.9 & 10.8 & 48.3 & 99.9 & 99.9 & 100.0 \\ 
$\mathcal{R}_{mc}$ & - & 100.0 & 99.9 & 100.0 & 75.6 & 79.3 & 99.6 & 5.3 & 
5.2 & 5.4 & 7.3 & 7.8 & 26.4 & 100.0 & 100.0 & 100.0 \\ 
&  &  &  &  &  &  &  &  &  &  &  &  &  &  &  &  \\ \hline\hline
\end{tabular}
\end{center}

\normalsize%

\end{sidewaystable}%

\subsection{\sectitlesize Performance of the exact Monte Carlo tests \label%
{sec: PerfMCE}}

We now examine the performance of the proposed exact Monte Carlo exogeneity
tests. Tables \ref{Table: Power of MCEtestsGau, T=50}\thinspace -\thinspace %
\ref{Table: Power of MCEtestsStudent, T=50} present the results for Gaussian
errors (Table \ref{Table: Power of MCEtestsGau, T=50}) and $t(3)$ errors
(Table \ref{Table: Power of MCEtestsStudent, T=50}). The results confirm our
theoretical findings.\emph{\ }

\emph{First}, the rejection frequencies under the null hypothesis of
exogeneity ($\lambda =0$) of all Monte Carlo tests are around $5\%$ whether
identification is weak ($\eta _{1},\eta _{2}\in \{0,\,0.01\}$), partial ($%
\eta _{1}\in \{0,\,0.01\}\,\text{and}\,\eta _{2}=0.5$ or \emph{vice versa}),
or strong ($\eta _{1}=\eta _{2}=0.5$), with or without Gaussian errors. This
represents a substantial improvement for the standard $\mathcal{T}_{3}$, $%
\mathcal{H}_{2}$ and \cite{Hausman(1978)} $\mathcal{H}_{1}$ statistics.

\emph{Second}, when $\lambda \neq 0$ (endogeneity), the rejection
frequencies of all tests improve in most cases. This is especially the case
for $\mathcal{T}_{3}$, $\mathcal{H}_{1}$ and $\mathcal{H}_{2}.$ For example,
with Gaussian errors and $k_{2}=5$ instruments, the rejection frequencies of 
$\mathcal{T}_{3}$, $\mathcal{H}_{1}$ and $\mathcal{H}_{2}$ have increased
from $34.1\%$, $20.9\%$ and $36.8\%$ (for the standard tests) to $60.7\%$, $%
56.5\%$ and $60.7\%$ (for the exact Monte Carlo tests); see the columns for $%
\lambda =1$ ($\eta _{1}=0.5$ and $\eta _{2}=0$) in Tables \ref{Table: Power
of exogeneity tests, T=50} and \ref{Table: Power of MCEtestsGau, T=50}. The
results are more remarkable with $t(3)$ errors and $k_{2}=5$ instruments. In
this case, the rejection frequencies of the exact Monte Carlo $\mathcal{T}%
_{3}$, $\mathcal{H}_{1}$ and $\mathcal{H}_{2}$ tests have tripled those of
their standard versions; see $\lambda =1$ ($\eta _{1}=0.5$ and $\eta _{2}=0$%
) in Tables \ref{Table: Power of exogeneity tests t(3), T=50} and \ref%
{Table: Power of MCEtestsStudent, T=50}. The results are essentially the
same for other values of $k_{2}$, $\lambda $ and IV strength ($\eta _{1}$
and $\eta _{2}$). Moreover, except for $\mathcal{T}_{1}$, the other exact
Monte Carlo tests exhibit power with or without Gaussian errors, including
when identification is very weak ($\eta _{1}=0.01,\,\eta _{2}=0$) and
endogeneity is large ($\lambda =100$ for example). Note that the standard
exogeneity tests (including $\mathcal{T}_{2}$ and $R$) perform poorly in
this case. Thus, size correction through the exact Monte Carlo test method
yields a substantial improvement for the exogeneity tests considered. In
addition, observe that after size correction, even the \cite{Hausman(1978)}
statistic ($\mathcal{H}_{1}$) becomes attractive in terms of power. This is
the case in particular for $t(3)$ errors when $k_{2}=10,\,20$ and $\lambda
=-5,\,1$; see Table \ref{Table: Power of MCEtestsStudent, T=50}.

\section{\sectitlesize Empirical illustrations \label{Sec: Empirical
illustrations}}

\resetcountersSection

We illustrate our theoretical results on exogeneity tests through two
empirical applications related to important issues in macroeconomics and
labor economics literature: (1) the relation between trade and growth [\cite%
{Irwin-Tervio(2002)}, \cite{Frankel-Romer(1999)}, \cite{Harrisson(1996)}, 
\cite{Mankiw-al(1992)}]; (2) the standard problem of measuring returns to
education [\cite{Dufour-Taamouti(2007)}, \cite{Angrist-Krueger(1991)}, \cite%
{Angrist-Krueger(1995)}, \cite{Angrist-al(1999)}, \cite{Mankiw-al(1992)}].

\subsection{\sectitlesize Trade and growth \label{Sec: Trade and growth}}

The trade and growth model studies the relationship between standards of
living and openness. \cite{Frankel-Romer(1999)} argued that trade share
(ratio of imports or exports to GDP) which is the commonly used indicator of
openness should be viewed as endogenous. So, instrumental variables method
should be used to estimate the income-trade relationship. The equation
studied is 
\begin{equation}
ln(\mathrm{Inc}_{i})=\beta _{0}+\beta _{1}\text{Trade}_{i}+\gamma _{1}ln(%
\text{Pop}_{i})+\gamma _{2}ln(\text{Area}_{i})+u_{i},\,i=1,\,\ldots ,\,T
\label{eq: empistruct2}
\end{equation}%
where Inc$_{i}$ is the income per capita in country $i,$ $\text{Trade}_{i}$
is the trade share (measured as a ratio of imports and exports to GDP), $%
\text{Pop}_{i}$ is the population of country $i,\,$ and $\text{Area}_{i}$ is
country $i$ area. The first stage model for $\text{Trade}$ variable is given
by 
\begin{equation}
\text{Trade}_{i}=a+bX_{i}+c_{1}ln(\text{Pop}_{i})+c_{2}ln(\text{Area}%
_{i})+V_{i},\,i=1,\,\ldots ,\,T  \label{eq: empireduced}
\end{equation}%
where $X_{i}$ is an instrument constructed on the basis of geographic
characteristics. In this paper, we use the sample of 150 countries and the
data include for each country: the trade share in 1985, the area and
population (1985), per capita income (1985), and the fitted trade share
(instrument).

We wish to assess the exogeneity of the trade share variable in (\ref{eq:
empistruct2}). The $F$-statistic in the first stage regression (\ref{eq:
empireduced}) is around 13 [see \cite[Table 2, p.385]{Frankel-Romer(1999)}
and \cite{Dufour-Taamouti(2007)}], so the fitted instrument $X$ does not
appear to be weak. Table \ref{Table: Trade and income} presents the $p$%
-values of the DWH and RH tests computed from the tabulated and exact Monte
Carlo critical values. The Monte Carlo critical values are computed for
Gaussian and $t(3)$ errors. Because the model contains one instrument and
one (supposedly) endogenous variable, the statistic $T_{1}$ is not well
defined and is omitted.

\begin{table}[h]%

\caption{Exogeneity in trade and growth model } \label{Table: Trade and
income}

\begin{center}
Table \thetable. Exogeneity in trade and growth model

\quad 
\small%

\begin{tabular}{|c|c|c|c|c|}
\hline\hline
Statistics \rule{0cm}{3ex} & Estimation & Standard $p$-value (\%) & MC $p$%
-value (\%) & MC $p$-value (\%) \\ 
&  &  & (Gaussian errors) & [$t(3)$-errors] \\ \hline
&  &  &  &  \\ 
$\mathcal{R}$ & 3.9221 & 4.95 & 4.98 & 5.38 \\ 
$\mathcal{H}_{1}$ & 2.3883 & 12.23 & 6.14 & 5.99 \\ 
$\mathcal{H}_{2}$ & 2.4269 & 11.93 & 6.12 & 5.96 \\ 
$\mathcal{H}_{3}$ & 3.9505 & 4.67 & 5.39 & 5.66 \\ 
$\mathcal{T}_{2}$ & 3.9221 & 4.95 & 5.39 & 5.66 \\ 
$\mathcal{T}_{3}$ & 2.3622 & 12.43 & 6.12 & 5.96 \\ 
$\mathcal{T}_{4}$ & 3.8451 & 4.99 & 5.49 & 5.66 \\ \hline\hline
\end{tabular}
\end{center}

\normalsize%

\end{table}%

\emph{First}, we note that the $p$-values based on the usual asymptotic
distributions are close to the $5\%$ nominal level for $\mathcal{H}_{3},$ $%
\mathcal{T}_{2},$ $\mathcal{T}_{4}$ and $\mathcal{R}$. So, there is evidence
against the exogeneity of the trade share (at nominal level of $5\%$) when
these statistics are applied. Meanwhile, the $p$-values of $\mathcal{H}_{1},$
$\mathcal{H}_{2},$ and $\mathcal{T}_{3}$ are relatively large (around 12\%)
so that there is little evidence against trade share exogeneity at $5\%$
nominal level using the latter statistics. Since the standard $\mathcal{H}%
_{1},$ $\mathcal{H}_{2},$ and $\mathcal{T}_{3}$ tests are conservative when
identification is weak, the latter result may be due to the fact that the
fitted instrument is not very strong.

\emph{Second}, we observe the exact Monte Carlo tests yield $p$-values close
to the $5\%$ level in all cases, thus indicating that there is evidence of
trade share endogeneity in this model. This is supported by the relatively
large discrepancy between the OLS estimate of $\beta _{1}$ (0.28) and the
2SLS estimate (2.03). Overall, our results underscore the importance of size
correction through the exact Monte Carlo procedures proposed.

\subsection{\sectitlesize Education and earnings \label{Sec: Education and
earnings}}

We now consider the well known example of estimating the returns to
education [see \cite{Angrist-Krueger(1991)}; \cite{Angrist-Krueger(1995)};
and \cite{Bound-Jaeger-Baker(1995)}]. The equation studies is a relationship
where the log-weekly earning ($y$) is explained by the number of years of
education ($E$) and several other covariates (age, age squared, 10 dummies
for birth of year): 
\begin{equation}
y=\beta _{0}+\beta _{1}E+\sum_{i=1}^{k_{1}}\gamma _{i}X_{i}+u.\,
\label{eq: empistructEducch2}
\end{equation}%
In this model, $\beta _{1}$ measures the return to education. Because
education can be viewed as endogenous, \cite{Angrist-Krueger(1991)} used
instrumental variables obtained by interacting quarter of birth with the
year of birth (in this application, we use 40 dummies instruments). The
basic idea is that individuals born in the first quarter of the year start
school at an older age, and can therefore drop out after completing less
schooling than individuals born near the end of the year. Consequently,
individuals born at the beginning of the year are likely to earn less than
those born during the rest of the year. The first stage model for $E$ is
then given by 
\begin{equation}
E=\pi _{0}+\sum_{i=1}^{k_{2}}\pi _{i}X_{i}+\sum_{i=1}^{k_{1}}\phi
_{i}X_{i}+V\,  \label{eq: empiReducEducch2}
\end{equation}%
where $X$ is the instrument matrix. It is well known that the instruments $X$
constructed in this way are very weak and explains very little of the
variation in education; see \cite{Bound-Jaeger-Baker(1995)}. The data set
consists of the $5\%$ public-use sample of the 1980 US census for men born
between 1930 and 1939. The sample contains 329 509 observations.

\begin{table}[tb]%

\caption{Exogeneity in education and earning model} \label{Table:
income-education}

\begin{center}
Table \thetable. Exogeneity in education and earning model

\quad 
\small%

\begin{tabular}{|c|c|c|c|c|}
\hline\hline
Statistics \rule{0cm}{3ex} & Estimation & Standard $p$-value (\%) & MC $p$%
-value (\%) & MC $p$-value (\%) \\ 
&  &  & (Gaussian errors) & [$t(3)$-errors] \\ \hline
&  &  &  &  \\ 
$\mathcal{R}$ & 0.68 & 93.99 & 49.91 & 49.93 \\ 
$\mathcal{H}_{1}$ & 1.34 & 24.76 & 24.26 & 24.30 \\ 
$\mathcal{H}_{2}$ & 1.34 & 24.76 & 24.26 & 24.30 \\ 
$\mathcal{H}_{3}$ & 1.35 & 24.54 & 24.26 & 24.30 \\ 
$\mathcal{T}_{1}$ & 2.04 & 16.11 & 22.49 & 22.99 \\ 
$\mathcal{T}_{2}$ & 1.35 & 24.54 & 24.26 & 24.30 \\ 
$\mathcal{T}_{3}$ & 1.35 & 22.48 & 24.26 & 24.30 \\ 
$\mathcal{T}_{4}$ & 1.35 & 24.54 & 24.26 & 24.30 \\ \hline\hline
\end{tabular}
\end{center}

\normalsize%

\end{table}%

As in Section \ref{Sec: Education and earnings}, we want to assess the
exogeneity of education in (\ref{eq: empistructEducch2})\thinspace
-\thinspace (\ref{eq: empiReducEducch2}). Table \ref{Table: income-education}
shows the results of the tests with both the usual and exact Monte Carlo
critical values. As seen, the $p$-values of all tests are quite large, thus
suggesting that there is little evidence against the exogeneity of the
education variable, even at 15\% nominal level. This means that either the
education variable is effectively exogenous or the instruments used are very
poor so that the power of the test is flat, as shown in Section \ref{Sec:
Power}. The latter scenario is highly plausible from the previous literature
[for example, see \cite{Bound-Jaeger-Baker(1995)}]. This viewed is
reinforced by the small discrepancy between the OLS estimate (0.07) and the
2SLS estimate (0.08) of $\beta _{1}.$

\section{\sectitlesize Conclusion \label{sec: Conclusion}}

\resetcountersSection

This paper develops a finite-sample theory of the distribution of standard
Durbin-Wu-Hausman and Revankar-Hartley specification tests under both the
null hypothesis of exogeneity (level) and the alternative hypothesis of
endogeneity (power), with or without identification. Our analysis provides
several new insights and extensions of earlier procedures.

Our study of the finite-sample distributions of the statistics under the
null hypothesis shows that all tests are robust to weak instruments, missing
instruments or misspecified reduced forms -- in the sense that level is
controlled. Indeed, we provided a general characterization of the structure
of the test statistics which allows one to perform exact Monte Carlo tests
under general parametric distributional assumptions, which are in no way
restricted to the Gaussian case, including heavy-tailed distributions
without moments. The tests so obtained are exact even in cases where
identification fails (or is weak) and conventional asymptotic theory breaks
down.

After proving a general invariance property, we provided a characterization
of the power of the tests that clearly exhibits the factors which determine
power. We showed that exogeneity tests have no power in the extreme case
where all IVs are weak [similar to \cite{Staiger-Stock(1997)}, and \cite%
{Guggenberger(2008)}], but typically have power as soon as we have one
strong instrument. Consequently, exogeneity tests can detect an exogeneity
problem even if not all model parameters are identified, provided at least
some parameters are identifiable.

Though the exact distributional theory given in this paper requires
relatively specific distributional assumptions, the \textquotedblleft
finite-sample\textquotedblright\ procedures provided remain asymptotically
valid in the same way (in the sense that test level is controlled) under
standard asymptotic assumptions. We study this problem in a separate paper [%
\cite{Doko-Dufour(2016b)}]. Further, even if exogeneity hypotheses can have
economic interest by themselves, we also show there how exogeneity tests can
be fruitfully applied to build pretest estimators which generally dominate
OLS and 2SLS estimators when the exogeneity of explanatory variables is in
uncertain.

\newpage

\appendix

\begin{center}
APPENDIX

\quad
\end{center}

\section{Wu and Hausman test statistics \label{sec: Wu and Hausman test
statistics}}

\resetcountersSection

We show here that Durbin-Wu statistics can be expressed in the same way as
alternative Hausman statistics. The statistics $T_{l},$ $l=1,\,2,\,3,\,4\,$
are defined in \cite[eqs. (2.1), (2.18), (3.16), and (3.20)]{Wu(1973)} as: 
\begin{eqnarray}
\mathcal{T}_{1} &=&\kappa _{1}\frac{Q^{\ast }}{Q_{1}},\,\mathcal{T}%
_{2}=\kappa _{2}\frac{Q^{\ast }}{Q_{2}},\,\mathcal{T}_{3}=\kappa _{3}\frac{%
Q^{\ast }}{Q_{3}},\,\mathcal{T}_{4}=\kappa _{4}\frac{Q^{\ast }}{Q_{4}},\,
\label{eq:AppWuT3T473} \\
Q^{\ast } &=&(b_{1}-b_{2})^{\prime }\left[ (Y^{\prime
}A_{2}Y)^{-1}-(Y^{\prime }A_{1}Y)^{-1}\right] ^{-1}(b_{1}-b_{2}),\,
\label{eq:AppWuQast} \\
Q_{1} &=&(y-Yb_{2})^{\prime }A_{2}(y-Yb_{2}),\,Q_{2}=Q_{4}-Q^{\ast },\,
\label{eq:AppWuQ1} \\
Q_{4} &=&(y-Yb_{1})^{\prime }A_{1}(y-Yb_{1}),\,Q_{3}=(y-Yb_{2})^{\prime
}A_{1}(y-Yb_{2}),\,  \label{eq:AppWuQ4} \\
b_{i} &=&(Y^{\prime }A_{i}Y)^{-1}Y^{\prime
}A_{i}y,\,i=1,\,2,\,A_{1}=M_{1},\,A_{2}=M-M_{1}\,,  \label{eq:AppWuA2}
\end{eqnarray}%
where $b_{1}$ is the ordinary least squares estimator of $\beta $, and $%
b_{2} $ is the instrumental variables method estimator of $\beta $. So, in
our notations, $b_{1}\equiv \hat{\beta}$ and $b_{2}\equiv \tilde{\beta}.$
From (\ref{eq: SigmaHat})\thinspace -\thinspace (\ref{eq: sigma2hat}), we
have: 
\begin{eqnarray}
Q^{\ast } &=&T\,(\tilde{\beta}-\hat{\beta})^{\prime }\hat{\Delta}^{-1}(%
\tilde{\beta}-\hat{\beta})=T\tilde{\sigma}^{2}(\tilde{\beta}-\hat{\beta}%
)^{\prime }\hat{\Sigma}_{2}^{-1}(\tilde{\beta}-\hat{\beta})\,, \\
Q_{1} &=&T\,\tilde{\sigma}_{1}^{2}\,,\text{\quad }Q_{3}=T\,\tilde{\sigma}%
^{2}\,,\text{\quad }Q_{4}=T\hat{\sigma}^{2}\,, \\
Q_{2} &=&Q_{4}-Q^{\ast }=T\hat{\sigma}^{2}-T\,(\tilde{\beta}-\hat{\beta}%
)^{\prime }\hat{\Delta}^{-1}(\tilde{\beta}-\hat{\beta})=T\,\tilde{\sigma}%
_{2}^{2}\,.\,
\end{eqnarray}%
Hence, we can write $\mathcal{T}_{l}\,$ as: 
\begin{equation*}
\mathcal{T}_{l}=\kappa _{l}(\tilde{\beta}-\hat{\beta})^{\prime }\tilde{\Sigma%
}_{l}^{-1}(\tilde{\beta}-\hat{\beta})\,,\text{\quad }l=1,\,2,\,3,\,4\,,\,
\end{equation*}%
where $\kappa _{l},\,$ and $\tilde{\Sigma}_{l}\,$ are defined in (\ref{eq:
SigmaHat})\thinspace -\thinspace (\ref{eq: sigma2hat}).

To obtain (\ref{eq: link T_4 T_2}), set $\mathcal{T}_{0}=(\tilde{\beta}-\hat{%
\beta})^{\prime }\hat{\Delta}^{-1}(\tilde{\beta}-\hat{\beta})$. Then $\tilde{%
\sigma}_{2}^{2}=\hat{\sigma}^{2}-\mathcal{T}_{0},$ $\mathcal{T}_{4}=\kappa
_{4}\mathcal{T}_{0}/\hat{\sigma}^{2}$, and%
\begin{equation}
\mathcal{T}_{2}=\kappa _{2}\frac{\mathcal{T}_{0}}{\tilde{\sigma}_{2}^{2}}%
=\kappa _{2}\frac{\mathcal{T}_{0}}{\hat{\sigma}^{2}-\mathcal{T}_{0}}=\kappa
_{2}\frac{(\mathcal{T}_{0}/\hat{\sigma}^{2})}{1-(\mathcal{T}_{0}/\hat{\sigma}%
^{2})}=\kappa _{2}\frac{(\mathcal{T}_{4}/\kappa _{4})}{1-(\mathcal{T}%
_{4}/\kappa _{4})}\,,
\end{equation}%
hence 
\begin{equation}
\frac{\mathcal{T}_{4}}{\kappa _{4}}=\frac{(\mathcal{T}_{2}/\kappa _{2})}{(%
\mathcal{T}_{2}/\kappa _{2})+1}=\frac{\mathcal{T}_{2}}{\mathcal{T}%
_{2}+\kappa _{2}}=\frac{1}{(\kappa _{2}/\mathcal{T}_{2})+1}\,.
\end{equation}

In the sequel of this appendix, we shall use the following matrix formulas
which are easily established by algebraic manipulations [on the
invertibility of matrix differences, see \cite[Theorem 18.2.4]%
{Harville(1997)}].

\begin{lemma}
\label{Th: Difference of matrix inverses} 
\captiontheorem{\lemmaname}{Difference of matrix inverses}
Let $A$ and $B$ be two nonsingular $r\times r$ matrices. Then 
\begin{eqnarray}
A^{-1}-B^{-1} &=&B^{-1}(B-A)A^{-1}=A^{-1}(B-A)B^{-1}  \notag \\
&=&A^{-1}(A-AB^{-1}A)A^{-1}=B^{-1}(BA^{-1}B-B)B^{-1}.
\label{eq: Inv(A) - Inv(B)}
\end{eqnarray}%
Furthermore, $A^{-1}-B^{-1}$ is nonsingular if and only if $B-A$ is
nonsingular. If $B-A$ is nonsingular, we have: 
\begin{eqnarray}
(A^{-1}-B^{-1})^{-1}
&=&A(B-A)^{-1}B=A-A(A-B)^{-1}A=A+A(B-A)^{-1}A=A[A^{-1}+(B-A)^{-1}]A  \notag
\\
&=&B(B-A)^{-1}A=B(B-A)^{-1}B-B=B[(B-A)^{-1}-B^{-1}]B  \notag \\
&=&A(A-AB^{-1}A)^{-1}A=B(BA^{-1}B-B)^{-1}B\,.
\label{eq: Inv[Inv(A) - Inv(B)]}
\end{eqnarray}
\end{lemma}

It is easy to see from condition (\ref{eq: Rank conditions}) that $\hat{%
\Omega}_{IV}$, $\hat{\Omega}_{LS}$ and $\hat{\Sigma}_{V}$ are nonsingular.
On setting $A=\hat{\Omega}_{IV}$ and $B=\hat{\Omega}_{LS}$, we get: 
\begin{equation}
B-A=\hat{\Omega}_{LS}-\hat{\Omega}_{IV}=\frac{1}{T}Y^{\prime }M_{1}Y-\frac{1%
}{T}Y^{\prime }N_{1}Y=\frac{1}{T}Y^{\prime }(M_{1}-N_{1})Y=\frac{1}{T}%
Y^{\prime }MY=\frac{1}{T}\hat{V}^{\prime }\hat{V}=\hat{\Sigma}_{V}\,,
\end{equation}%
so $\hat{\Omega}_{LS}-\hat{\Omega}_{IV}$ is nonsingular. By Lemma \ref{Th:
Difference of matrix inverses}, $\hat{\Delta}=\hat{\Omega}_{IV}^{-1}-\hat{%
\Omega}_{LS}^{-1}=A^{-1}-B^{-1}$ is also nonsingular, and 
\begin{eqnarray}
\hat{\Delta}^{-1} &=&A+A(B-A)^{-1}A=\hat{\Omega}_{IV}+\hat{\Omega}_{IV}(\hat{%
\Omega}_{LS}-\hat{\Omega}_{IV})^{-1}\hat{\Omega}_{IV}=\hat{\Omega}_{IV}+\hat{%
\Omega}_{IV}\,\hat{\Sigma}_{V}^{-1}\,\hat{\Omega}_{IV}  \notag \\
&=&\frac{1}{T}\left[ Y^{\prime }N_{1}Y+Y^{\prime }N_{1}Y(Y^{\prime
}MY)^{-1}Y^{\prime }N_{1}Y\right] =\frac{1}{T}Y^{\prime }N_{1}\left[
I_{T}+Y(Y^{\prime }MY)^{-1}Y^{\prime }\right] N_{1}Y\,.\quad \quad
\end{eqnarray}%
From the above form, it is clear that $\hat{\Delta}^{-1}$ is positive
definite. Note also that 
\begin{eqnarray}
\hat{\Delta}^{-1} &=&B(B-A)^{-1}B-B=\hat{\Omega}_{LS}(\hat{\Omega}_{LS}-\hat{%
\Omega}_{IV})^{-1}\hat{\Omega}_{LS}-\hat{\Omega}_{LS}=\hat{\Omega}_{LS}\,%
\hat{\Sigma}_{V}^{-1}\,\hat{\Omega}_{LS}-\hat{\Omega}_{LS}  \notag \\
&=&\frac{1}{T}[(Y^{\prime }M_{1}Y)(Y^{\prime }MY)^{-1}(Y^{\prime
}M_{1}Y)-(Y^{\prime }M_{1}Y)]=\frac{1}{T}Y^{\prime }M_{1}[Y(Y^{\prime
}MY)^{-1}Y^{\prime }-I_{T}]M_{1}Y\,.\quad \quad \quad \quad
\end{eqnarray}%
The latter shows that $\hat{\Delta}^{-1}$ only depends on the least-squares
residuals $M_{1}Y$ and $MY$.

\section{Regression interpretation of DWH test statistics \label{sec:
Appendix : Regression Formula}}

\resetcountersSection

Let us now consider the regressions (\ref{eq: StructPojVhat})\thinspace
-\thinspace (\ref{eq: StructYZ}). Using $Y=\hat{Y}+\hat{V}$, $\hat{Y}=X\hat{%
\Pi}$ and $\hat{\Pi}=(X^{\prime }X)^{-1}X^{\prime }Y$, we see that the 2SLS
residual vector $\tilde{u}$ for model (\ref{eq: y StrucEqch2}) based on the
instrument matrix $X=[X_{1},\,X_{2}]$ can be written as 
\begin{eqnarray}
\tilde{u} &=&y-Y\tilde{\beta}-X_{1}\tilde{\gamma}=(y-\hat{Y}\tilde{\beta}%
-X_{1}\tilde{\gamma})-\hat{V}\tilde{\beta}=M_{1}(y-\hat{Y}\tilde{\beta})-%
\hat{V}\tilde{\beta}  \notag \\
&=&M_{1}(y-\hat{Y}\tilde{\beta}-\hat{V}\tilde{\beta})=M_{1}(y-Y\tilde{\beta})
\label{eq:  u_2SLS}
\end{eqnarray}%
where $\tilde{\beta}$ and $\tilde{\gamma}$ are the 2SLS estimators of $\beta 
$ and $\gamma $, and the different sum-of-squares functions satisfy: 
\begin{gather}
S(\hat{\theta})=S_{\ast }(\hat{\theta}_{\ast })\,,\quad \tilde{u}^{\prime }%
\tilde{u}=S(\hat{\theta}^{0})=S_{\ast }(\hat{\theta}_{\ast }^{0})=\tilde{S}(%
\hat{\theta}_{\ast \ast }^{0})\,,\quad \tilde{S}(\hat{\theta}_{\ast \ast
})=(y-Y\tilde{\beta})^{\prime }M(y-Y\tilde{\beta})\,, \\
S(\hat{\theta}^{0})-S(\hat{\theta})=S_{\ast }(\hat{\theta}_{\ast
}^{0})-S_{\ast }(\hat{\theta}_{\ast })\,.
\end{gather}%
Let $R=\left[ 
\begin{array}{ccc}
0 & 0 & I_{G}%
\end{array}%
\right] ,$ and $R_{\ast }=\left[ 
\begin{array}{ccc}
I_{G} & 0 & -I_{G}%
\end{array}%
\right] ,$ so that $Rb=a$ and $R_{\ast }\theta _{\ast }=\beta -a.$ The null
hypotheses $H_{0}:a=0$ and $H_{0}^{\ast }:\beta =b$ can thus be written as 
\begin{equation}
H_{0}:R\theta =0\,,\quad H_{0}^{\ast }:R_{\ast }\theta _{\ast }=0.
\end{equation}%
Further, $\hat{\theta}_{\ast }=[\tilde{\beta}^{\prime },\,\tilde{\gamma}%
^{\prime },\,\tilde{b}^{\prime }]^{\prime }$ and $\hat{\theta}_{\ast }^{0}=[%
\hat{\beta}^{\prime },\,\hat{\gamma}^{\prime },\,\hat{\beta}^{\prime
}]^{\prime },$ where $\hat{\beta}$ and $\hat{\gamma}$ are the OLS estimators
of $\beta $ and $\gamma $ based on the model (\ref{eq: y StrucEqch2}), and 
\begin{eqnarray}
R_{\ast }\hat{\theta} &=&\left[ 
\begin{array}{ccc}
I_{G} & 0 & -I_{G}%
\end{array}%
\right] \left[ 
\begin{array}{c}
\tilde{\beta} \\ 
\tilde{\gamma} \\ 
\tilde{b}%
\end{array}%
\right] =\tilde{\beta}-\tilde{b}\,, \\
\hat{\theta}_{\ast }^{0} &=&\hat{\theta}_{\ast }+(Z_{\ast }^{\prime }Z_{\ast
})^{-1}R_{\ast }^{\prime }\left[ R_{\ast }(Z_{\ast }^{\prime }Z_{\ast
})^{-1}R_{\ast }^{\prime }\right] ^{-1}(-R_{\ast }\hat{\theta}_{\ast })\,, \\
S(\hat{\theta}_{\ast }^{0})-S(\hat{\theta}_{\ast }) &=&(\hat{\theta}_{\ast
}^{0}-\hat{\theta}_{\ast })^{\prime }Z_{\ast }^{\prime }Z_{\ast }(\hat{\theta%
}_{\ast }^{0}-\hat{\theta}_{\ast })=(R_{\ast }\hat{\theta}_{\ast })^{\prime }%
\left[ R_{\ast }(Z_{\ast }^{\prime }Z_{\ast })^{-1}R_{\ast }^{\prime }\right]
^{-1}(R_{\ast }\hat{\theta}_{\ast })\,,
\end{eqnarray}%
where $Z_{\ast }=[\hat{Y},\,X_{1},\,\hat{V}].$ On writing $Z_{\ast }=[\hat{X}%
_{1},$\thinspace $\hat{V}],$ where $\hat{X}_{1}=[\hat{Y},\,X_{1}]$, we get:%
\begin{gather}
Z_{\ast }^{\prime }Z_{\ast }=\left[ 
\begin{array}{cc}
(\hat{X}_{1}^{\prime }\hat{X}_{1}) & 0 \\ 
0 & (\hat{V}^{\prime }\hat{V})%
\end{array}%
\right] ,\quad (Z_{\ast }^{\prime }Z_{\ast })^{-1}=\left[ 
\begin{array}{cc}
(\hat{X}_{1}^{\prime }\hat{X}_{1})^{-1} & 0 \\ 
0 & (\hat{V}^{\prime }\hat{V})^{-1}%
\end{array}%
\right] \,, \\
(\hat{X}_{1}^{\prime }\hat{X}_{1})^{-1}=\left[ 
\begin{array}{cc}
\hat{Y}^{\prime }\hat{Y} & \hat{Y}^{\prime }X_{1} \\ 
X_{1}^{\prime }\hat{Y} & X_{1}^{\prime }X_{1}%
\end{array}%
\right] ^{-1}=\left[ 
\begin{array}{cc}
W_{YY} & W_{Y1} \\ 
W_{1Y} & W_{11}%
\end{array}%
\right] \,,
\end{gather}%
where $W_{YY}=\left[ (\hat{Y}^{\prime }\hat{Y})-\hat{Y}^{\prime
}X_{1}(X_{1}^{\prime }X_{1})^{-1}X_{1}^{\prime }\hat{Y}\right] ^{-1}=\left[ 
\hat{Y}^{\prime }M_{1}\hat{Y}\right] ^{-1}=[Y^{\prime }(M_{1}-M)Y]^{-1}$, 
\begin{gather}
(Z_{\ast }^{\prime }Z_{\ast })^{-1}R_{\ast }^{\prime }=\left[ 
\begin{array}{ccc}
W_{YY} & W_{Y1} & 0 \\ 
W_{1Y} & W_{11} & 0 \\ 
0 & 0 & (\hat{V}^{\prime }\hat{V})^{-1}%
\end{array}%
\right] \left[ 
\begin{array}{c}
I_{G} \\ 
0 \\ 
-I_{G}%
\end{array}%
\right] =\left[ 
\begin{array}{c}
W_{YY} \\ 
W_{1Y} \\ 
-(\hat{V}^{\prime }\hat{V})^{-1}%
\end{array}%
\right] \,, \\
R_{\ast }(Z_{\ast }^{\prime }Z_{\ast })^{-1}R_{\ast }^{\prime }=W_{YY}+(\hat{%
V}^{\prime }\hat{V})^{-1}\,, \\
\hat{\theta}_{\ast }^{0}-\hat{\theta}_{\ast }=\left[ 
\begin{array}{c}
\hat{\beta}-\tilde{\beta} \\ 
\hat{\gamma}-\tilde{\gamma} \\ 
\hat{\beta}-\tilde{b}%
\end{array}%
\right] =\left[ 
\begin{array}{c}
W_{YY} \\ 
W_{1Y} \\ 
-(\hat{V}^{\prime }\hat{V})^{-1}%
\end{array}%
\right] \left[ W_{YY}+(\hat{V}^{\prime }\hat{V})^{-1}\right] ^{-1}(\tilde{b}-%
\tilde{\beta})\,.
\end{gather}%
From the latter equation, we see that 
\begin{equation}
\hat{\beta}-\tilde{\beta}=W_{YY}\left[ W_{YY}+(\hat{V}^{\prime }\hat{V})^{-1}%
\right] ^{-1}(\tilde{b}-\tilde{\beta})=W_{YY}\left[ W_{YY}+(\hat{V}^{\prime }%
\hat{V})^{-1}\right] ^{-1}\tilde{a}\,,
\end{equation}%
where $\tilde{a}=\tilde{b}-\tilde{\beta}$ is the OLS estimate of $a$ in (\ref%
{eq: StructVhat}). Hence, we have 
\begin{eqnarray}
\tilde{a}=\tilde{b}-\tilde{\beta} &=&\left[ W_{YY}+(\hat{V}^{\prime }\hat{V}%
)^{-1}\right] W_{YY}^{-1}(\hat{\beta}-\tilde{\beta})  \notag \\
&=&\left\{ [Y^{\prime }(M_{1}-M)Y]^{-1}+(\hat{V}^{\prime }\hat{V}%
)^{-1}\right\} [Y^{\prime }(M_{1}-M)Y](\hat{\beta}-\tilde{\beta})\,,\,
\label{eq: AppRegresstilde(a)fin}
\end{eqnarray}%
which entails that 
\begin{eqnarray}
S(\hat{\theta}_{\ast }^{0})-S(\hat{\theta}_{\ast }) &=&(R_{\ast }\hat{\theta}%
_{\ast })^{\prime }\left[ R_{\ast }(Z_{\ast }^{\prime }Z_{\ast
})^{-1}R_{\ast }^{\prime }\right] ^{-1}(R_{\ast }\hat{\theta}_{\ast }) 
\notag \\
&=&(\tilde{b}-\tilde{\beta})^{\prime }\left\{ [Y^{\prime }(M_{1}-M)Y]^{-1}+(%
\hat{V}^{\prime }\hat{V})^{-1}\right\} ^{-1}(\tilde{b}-\tilde{\beta})  \notag
\\
&=&(\hat{\beta}-\tilde{\beta})^{\prime }[Y^{\prime }(M_{1}-M)Y]\left\{
[Y^{\prime }(M_{1}-M)Y]^{-1}+(\hat{V}^{\prime }\hat{V})^{-1}\right\}
[Y^{\prime }(M_{1}-M)Y](\hat{\beta}-\tilde{\beta})  \notag \\
&=&(\hat{\beta}-\tilde{\beta})^{\prime }W_{YY}^{-1}\left[ W_{YY}+(Y^{\prime
}MY)^{-1}\right] W_{YY}^{-1}(\hat{\beta}-\tilde{\beta})  \notag \\
&=&(\hat{\beta}-\tilde{\beta})^{\prime }W_{YY}^{-1}\left[ W_{YY}+(Y^{\prime
}M_{1}Y-W_{YY}^{-1})^{-1}\right] W_{YY}^{-1}(\hat{\beta}-\tilde{\beta})\,.
\label{eq: AppRegresstildeSthetafin}
\end{eqnarray}%
Using Lemma \ref{Th: Difference of matrix inverses} with $\,A=W_{YY}^{-1}$
and $B=Y^{\prime }M_{1}Y$ in (\ref{eq: AppRegresstildeSthetafin}), we then
get: 
\begin{eqnarray}
S(\hat{\theta}_{\ast }^{0})-S(\hat{\theta}_{\ast }) &=&(\hat{\beta}-\tilde{%
\beta})^{\prime }W_{YY}^{-1}\left[ W_{YY}+(Y^{\prime
}M_{1}Y-W_{YY}^{-1})^{-1}\right] W_{YY}^{-1}(\hat{\beta}-\tilde{\beta}) 
\notag \\
&=&(\hat{\beta}-\tilde{\beta})^{\prime }A\left[ A^{-1}+(B-A)^{-1}\right] A(%
\hat{\beta}-\tilde{\beta})=(\hat{\beta}-\tilde{\beta})^{\prime
}(B^{-1}-A^{-1})^{-1}(\hat{\beta}-\tilde{\beta})  \notag \\
&=&(\hat{\beta}-\tilde{\beta})^{\prime }\{[Y^{\prime
}(M_{1}-M)Y]^{-1}-(Y^{\prime }M_{1}Y)^{-1}\}^{-1}(\hat{\beta}-\tilde{\beta})
\notag \\
&=&T(\tilde{\beta}-\hat{\beta})^{\prime }[\hat{\Omega}_{IV}^{-1}-\hat{\Omega}%
_{LS}^{-1}]^{-1}(\tilde{\beta}-\hat{\beta})=T(\tilde{\beta}-\hat{\beta}%
)^{\prime }\hat{\Delta}^{-1}(\tilde{\beta}-\hat{\beta})
\label{eq: AppRegresstildeSthetalemma}
\end{eqnarray}%
where $\hat{\Omega}_{IV}=\frac{1}{T}Y^{\prime }(M_{1}-M)Y$ and $\hat{\Omega}%
_{LS}=\frac{1}{T}Y^{\prime }M_{1}Y.$ Since we have $S_{\ast }(\hat{\theta}%
_{\ast }^{0})-S_{\ast }(\hat{\theta}_{\ast })=S(\hat{\theta}^{0})-S(\hat{%
\theta})$, we get from (\ref{eq: AppRegresstildeSthetalemma}), (\ref{eq:
sigma2hat}) and (\ref{eq: SumsquareResiduals}): 
\begin{equation}
S(\hat{\theta})=S(\hat{\theta}^{0})-[S_{\ast }(\hat{\theta}_{\ast
}^{0})-S_{\ast }(\hat{\theta}_{\ast })]=S(\hat{\theta}^{0})-T(\tilde{\beta}-%
\hat{\beta})^{\prime }\hat{\Delta}^{-1}(\tilde{\beta}-\hat{\beta})=T\,\tilde{%
\sigma}_{2}^{2}\,.
\end{equation}%
It is also clear from (\ref{eq: sigma2hat}) and (\ref{eq: SumsquareResiduals}%
) that 
\begin{equation}
S(\hat{\theta}^{0})=T\hat{\sigma}^{2},\,\,S_{\ast }(\hat{\theta}_{\ast
}^{0})=T\tilde{\sigma}^{2}\,.
\end{equation}%
Hence, except for $\mathcal{H}_{1},$ the other statistics can be expressed
as: 
\begin{gather}
\mathcal{H}_{2}=T\,\left( \frac{S(\hat{\theta}^{0})-S(\hat{\theta})}{S_{\ast
}(\hat{\theta}_{\ast }^{0})}\right) \,,\,\mathcal{H}_{3}=T\,\left( \frac{S(%
\hat{\theta}^{0})-S(\hat{\theta})}{S(\hat{\theta}^{0})}\right) \,,
\label{eq: AppH2H3auxiliary} \\
\mathcal{T}_{1}=\kappa _{1}\,\left( \frac{S(\hat{\theta}^{0})-S(\hat{\theta})%
}{S_{\ast }(\hat{\theta}_{\ast }^{0})-\tilde{S}(\hat{\theta}_{\ast \ast })}%
\right) =\kappa _{1}\,\left( \frac{S(\hat{\theta}^{0})-S(\hat{\theta})}{%
\tilde{S}(\hat{\theta}_{\ast \ast }^{0})-\tilde{S}(\hat{\theta}_{\ast \ast })%
}\right) \,,  \label{eq: AppT1T2auxiliary} \\
\mathcal{T}_{2}=\kappa _{2}\,\left( \frac{S(\hat{\theta}^{0})-S(\hat{\theta})%
}{S(\hat{\theta})}\right) \,,\quad \mathcal{T}_{3}=\kappa _{3}\,\left( \frac{%
S(\hat{\theta}^{0})-S(\hat{\theta})}{S_{\ast }(\hat{\theta}_{\ast }^{0})}%
\right) \,,\quad \mathcal{T}_{4}=\kappa _{4}\,\left( \frac{S(\hat{\theta}%
^{0})-S(\hat{\theta})}{S(\hat{\theta}^{0})}\right) \,,
\label{eq: AppT3T4auxiliary} \\
\mathcal{R}=\kappa _{R}\,\left( \frac{\bar{S}(\check{\theta}^{0})-\bar{S}(%
\check{\theta})}{\bar{S}(\check{\theta})}\right) \,.
\label{eq: ppstatRevreg}
\end{gather}

\section{Proofs \label{sec: Appendix : Proofs2}}

\resetcountersSection

To establish Proposition \ref{Th: Quadratic-form representations of
exogeneity statistics}, it will be useful to state some basic identities for
the different components of alternative exogeneity test statistics.

\begin{lemma}
\label{Th: Properties of exogeneity statistics components} 
\captiontheorem{\lemmaname}{Properties of exogeneity statistics components}
The random vectors and matrices in $(\ref{eq: statT1})$\thinspace
-\thinspace $(\ref{eq: Psi_R})$ satisfy the following identities: setting 
\begin{equation}
B_{1}=:(Y^{\prime }M_{1}Y)^{-1}Y^{\prime }M_{1}\,,\quad B_{2}=:(Y^{\prime
}N_{1}Y)^{-1}Y^{\prime }N_{1}\,,  \label{eq: A_1}
\end{equation}%
\begin{equation}
C_{1}=:B_{2}-B_{1}\,,\quad \Psi _{_{0}}=:C_{1}^{\prime }\hat{\Delta}%
^{-1}C_{1}\,,\quad N_{2}=:I_{T}-M_{1}YA_{2}\,,  \label{eq: C_1}
\end{equation}%
we have%
\begin{equation}
B_{1}\,M_{1}=B_{1}\,,\quad B_{2}\,M_{1}=B_{2}\,N_{1}=B_{2}\,,\quad
B_{1}\,Y=B_{2}\,Y=I_{G},  \label{eq: N_1}
\end{equation}%
\begin{equation}
C_{1}Y=0\,,\quad C_{1}X_{1}=0\,,\quad C_{1}\,\mathrm{\bar{P}}%
[M_{1}Y]=0\,,\quad C_{1}\,M_{1}=C_{1}\,\mathrm{\bar{M}}[M_{1}Y]=C_{1}\,,
\label{eq: C_1*Y}
\end{equation}%
\begin{equation}
M_{1}YA_{1}=\mathrm{\bar{P}}[M_{1}Y]\,,\quad M_{1}\,\Psi
_{_{0}}\,M_{1}=M_{1}\,\Psi _{_{0}}=\Psi _{_{0}}\,M_{1}=\Psi _{_{0}}\,,
\label{eq: M_1*Y*A_1}
\end{equation}%
\begin{equation}
M_{1}\Psi _{R}\,M_{1}=\Psi _{R}\,,\quad M_{1}\,\Lambda
_{_{R}}\,M_{1}=M\,\Lambda _{_{R}}\,M=\Lambda _{_{R}}\,,
\label{eq: M_1*Psi_R*M_1}
\end{equation}%
\begin{equation}
B_{1}\,B_{1}^{\prime }=B_{1}\,B_{2}^{\prime }=B_{2}\,B_{1}^{\prime }=\frac{1%
}{T}\hat{\Omega}_{LS}^{-1}\,,\quad B_{2}\,B_{2}^{\prime }=\frac{1}{T}\hat{%
\Omega}_{IV}^{-1}\,,  \label{eq: A_1*A_1'}
\end{equation}%
\begin{equation}
C_{1}C_{1}^{\prime }=\frac{1}{T}\big(\hat{\Omega}_{IV}^{-1}-\hat{\Omega}%
_{LS}^{-1}\big)=\frac{1}{T}\hat{\Delta}\,,\quad C_{1}\Psi _{_{0}}=\frac{1}{T}%
C_{1}\,,\quad \Psi _{_{0}}\Psi _{_{0}}=\frac{1}{T}\Psi _{0}\,,
\label{eq: C_1*C_1'}
\end{equation}%
\begin{equation}
\tilde{\beta}-\hat{\beta}=(B_{2}-B_{1})\,y=C_{1}\,y=C_{1}\,(M_{1}\,y)\,,
\label{eq: b_2sls - b_OLS}
\end{equation}%
\begin{equation}
(\tilde{\beta}-\hat{\beta})^{\prime }\hat{\Delta}^{-1}(\tilde{\beta}-\hat{%
\beta})=y^{\prime }\,\Psi _{_{0}}\,y=(M_{1}\,y)^{\prime }\,\Psi
_{_{0}}\,(M_{1}\,y)\,,  \label{eq: (b_2sls - b_OLS)D(eq: b_2sls - b_OLS)}
\end{equation}%
\begin{equation}
y-Y\hat{\beta}=[I_{T}-Y\,B_{1}]y\,,\quad y-Y\tilde{\beta}=[I_{T}-Y\,B_{2}]y%
\,,  \label{eq: M_1(y-b_OLS)}
\end{equation}%
\begin{equation}
\hat{u}=M_{1}(y-Y\hat{\beta})=\mathrm{\bar{M}}[\bar{Y}]y=M_{1}\mathrm{\bar{M}%
}[M_{1}Y]y=\mathrm{\bar{M}}[M_{1}Y](M_{1}\,y)\,,
\label{eq: M_1(y - Y*b_OLS)}
\end{equation}%
\begin{equation}
M(y-Y\hat{\beta})=M\,\mathrm{\bar{M}}[M_{1}Y]y=M\,\mathrm{\bar{M}}%
[M_{1}Y](M_{1}\,y)\,,  \label{eq: M(y - Y*b_OLS)}
\end{equation}%
\begin{eqnarray}
N_{1}\,(y-Y\tilde{\beta}) &=&M_{1}\,P(y-Y\tilde{\beta})=M_{1}\,\mathrm{\bar{M%
}}[M_{1}PY]Py=\mathrm{\bar{M}}[N_{1}Y]N_{1}\,y  \notag \\
&=&P\,M_{1}(y-Y\tilde{\beta})=\mathrm{\bar{M}}[P\,M_{1}Y]P\,(M_{1}\,y)\,,
\label{eq: M_1P(y - Yb_2SLS)}
\end{eqnarray}%
\begin{equation}
\tilde{u}=M_{1}(y-Y\tilde{\beta})=N_{2}(M_{1}\,y)\,,\quad M(y-Y\tilde{\beta}%
)=M\,N_{2}\,(M_{1}\,y)\,,  \label{eq: M_1(y - Y*b_2SLS)}
\end{equation}%
\begin{equation}
\tilde{\sigma}^{2}=\frac{1}{T}(M_{1}\,y)^{\prime }\,N_{2}^{\prime
}\,N_{2}\,(M_{1}\,y)\,,  \label{eq: sigmatilde^2}
\end{equation}%
\begin{equation}
\hat{\sigma}^{2}=\frac{1}{T}\,y^{\prime }\mathrm{\bar{M}}[\bar{Y}]\,y=\frac{1%
}{T}\,y^{\prime }M_{1}\mathrm{\bar{M}}[M_{1}Y]\,y=\frac{1}{T}%
\,(M_{1}\,y)^{\prime }\mathrm{\bar{M}}[M_{1}Y]\,(M_{1}\,y)\,\,,
\label{eq: sigmahat^2}
\end{equation}%
\begin{equation}
\tilde{\sigma}_{1}^{2}=\frac{1}{T}y^{\prime }N_{1}\,\mathrm{\bar{M}}%
[N_{1}Y]\,N_{1}\,y=\frac{1}{T}(M_{1}\,y)^{\prime }P\mathrm{\bar{M}}%
[P\,M_{1}Y]P\,(M_{1}\,y)\,,  \label{eq: sigma_1tilde^2}
\end{equation}%
\begin{equation}
\tilde{\sigma}_{2}^{2}=(M_{1}\,y)^{\prime }\left\{ \frac{1}{T}\mathrm{\bar{M}%
}[M_{1}Y]-\Psi _{_{0}}\right\} (M_{1}\,y)\,,  \label{eq: sigma_2tilde^2}
\end{equation}%
\begin{equation}
y^{\prime }\,\Psi _{R}\,y=\frac{1}{T}y^{\prime }\mathrm{\bar{P}}\big[\mathrm{%
\bar{M}}[\bar{Y}]X_{2}\big]\mathrm{\bar{M}}[\bar{Y}]y=\frac{1}{T}%
(M_{1}\,y)^{\prime }\mathrm{\bar{P}}\big[\mathrm{\bar{M}}[\bar{Y}]X_{2}\big]%
(M_{1}\,y)\,,  \label{eq: y'Psi_Ry}
\end{equation}%
\begin{equation}
\hat{\sigma}_{R}^{2}=\frac{1}{T}y^{\prime }\,\mathrm{\bar{M}}[Z]\,y=\frac{1}{%
T}(M_{1}\,y)^{\prime }\mathrm{\bar{M}}[Z](M_{1}\,y)\,.  \label{eq:sigma_R^2}
\end{equation}
\end{lemma}

\begin{proofflexc}
\captionproofflexc{\lemmaname}{Th: Properties of exogeneity statistics components}
Using the idempotence of $M_{1}$ and (\ref{eq: N_1 forms}), we see that:%
\begin{equation}
B_{1}\,M_{1}=(Y^{\prime }M_{1}Y)^{-1}Y^{\prime }M_{1}M_{1}=(Y^{\prime
}M_{1}Y)^{-1}Y^{\prime }M_{1}=B_{1}\,,
\end{equation}%
\begin{equation}
B_{2}\,M_{1}=[Y^{\prime }N_{1}Y]^{-1}Y^{\prime }N_{1}M_{1}=[Y^{\prime
}N_{1}Y]^{-1}Y^{\prime }N_{1}=B_{2}=B_{2}\,N_{1}=B_{2}(M_{1}-M)\,,
\end{equation}%
\begin{equation}
M_{1}YA_{1}=M_{1}Y(Y^{\prime }M_{1}Y)^{-1}Y^{\prime }M_{1}=\mathrm{\bar{P}}%
(M_{1}Y)\,,
\end{equation}%
\begin{equation}
C_{1}\,M_{1}=B_{2}\,M_{1}-B_{1}\,M_{1}=B_{2}-B_{1}=C_{1}\,,\quad
C_{1}X_{1}=C_{1}M_{1}\,X_{1}=0\,,
\end{equation}%
\begin{equation}
B_{1}Y=(Y^{\prime }M_{1}Y)^{-1}Y^{\prime }M_{1}Y=I_{G}=(Y^{\prime
}N_{1}Y)^{-1}Y^{\prime }N_{1}Y=B_{2}Y\,,
\end{equation}%
\begin{equation}
C_{1}Y=B_{2}Y-B_{1}Y=0\,,
\end{equation}%
\begin{eqnarray}
C_{1}\,\mathrm{\bar{P}}[M_{1}Y] &=&[(Y^{\prime }N_{1}Y)^{-1}Y^{\prime
}N_{1}-(Y^{\prime }M_{1}Y)^{-1}Y^{\prime }M_{1}]\,M_{1}Y(Y^{\prime
}M_{1}Y)^{-1}Y^{\prime }M_{1}  \notag \\
&=&[(Y^{\prime }N_{1}Y)^{-1}Y^{\prime }N_{1}Y-(Y^{\prime
}M_{1}Y)^{-1}Y^{\prime }M_{1}Y]\,(Y^{\prime }M_{1}Y)^{-1}Y^{\prime }M_{1} 
\notag \\
&=&(I_{G}-I_{G})(Y^{\prime }M_{1}Y)^{-1}Y^{\prime }M_{1}=0\,,
\end{eqnarray}%
\begin{equation}
C_{1}\,\mathrm{\bar{M}}[M_{1}Y]=C_{1}\,[I_{T}-\mathrm{\bar{P}}%
[M_{1}Y]=C_{1}\,,
\end{equation}%
\begin{equation}
M_{1}\,\mathrm{\bar{M}}[\bar{Y}]\,M_{1}=\mathrm{\bar{M}}[\bar{Y}]\,\,,\quad
M_{1}\,\mathrm{\bar{M}}[Z]\,M_{1}=\mathrm{\bar{M}}[Z]\,,
\end{equation}%
\begin{equation}
M_{1}\Psi _{R}\,M_{1}=\frac{1}{T}\{M_{1}\,\mathrm{\bar{M}}[\bar{Y}%
]\,M_{1}-M_{1}\,\mathrm{\bar{M}}[Z]\,M_{1}\}=\Psi _{R}\,,\quad M_{1}\Lambda
_{_{R}}M_{1}=\frac{1}{T}M_{1}\mathrm{\bar{M}}[Z]M_{1}=\Lambda _{_{R}}\,,
\end{equation}%
so (\ref{eq: N_1})\thinspace -\thinspace (\ref{eq: M_1*Psi_R*M_1}) are
established. (\ref{eq: A_1*A_1'}) and (\ref{eq: C_1*C_1'}) follow directly
from (\ref{eq: N_1 forms}) and the definitions of $B_{1}$, $B_{2}$, $C_{1}$
and $\Psi _{_{0}}$. We get (\ref{eq: b_2sls - b_OLS}) and (\ref{eq: (b_2sls
- b_OLS)D(eq: b_2sls - b_OLS)}) by using the definitions of $\hat{\beta}$
and $\tilde{\beta}$ in (\ref{eq: b_OLS})\thinspace -\thinspace (\ref{eq:
b_2SLS}). (\ref{eq: M_1(y-b_OLS)}) follows on using (\ref{eq: b_OLS}) and (%
\ref{eq: b_2SLS}). (\ref{eq: M_1(y - Y*b_OLS)}) comes from the fact that the
residuals $M_{1}(y-Y\hat{\beta})$ are obtained by minimizing $\Vert y-Y\hat{%
\beta}-X_{1}\gamma \Vert ^{2}$ with respect to $\gamma $, or equivalently $%
\Vert y-Y\beta -X_{1}\gamma \Vert ^{2}$ with respect to $\beta $ and $\gamma 
$. (\ref{eq: M(y - Y*b_OLS)}) follows from (\ref{eq: M_1(y - Y*b_OLS)}) and
noting that $M$\thinspace $=M\,M_{1}$. Similarly, the first identity in (\ref%
{eq: M_1P(y - Yb_2SLS)}) comes from the fact that the residuals $M_{1}\,P(y-Y%
\tilde{\beta})=M_{1}(y-PY\tilde{\beta})$ are obtained by minimizing $\Vert
y-PY\tilde{\beta}-X_{1}\gamma \Vert ^{2}$ with respect to $\gamma $, or
equivalently by minimizing $\Vert y-PY\beta -X_{1}\gamma \Vert ^{2}$ with
respect to $\beta $ and $\gamma $. The others follow on noting that $%
N_{1}=M_{1}\,P=P\,M_{1}$ and 
\begin{equation}
M_{1}\,\mathrm{\bar{M}}[M_{1}PY]P=\mathrm{\bar{M}}[PM_{1}Y]M_{1}P=\mathrm{%
\bar{M}}[PM_{1}Y]PM_{1}\,.
\end{equation}%
To get (\ref{eq: M_1(y - Y*b_2SLS)}) and (\ref{eq: sigmatilde^2}), we note
that%
\begin{equation}
\tilde{u}=y-Y\tilde{\beta}-X_{1}\tilde{\gamma}=M_{1}(y-Y\tilde{\beta}%
)=M_{1}\,[I_{T}-YA_{2}]y=[I_{T}-M_{1}YA_{2}](M_{1}\,y)=N_{2}(M_{1}\,y)
\end{equation}%
hence 
\begin{equation}
\tilde{\sigma}^{2}=\frac{1}{T}\tilde{u}^{\prime }\tilde{u}=\frac{1}{T}(y-Y%
\tilde{\beta})^{\prime }M_{1}M_{1}(y-Y\tilde{\beta})=\frac{1}{T}%
(M_{1}\,y)^{\prime }N_{2}^{\prime }N_{2}(M_{1}\,y)\,.
\end{equation}%
Further, using (\ref{eq: sigma hat})\thinspace -\thinspace (\ref{eq: statRH}
), (\ref{eq: M_1(y - Y*b_OLS)}) and (\ref{eq: M_1P(y - Yb_2SLS)}), we see
that: 
\begin{equation}
\hat{\sigma}^{2}=\frac{1}{T}(y-Y\hat{\beta})^{\prime }M_{1}(y-Y\hat{\beta})=%
\frac{1}{T}y^{\prime }\mathrm{\bar{M}}[\bar{Y}]y=\frac{1}{T}y^{\prime
}M_{1}\,\mathrm{\bar{M}}[M_{1}Y]\,y=\frac{1}{T}(M_{1}\,y)^{\prime }\mathrm{%
\bar{M}}[M_{1}Y](M_{1}\,y)\,,
\end{equation}%
\begin{eqnarray}
\tilde{\sigma}_{1}^{2} &=&\frac{1}{T}(y-Y\tilde{\beta})^{\prime }N_{1}(y-Y%
\tilde{\beta})=\frac{1}{T}(y-Y\tilde{\beta})^{\prime }PM_{1}\,P(y-Y\tilde{%
\beta})  \notag \\
&=&\frac{1}{T}y^{\prime }N_{1}^{\prime }\mathrm{\bar{M}}[N_{1}Y]N_{1}\,y=%
\frac{1}{T}(M_{1}\,y)^{\prime }P\mathrm{\bar{M}}[P\,M_{1}Y]P\,(M_{1}\,y)\,,
\end{eqnarray}%
\begin{eqnarray}
\tilde{\sigma}_{2}^{2} &=&\hat{\sigma}^{2}-(\tilde{\beta}-\hat{\beta}%
)^{\prime }\hat{\Delta}^{-1}(\tilde{\beta}-\hat{\beta})=\frac{1}{T}%
\{y^{\prime }M_{1}\,\mathrm{\bar{M}}[M_{1}Y]\,y\}-y^{\prime }\,\Psi
_{_{0}}\,y  \notag \\
&=&(M_{1}\,y)^{\prime }\left\{ \frac{1}{T}\mathrm{\bar{M}}[M_{1}Y]-\Psi
_{_{0}}\right\} (M_{1}\,y)\,,
\end{eqnarray}%
so (\ref{eq: sigma hat})\thinspace -\thinspace (\ref{eq: sigma2hat}) are
established. Finally, (\ref{eq: y'Psi_Ry}) and (\ref{eq:sigma_R^2}) follow
by observing that $M_{1}\mathrm{\bar{M}}[\bar{Y}]=\mathrm{\bar{M}}[\bar{Y}%
]M_{1}=\mathrm{\bar{M}}[\bar{Y}]M_{1}$ and $M_{1}\mathrm{\bar{M}}[Z]=M_{1}%
\mathrm{\bar{M}}[Z]=\mathrm{\bar{M}}[Z]$, so that $M_{1}\mathrm{\bar{P}}\big[%
\mathrm{\bar{M}}[\bar{Y}]X_{2}\big]M_{1}=\mathrm{\bar{P}}\big[\mathrm{\bar{M}%
}[\bar{Y}]X_{2}\big]$ and $M_{1}\mathrm{\bar{M}}[Z]M_{1}=\mathrm{\bar{M}}[Z]$%
.
\end{proofflexc}

Using Lemma \ref{Th: Properties of exogeneity statistics components}, we can
now prove Proposition \ref{Th: Quadratic-form representations of exogeneity
statistics}.

\quad

\begin{proofflexc}
\captionproofflexc{\propositionname}{Th: Quadratic-form representations of exogeneity statistics}
We first note that 
\begin{equation}
\tilde{\beta}-\hat{\beta}=(B_{2}-B_{1})\,y=C_{1}\,y\,,
\end{equation}%
\begin{equation}
(\tilde{\beta}-\hat{\beta})^{\prime }\hat{\Delta}^{-1}(\tilde{\beta}-\hat{%
\beta})=y^{\prime }\,C_{1}^{\prime }\,\hat{\Delta}^{-1}C_{1}\,y=y^{\prime
}\,\Psi _{_{0}}\,y\,,
\end{equation}%
so that, by the definitions (\ref{eq: statT1})\thinspace -\thinspace (\ref%
{eq: statRH}), 
\begin{equation}
\mathcal{T}_{l}=\kappa _{l}(\tilde{\beta}-\hat{\beta})^{\prime }\tilde{\Sigma%
}_{l}^{-1}(\tilde{\beta}-\hat{\beta})=\kappa _{l}\frac{(\tilde{\beta}-\hat{%
\beta})^{\prime }\hat{\Delta}^{-1}(\tilde{\beta}-\hat{\beta})}{\tilde{\sigma}%
_{l}^{2}}=\frac{y^{\prime }\,\Psi _{_{0}}\,y}{\tilde{\sigma}_{l}^{2}}\,,%
\text{\quad }l=1,\,2,\,3,\,4,
\end{equation}%
\begin{equation}
\mathcal{H}_{i}=T(\tilde{\beta}-\hat{\beta})^{\prime }\hat{\Sigma}_{i}^{-1}(%
\tilde{\beta}-\hat{\beta})=T\frac{(\tilde{\beta}-\hat{\beta})^{\prime }\hat{%
\Delta}^{-1}(\tilde{\beta}-\hat{\beta})}{\hat{\sigma}_{i}^{2}}=\frac{%
y^{\prime }\,\Psi _{_{0}}\,y}{\hat{\sigma}_{i}^{2}}\,,\text{\quad }i=2,\,3,
\end{equation}%
where, using Lemma \ref{Th: Properties of exogeneity statistics components}, 
\begin{equation}
\tilde{\sigma}_{1}^{2}=\frac{1}{T}(y-Y\tilde{\beta})^{\prime }N_{1}(y-Y%
\tilde{\beta})=\frac{1}{T}y^{\prime }N_{1}\,\mathrm{\bar{M}}%
[N_{1}Y]\,N_{1}\,y=y^{\prime }\Lambda _{1}\,y\,,
\end{equation}%
\begin{equation}
\tilde{\sigma}_{2}^{2}=y^{\prime }\,M_{1}\left\{ \frac{1}{T}\mathrm{\bar{M}}%
[M_{1}Y]-\Psi _{_{0}}\right\} (M_{1}\,y)=y^{\prime }\,\Lambda _{2}\,y\,,
\end{equation}%
\begin{equation}
\tilde{\sigma}_{3}^{2}=\tilde{\sigma}^{2}=\frac{1}{T}y^{\prime
}\,M_{1}N_{2}^{\prime }N_{2}M_{1}\,y=y^{\prime }\,\Lambda _{3}\,y\,,
\end{equation}%
\begin{equation}
\tilde{\sigma}_{4}^{2}=\hat{\sigma}^{2}=\frac{1}{T}\,y^{\prime }\mathrm{\bar{%
M}}[\bar{Y}]\,y=\frac{1}{T}\,y^{\prime }\,M_{1}\,\mathrm{\bar{M}}%
[M_{1}Y]\,M_{1}\,y=y^{\prime }\,\Lambda _{4}\,y\,,
\end{equation}%
\begin{equation}
\hat{\sigma}_{2}^{2}=\tilde{\sigma}^{2}=y^{\prime }\,\Lambda _{3}\,y\,,\quad 
\hat{\sigma}_{3}^{2}=\hat{\sigma}^{2}=y^{\prime }\,\Lambda _{4}\,y\,.
\end{equation}%
For $\mathcal{H}_{1}$, we have%
\begin{equation}
\mathcal{H}_{1}=T(\tilde{\beta}-\hat{\beta})^{\prime }\hat{\Sigma}_{1}^{-1}(%
\tilde{\beta}-\hat{\beta})=T\,y^{\prime }\,C_{1}^{\prime }\,\hat{\Sigma}%
_{1}^{-1}C_{1}\,y=T\,(y^{\prime }\,\Psi _{_{1}}[y]\,y)
\end{equation}%
where 
\begin{equation}
\hat{\Sigma}_{1}=\tilde{\sigma}^{2}\hat{\Omega}_{IV}^{-1}-\hat{\sigma}^{2}%
\hat{\Omega}_{LS}^{-1}=(y^{\prime }\,\Lambda _{3}\,y)\,\hat{\Omega}%
_{IV}^{-1}-(y^{\prime }\,\Lambda _{4}\,y)\,\hat{\Omega}_{LS}^{-1}\,.
\end{equation}%
The result for $\mathcal{R}$ follows directly by using (\ref{eq: statRH}).
\end{proofflexc}

In order to characterize the null distributions of the test statistics
(Theorem \ref{Th: Null distributions of exogeneity test statistics}), it
will be useful to first spell out some algebraic properties of the weighting
matrices in Proposition \ref{Th: Quadratic-form representations of
exogeneity statistics}. This is done by the following lemma.

\begin{lemma}
\label{Th: Properties of weighting matrices in exogeneity statistics} 
\captiontheorem{\lemmaname}{Properties of weighting matrices in exogeneity statistics}
The matrices $\Psi _{_{0}}$, $\Lambda _{1}$, $\Lambda _{2}$, $\Lambda _{4}$, 
$\Psi _{R}$ and $\Lambda _{_{R}}$ in $(\ref{eq: T_l y})$\thinspace
-\thinspace $(\ref{eq: Psi1})$ satisfy the following identities: 
\begin{equation}
\Lambda _{2}=\Lambda _{4}-\Psi _{_{0}}\,,\;C_{1}\,\Lambda
_{1}=C_{1}\,\Lambda _{2}=\Psi _{_{0}}\,\Lambda _{1}=\Psi _{_{0}}\,\Lambda
_{2}=\Psi _{R}\,\Lambda _{_{R}}=0\,,  \label{eq: C_1*Lambda_1 = 0}
\end{equation}%
\begin{equation}
C_{1}\,\Lambda _{4}=\frac{1}{T}C_{1}\,,\;\Psi _{_{0}}\,\Lambda _{4}=\frac{1}{%
T}\Psi _{_{0}}\,,  \label{eq: C_1*Lambda_4}
\end{equation}%
\begin{equation}
M_{1}\,\Lambda _{l}\,M_{1}=\Lambda _{l}\,,\;l=1,\ldots ,\,4\,.
\label{eq: M_1*Lambda_1*M_1 = Lambda_1}
\end{equation}%
Further, the matrices $T\Psi _{_{0}}$, $T\Lambda _{1}$, $T\Lambda _{2}$, $%
T\Lambda _{4}$, $T\Psi _{R}$ and $T\Lambda _{_{R}}$ are symmetric idempotent.
\end{lemma}

\begin{proofflexc}
\captionproofflexc{\lemmaname}{Th: Properties of weighting matrices in exogeneity statistics}
To get (\ref{eq: C_1*Lambda_1 = 0})\thinspace -\thinspace (\ref{eq:
C_1*Lambda_4}), we observe that:%
\begin{equation}
\Lambda _{2}=M_{1}\left( \frac{1}{T}\mathrm{\bar{M}}[M_{1}Y]-\Psi
_{_{0}}\right) M_{1}=\Lambda _{4}-M_{1}\Psi _{_{0}}M_{1}=\Lambda _{4}-\Psi
_{_{0}}\,,
\end{equation}%
\begin{eqnarray}
C_{1}N_{1}\,\mathrm{\bar{P}}[N_{1}Y] &=&\frac{1}{T}[B_{2}-B_{1}]N_{1}N_{1}Y\,%
\hat{\Omega}_{IV}^{-1}\,Y^{\prime }N_{1}=\frac{1}{T}[\hat{\Omega}%
_{IV}^{-1}\,Y^{\prime }N_{1}-\hat{\Omega}_{LS}^{-1}\,Y^{\prime
}M_{1}]N_{1}Y\,\hat{\Omega}_{IV}^{-1}\,Y^{\prime }N_{1}  \notag \\
&=&\frac{1}{T}[\hat{\Omega}_{IV}^{-1}\,Y^{\prime }N_{1}Y\,\hat{\Omega}%
_{IV}^{-1}\,Y^{\prime }-\hat{\Omega}_{LS}^{-1}\,Y^{\prime }N_{1}Y\,\hat{%
\Omega}_{IV}^{-1}\,Y^{\prime }]N_{1}=\frac{1}{T}[\hat{\Omega}%
_{IV}^{-1}\,Y^{\prime }-\hat{\Omega}_{LS}^{-1}\,Y^{\prime }]N_{1}  \notag \\
&=&\frac{1}{T}[\hat{\Omega}_{IV}^{-1}\,Y^{\prime }N_{1}-\hat{\Omega}%
_{LS}^{-1}\,Y^{\prime }M_{1}]N_{1}=[B_{2}-B_{1}]N_{1}=C_{1}N_{1}\,,
\end{eqnarray}%
\begin{equation}
C_{1}M_{1}\,\mathrm{\bar{P}}[M_{1}Y]=C_{1}M_{1}Y(Y^{\prime
}M_{1}Y)^{-1}Y^{\prime }M_{1}=0\,,
\end{equation}%
\begin{equation}
\mathrm{\bar{M}}[\bar{Y}]\mathrm{\bar{M}}[Z]=\mathrm{\bar{M}}[Z]\,,
\end{equation}%
hence%
\begin{equation}
C_{1}\Lambda _{1}=C_{1}\left( \frac{1}{T}N_{1}\,\mathrm{\bar{M}}%
[N_{1}Y]\,N_{1}\right) =\frac{1}{T}C_{1}N_{1}\,\mathrm{\bar{M}}%
[N_{1}Y]\,N_{1}=\frac{1}{T}C_{1}N_{1}\,\big(I_{T}-\mathrm{\bar{P}}[N_{1}Y]%
\big)\,N_{1}=0\,,
\end{equation}%
\begin{eqnarray}
C_{1}\Lambda _{2} &=&C_{1}M_{1}\left( \frac{1}{T}\mathrm{\bar{M}}%
[M_{1}Y]-\Psi _{_{0}}\right) M_{1}=\frac{1}{T}C_{1}M_{1}\mathrm{\bar{M}}%
[M_{1}Y]M_{1}-C_{1}M_{1}\Psi _{_{0}}M_{1}  \notag \\
&=&\frac{1}{T}C_{1}M_{1}\big(I_{T}-\mathrm{\bar{P}}[M_{1}Y]\big)%
M_{1}-C_{1}\Psi _{_{0}}=\frac{1}{T}C_{1}-\frac{1}{T}C_{1}=0\,,
\end{eqnarray}%
\begin{equation}
C_{1}\Lambda _{4}=\frac{1}{T}C_{1}M_{1}\mathrm{\bar{M}}[M_{1}Y]M_{1}=\frac{1%
}{T}C_{1}M_{1}\mathrm{\bar{M}}[M_{1}Y]=\frac{1}{T}C_{1}\,,
\end{equation}%
\begin{equation}
\Psi _{_{0}}\Lambda _{4}=\frac{1}{T}C_{1}^{\prime }\hat{\Delta}%
^{-1}C_{1}\,M_{1}\mathrm{\bar{M}}[M_{1}Y]M_{1}=\frac{1}{T}C_{1}^{\prime }%
\hat{\Delta}^{-1}C_{1}\,M_{1}\mathrm{\bar{M}}[M_{1}Y]=\frac{1}{T}%
C_{1}^{\prime }\hat{\Delta}^{-1}C_{1}\,=\frac{1}{T}\Psi _{_{0}}\,,
\end{equation}%
\begin{equation}
\Psi _{_{0}}\Lambda _{2}=\Psi _{_{0}}\,M_{1}\left( \frac{1}{T}\mathrm{\bar{M}%
}[M_{1}Y]-\Psi _{_{0}}\right) M_{1}=\Psi _{_{0}}\,\left( \Lambda _{4}-\Psi
_{_{0}}\right) =\frac{1}{T}\Psi _{_{0}}-\frac{1}{T}\Psi _{0}=0\,,
\end{equation}%
\begin{equation}
\Psi _{R}\Lambda _{_{R}}=\frac{1}{T^{2}}\{\mathrm{\bar{M}}[\bar{Y}]-\mathrm{%
\bar{M}}[Z]\}\mathrm{\bar{M}}[Z]=0\,.
\end{equation}%
(\ref{eq: M_1*Lambda_1*M_1 = Lambda_1}) follow directly from the idempotence
of $M_{1}$ and the definitions of $\Lambda _{l}\,,$\ $l=1,\ldots ,\,4\,$.
Finally, the idempotence and symmetry of the weight matrices can be checked
as follows:%
\begin{eqnarray}
(T\,\Psi _{_{0}})(T\,\Psi _{_{0}}) &=&T\,C_{1}^{\prime }\hat{\Delta}%
^{-1}C_{1}C_{1}^{\prime }\hat{\Delta}^{-1}C_{1}=T^{2}\,C_{1}^{\prime }\hat{%
\Delta}^{-1}\left( \frac{1}{T}\hat{\Delta}\right) \hat{\Delta}%
^{-1}C_{1}=T\,C_{1}^{\prime }\hat{\Delta}^{-1}C_{1}  \notag \\
&=&T\,\Psi _{_{0}}=T\,\Psi _{_{0}}^{\prime }\,,
\end{eqnarray}%
\begin{equation}
(T\,\Lambda _{1})(T\,\Lambda _{1})=\left( N_{1}\,\mathrm{\bar{M}}%
[N_{1}Y]\,N_{1}\right) \left( N_{1}\,\mathrm{\bar{M}}[N_{1}Y]\,N_{1}\right)
=N_{1}\,\mathrm{\bar{M}}[N_{1}Y]\,N_{1}=T\,\Lambda _{1}=T\,\Lambda
_{1}^{\prime }\,,
\end{equation}%
\begin{equation}
(T\,\Lambda _{4})(T\,\Lambda _{4})=M_{1}\mathrm{\bar{M}}[M_{1}Y]M_{1}\,M_{1}%
\,\mathrm{\bar{M}}[M_{1}Y]M_{1}=M_{1}\mathrm{\bar{M}}[M_{1}Y]M_{1}=T\,%
\Lambda _{4}=T\,\Lambda _{4}^{\prime }\,,
\end{equation}%
\begin{eqnarray}
(T\,\Lambda _{2})(T\,\Lambda _{2}) &=&T^{2}\left( \Lambda _{4}-\Psi
_{_{0}}\right) \left( \Lambda _{4}-\Psi _{_{0}}\right) =T^{2}\left( \Lambda
_{4}\,\Lambda _{4}-\Lambda _{4}\Psi _{_{0}}-\Psi _{_{0}}\Lambda _{4}+\Psi
_{_{0}}\Psi _{_{0}}\right)  \notag \\
&=&T^{2}\left( \frac{1}{T}\Lambda _{4}-\frac{2}{T}\,\Psi _{_{0}}+\frac{1}{T}%
\Psi _{0}\right) =T(\Lambda _{4}-\Psi _{_{0}})=T\,\Lambda _{2}=T\,\Lambda
_{2}^{\prime }\,,
\end{eqnarray}%
\begin{equation}
(T\,\Psi _{R})(T\,\Psi _{R})=\{\mathrm{\bar{M}}[\bar{Y}]-\mathrm{\bar{M}}%
[Z]\}\{\mathrm{\bar{M}}[\bar{Y}]-\mathrm{\bar{M}}[Z]\}=\mathrm{\bar{M}}[\bar{%
Y}]-\mathrm{\bar{M}}[Z]=T\,\Psi _{R}=T\,\Psi _{R}^{\prime }\,,
\end{equation}%
\begin{equation}
(T\,\Lambda _{_{R}})(T\,\Lambda _{_{R}})=\mathrm{\bar{M}}[Z]\,\mathrm{\bar{M}%
}[Z]=\mathrm{\bar{M}}[Z]=T\,\Lambda _{_{R}}=T\,\Lambda _{_{R}}^{\prime }\,.
\end{equation}
\end{proofflexc}

\quad

\begin{proofflexc}
\captionproofflexc{\theoremname}{Th: Null distributions of exogeneity test statistics}
Using Lemma \ref{Th: Properties of exogeneity statistics components}, we
first note the following identities: 
\begin{equation}
B_{1}Y=(Y^{\prime }M_{1}Y)^{-1}Y^{\prime }M_{1}Y=I_{G}=(Y^{\prime
}N_{1}Y)^{-1}Y^{\prime }N_{1}Y\,=B_{2}Y\,,  \label{eq: A_1*Y}
\end{equation}%
\begin{equation}
\mathrm{\bar{M}}[M_{1}Y]\,M_{1}\,Y=\mathrm{\bar{M}}[N_{1}Y]\,N_{1}\,Y=0\,,%
\quad B_{1}X_{1}=B_{2}X_{1}=0\,,\quad N_{1}\,X_{1}=M_{1}\,X_{1}=0\,,
\label{eq: M[M_1*y]M_1*Y}
\end{equation}%
\begin{equation}
N_{2}M_{1}Y=(I_{T}-M_{1}YA_{2})M_{1}Y=(M_{1}-M_{1}YA_{2})Y=M_{1}(Y-YA_{2}Y)=0\,,\quad N_{2}M_{1}X_{1}=0\,,
\label{eq: N_2*M_1*Y}
\end{equation}%
\begin{equation}
\mathrm{\bar{M}}[\bar{Y}]Y=\mathrm{\bar{M}}[Z]Y=0\,,\;\mathrm{\bar{M}}[\bar{Y%
}]X_{1}=\mathrm{\bar{M}}[Z]X_{1}=0\,,\;\mathrm{\bar{P}}\big[\mathrm{\bar{M}}[%
\bar{Y}]X_{2}\big]\mathrm{\bar{M}}[\bar{Y}]=\mathrm{\bar{M}}[\bar{Y}]\,%
\mathrm{\bar{P}}\big[\mathrm{\bar{M}}[\bar{Y}]X_{2}\big]\mathrm{\bar{M}}[%
\bar{Y}]\,\,.  \label{eq:M[Ybar]Y = 0}
\end{equation}%
Then%
\begin{equation}
C_{1}\,y=(B_{2}-B_{1})(Y\beta +X_{1}\gamma +u)=C_{1}u\,,  \label{eq: C_1*u}
\end{equation}%
\begin{equation}
y^{\prime }\,\Psi _{_{0}}\,y=y^{\prime }\,C_{1}^{\prime }\hat{\Delta}%
^{-1}C_{1}\,y=u^{\prime }\,C_{1}^{\prime }\hat{\Delta}^{-1}C_{1}\,u=u^{%
\prime }\,\Psi _{_{0}}\,u\,,
\end{equation}%
\begin{equation}
y^{\prime }\,\Lambda _{1}\,y=\frac{1}{T}y^{\prime }\,N_{1}\,\mathrm{\bar{M}}%
[N_{1}Y]\,N_{1}\,y=\frac{1}{T}u^{\prime }\,N_{1}\,\mathrm{\bar{M}}%
[N_{1}Y]\,N_{1}\,u=u^{\prime }\,\Lambda _{1}\,u\,,
\end{equation}%
\begin{equation}
y^{\prime }\,\Lambda _{2}\,y=\frac{1}{T}y^{\prime }\,M_{1}\,(\mathrm{\bar{M}}%
[M_{1}Y]-\Psi _{_{0}})\,M_{1}\,y=\frac{1}{T}u^{\prime }\,M_{1}\,(\mathrm{%
\bar{M}}[M_{1}Y]-\Psi _{_{0}})\,M_{1}\,u=u^{\prime }\,\Lambda _{2}\,u\,,
\end{equation}%
\begin{equation}
y^{\prime }\,\Lambda _{3}\,y=\frac{1}{T}y^{\prime }\,M_{1}\,N_{2}^{\prime
}N_{2}M_{1}\,y=\frac{1}{T}u^{\prime }\,M_{1}\,N_{2}^{\prime }N_{2}M_{1}\,u\,,
\end{equation}%
\begin{equation}
y^{\prime }\,\Lambda _{4}\,y=\frac{1}{T}y^{\prime }\,\mathrm{\bar{M}}[\bar{Y}%
]\,y=\frac{1}{T}u^{\prime }\,\mathrm{\bar{M}}[\bar{Y}]\,u=u^{\prime
}\,\Lambda _{4}\,u\,,
\end{equation}%
\begin{eqnarray}
y^{\prime }\,\Psi _{R}\,y &=&\frac{1}{T}y^{\prime }\,\mathrm{\bar{P}}\big[%
\mathrm{\bar{M}}[\bar{Y}]X_{2}\big]\,\mathrm{\bar{M}}[\bar{Y}]\,y=\frac{1}{T}%
y^{\prime }\,\mathrm{\bar{M}}[\bar{Y}]\,\mathrm{\bar{P}}\big[\mathrm{\bar{M}}%
[\bar{Y}]X_{2}\big]\,\mathrm{\bar{M}}[\bar{Y}]\,y  \notag \\
&=&\frac{1}{T}u^{\prime }\,\mathrm{\bar{M}}[\bar{Y}]\,\mathrm{\bar{P}}\big[%
\mathrm{\bar{M}}[\bar{Y}]X_{2}\big]\,\mathrm{\bar{M}}[\bar{Y}]\,u=\frac{1}{T}%
u^{\prime }\,\mathrm{\bar{P}}\big[\mathrm{\bar{M}}[\bar{Y}]X_{2}\big]\,%
\mathrm{\bar{M}}[\bar{Y}]\,u=u^{\prime }\,\Psi _{R}\,u\,,
\end{eqnarray}%
\begin{equation}
\hat{\sigma}_{R}^{2}=\frac{1}{T}y^{\prime }\,\mathrm{\bar{M}}[Z]y=\frac{1}{T}%
u^{\prime }\,\mathrm{\bar{M}}[Z]\,u\,.  \label{eq: sigma_R u}
\end{equation}%
Further, when $a=0$, we have $u=\sigma _{1}(\bar{X})\,\varepsilon $%
\thinspace , and the expressions in (\ref{eq: T_l e})\thinspace -\thinspace (%
\ref{eq: H_i e}) follow from (\ref{eq: T_l y})\thinspace -\thinspace (\ref%
{eq:H_2 y}) in Proposition \ref{Th: Quadratic-form representations of
exogeneity statistics} once $u$ is replaced by $\sigma _{1}(\bar{X}%
)\,\varepsilon $ in (\ref{eq: C_1*u})\thinspace - (\ref{eq: sigma_R u}). $%
\sigma _{1}(\bar{X})$ disappears because it can be factorized in both the
numerator and the denominator of each statistic.
\end{proofflexc}

\quad

\begin{proofflexc}
\captionproofflexc{\propositionname}{Th: Block-triangular invariance of exogeneity tests}
We must study how the statistics defined in (\ref{eq: statT1})\thinspace
-\thinspace (\ref{eq: statRH}) change when $y$ and $Y$ are replaced by $%
y^{\ast }=yR_{11}+YR_{21}$ and $Y^{\ast }=YR_{22}$. This can be done by
looking at the way the relevant variables in (\ref{eq: b_OLS})\thinspace
-\thinspace (\ref{eq: Psi_R}) change. We first note that 
\begin{equation}
\hat{\Omega}_{IV}^{\ast }=\frac{1}{T}Y^{\ast \prime }N_{1}Y^{\ast
}=(YR_{22})^{\prime }N_{1}(YR_{22})=R_{22}^{\prime }\hat{\Omega}%
_{IV}R_{22}\,,\quad \hat{\Omega}_{LS}^{\ast }=\frac{1}{T}Y^{\ast \prime
}M_{1}Y^{\ast }=R_{22}^{\prime }\hat{\Omega}_{LS}R_{22}\,,
\end{equation}%
hence 
\begin{equation}
\hat{\Delta}^{\ast }=(\hat{\Omega}_{IV}^{\ast })^{-1}-(\hat{\Omega}%
_{LS}^{\ast })^{-1}=R_{22}^{-1}(\hat{\Omega}_{IV}^{-1}-\hat{\Omega}%
_{LS}^{-1})(R_{22}^{-1})^{\prime }=R_{22}^{-1}\hat{\Delta}%
(R_{22}^{-1})^{\prime }\,.
\end{equation}%
Using Lemma \ref{Th: Properties of exogeneity statistics components}, we
also get:%
\begin{eqnarray}
B_{1}^{\ast } &=&(Y^{\ast \prime }M_{1}Y^{\ast })^{-1}Y^{\ast \prime
}M_{1}=[(YR_{22})^{\prime }M_{1}(YR_{22})]^{-1}(YR_{22})^{\prime
}M_{1}=R_{22}^{-1}(Y^{\prime }M_{1}Y)^{-1}Y^{\prime }M_{1}\quad \quad \quad 
\notag \\
&=&R_{22}^{-1}B_{1}\,,
\end{eqnarray}%
\begin{equation}
B_{2}^{\ast }=(Y^{\ast \prime }N_{1}Y^{\ast })^{-1}Y^{\ast \prime
}N_{1}=R_{22}^{-1}(Y^{\prime }N_{1}Y)^{-1}Y^{\prime
}N_{1}=R_{22}^{-1}B_{2}\,,
\end{equation}%
\begin{equation}
C_{1}^{\ast }=B_{2}^{\ast }-B_{1}^{\ast }=R_{22}^{-1}C_{1}\,,\quad
C_{1}^{\ast }Y=R_{22}^{-1}C_{1}Y=0\,,
\end{equation}%
\begin{equation}
\hat{\beta}^{\ast }=B_{1}^{\ast }y^{\ast
}=R_{22}^{-1}B_{1}(yR_{11}+YR_{21})=R_{11}R_{22}^{-1}\hat{\beta}%
+R_{22}^{-1}R_{21}\,,
\end{equation}%
\begin{equation}
\tilde{\beta}^{\ast }=B_{2}^{\ast }y^{\ast }=R_{11}R_{22}^{-1}\tilde{\beta}%
+R_{22}^{-1}R_{21}\,,
\end{equation}%
\begin{equation}
\tilde{\beta}^{\ast }-\hat{\beta}^{\ast }=C_{1}^{\ast }\,y^{\ast
}=R_{11}R_{22}^{-1}(\tilde{\beta}-\hat{\beta})\,,
\end{equation}%
\begin{eqnarray}
\hat{u}^{\ast } &=&M_{1}(y^{\ast }-Y^{\ast }\hat{\beta}^{\ast })=M_{1}\big(%
yR_{11}+YR_{21}-YR_{22}(R_{11}R_{22}^{-1}\hat{\beta}+R_{22}^{-1}R_{21})\big)
\notag \\
&=&R_{11}\,M_{1}(y-Y\hat{\beta})=R_{11}\,\hat{u}\,,
\end{eqnarray}%
\begin{equation}
\tilde{u}^{\ast }=M_{1}(y^{\ast }-Y^{\ast }\tilde{\beta}^{\ast })=M_{1}\big(%
yR_{11}+YR_{21}-YR_{22}(R_{11}R_{22}^{-1}\tilde{\beta}+R_{22}^{-1}R_{21})%
\big)=R_{11}\,\tilde{u}\,,
\end{equation}%
hence, since $N_{1}X_{1}=0,$ 
\begin{equation}
\hat{\sigma}^{\ast 2}=\frac{1}{T}\hat{u}^{\ast \prime }\hat{u}^{\ast
}=R_{11}^{2}\,\hat{\sigma}^{2}\,,\quad \tilde{\sigma}^{\ast 2}=\frac{1}{T}%
\tilde{u}^{\ast \prime }\tilde{u}^{\ast }=R_{11}^{2}\,\tilde{\sigma}^{2}\,,
\end{equation}%
\begin{eqnarray}
\tilde{\sigma}_{1}^{\ast 2} &=&\frac{1}{T}(y^{\ast }-Y^{\ast }\tilde{\beta}%
^{\ast })^{\prime }N_{1}(y^{\ast }-Y^{\ast }\tilde{\beta}^{\ast })=\frac{1}{T%
}(y^{\ast }-Y^{\ast }\tilde{\beta}^{\ast }-X_{1}\tilde{\gamma}^{\ast
})^{\prime }N_{1}(y^{\ast }-Y^{\ast }\tilde{\beta}^{\ast }-X_{1}\tilde{\gamma%
}^{\ast })\quad \quad  \notag \\
&=&\frac{1}{T}\tilde{u}^{\ast \prime }N_{1}\,\tilde{u}^{\ast }=R_{11}^{2}%
\frac{1}{T}\tilde{u}^{\prime }N_{1}\,\tilde{u}=R_{11}^{2}\tilde{\sigma}%
_{1}^{2}\,,
\end{eqnarray}%
\begin{eqnarray}
\tilde{\sigma}_{2}^{\ast 2} &=&\hat{\sigma}^{\ast 2}-(\tilde{\beta}^{\ast }-%
\hat{\beta}^{\ast })^{\prime }(\hat{\Delta}^{\ast })^{-1}(\tilde{\beta}%
^{\ast }-\hat{\beta}^{\ast })  \notag \\
&=&R_{11}^{2}\,\hat{\sigma}^{2}-(\tilde{\beta}-\hat{\beta})^{\prime
}(R_{11}R_{22}^{-1})^{\prime }R_{22}^{\prime }\hat{\Delta}%
^{-1}R_{22}(R_{11}R_{22}^{-1})(\tilde{\beta}-\hat{\beta})  \notag \\
&=&R_{11}^{2}[\hat{\sigma}^{2}-(\tilde{\beta}-\hat{\beta})^{\prime }\hat{%
\Delta}^{-1}(\tilde{\beta}-\hat{\beta})]=R_{11}^{2}\,\tilde{\sigma}%
_{2}^{2}\,,
\end{eqnarray}%
\begin{eqnarray}
\tilde{\Sigma}_{i}^{\ast } &=&\tilde{\sigma}_{i}^{\ast 2}\hat{\Delta}^{\ast
}=(R_{11}^{2}\tilde{\sigma}_{i}^{2})R_{22}^{-1}\hat{\Delta}%
(R_{22}^{-1})^{\prime }=R_{11}^{2}\,R_{22}^{-1}(\tilde{\sigma}_{i}^{2}\hat{%
\Delta})(R_{22}^{-1})^{\prime }  \notag \\
&=&R_{11}^{2}\,R_{22}^{-1}\,\tilde{\Sigma}_{i}\,(R_{22}^{-1})^{\prime },%
\text{\quad }i=1,\,2,\,3,\,4\,,
\end{eqnarray}%
\begin{equation}
\hat{\Sigma}_{j}^{\ast }=R_{11}^{2}\,R_{22}^{-1}\hat{\Sigma}%
_{j}(R_{22}^{-1})^{\prime }\,,\quad j=1,2,\,3.
\end{equation}%
It follows that the $\mathcal{T}_{i}$ and $\mathcal{H}_{j}$ exogeneity test
statistics based on the transformed data are identical to those based on the
original data: 
\begin{eqnarray}
\mathcal{T}_{i}^{\ast } &=&\kappa _{i}(\tilde{\beta}^{\ast }-\hat{\beta}%
^{\ast })^{\prime }(\tilde{\Sigma}_{i}^{\ast })^{-1}(\tilde{\beta}^{\ast }-%
\hat{\beta}^{\ast })  \notag \\
&=&(\tilde{\beta}-\hat{\beta})^{\prime }(R_{11}R_{22}^{-1})^{\prime
}[R_{11}^{2}\,R_{22}^{-1}\,\tilde{\Sigma}_{i}\,(R_{22}^{-1})^{\prime
}]^{-1}(R_{11}R_{22}^{-1})(\tilde{\beta}-\hat{\beta})  \notag \\
&=&\kappa _{i}(\tilde{\beta}-\hat{\beta})^{\prime }\tilde{\Sigma}_{i}^{-1}(%
\tilde{\beta}-\hat{\beta})=\mathcal{T}_{i}\,,\;i=1,\,2,\,3,\,4,
\end{eqnarray}%
\begin{eqnarray}
\mathcal{H}_{j}^{\ast } &=&T(\tilde{\beta}^{\ast }-\hat{\beta}^{\ast
})^{\prime }(\hat{\Sigma}_{j}^{\ast })^{-1}(\tilde{\beta}^{\ast }-\hat{\beta}%
^{\ast })  \notag \\
&=&T(\tilde{\beta}-\hat{\beta})^{\prime }(R_{11}R_{22}^{-1})^{\prime
}[R_{11}^{2}\,R_{22}^{-1}\,\hat{\Sigma}_{j}\,(R_{22}^{-1})^{\prime
}]^{-1}(R_{11}R_{22}^{-1})(\tilde{\beta}-\hat{\beta})=\mathcal{H}%
_{j}\,,\;j=1,\,2,\,3.\quad \quad \quad
\end{eqnarray}%
Finally, the invariance of the statistic $\mathcal{R}$ is obtained by
observing that 
\begin{equation}
y^{\ast \prime }\mathrm{\bar{M}}[Z^{\ast }]y^{\ast }=R_{11}^{2}\,y^{\prime }%
\mathrm{\bar{M}}[Z]y\,,\quad y^{\ast \prime }\mathrm{\bar{M}}[\bar{Y}^{\ast
}]y^{\ast }=R_{11}^{2}\,y^{\prime }\mathrm{\bar{M}}[\bar{Y}]y\,,
\end{equation}%
where $Z^{\ast }=[Y^{\ast },\,X_{1},\,X_{2}]$ and $\bar{Y}^{\ast }=[Y^{\ast
},\,X_{1}]$, so $R_{11}^{2}$ cancels out in $\mathcal{R}$
\end{proofflexc}

\quad

\begin{proofflexc}
\captionproofflexc{\theoremname}{Th: Exogeneity test distributions under the alternative hypothesis}
Since $u=Va+\sigma _{1}(\bar{X})\,\varepsilon $\thinspace , we can use the
identities (\ref{eq: C_1*u})\thinspace - (\ref{eq: sigma_R u}) and replace $%
y $ by $Va+\sigma _{1}(\bar{X})\,\varepsilon $ in (\ref{eq: T_l y}%
)\thinspace -\thinspace (\ref{eq: T_l y}). The expressions (\ref{eq: T_l
u(a)})\thinspace - (\ref{eq: RH u(a)}) then follow through division of the
numerator and denominator of each statistic by $\sigma _{1}(\bar{X})$.
\end{proofflexc}

\quad

\begin{proofflexc}
\captionproofflexc{\theoremname}{Th: Invariance-based distributions of exogeneity test statistics}
This result follows by applying the invariance property of Proposition \ref%
{Th: Block-triangular invariance of exogeneity tests} with $R$ defined as in
(\ref{eq: R standard}). $y$ is then replaced by $y^{\ast }=X_{1}\gamma
+[V-g(X_{1},\,X_{2},\,X_{3},\,V,\,\bar{\Pi})]a+e$ [see (\ref{eq: mu_y*})],
and the identities (\ref{eq: C_1*u})\thinspace - (\ref{eq: sigma_R u}) hold
with $u$ replaced by 
\begin{equation}
u_{\ast }=[V-g(X_{1},\,X_{2},\,X_{3},\,V,\,\bar{\Pi})]a+e\,.
\end{equation}%
Further, in view of (\ref{eq: M_1*Y*A_1}) and (\ref{eq: Lambda_1})\thinspace
-\thinspace (\ref{eq: Psi_R}), each one of the matrices $\Psi _{_{0}}$, $%
\Lambda _{1},\ldots ,\,\Lambda _{4}$, $\Psi _{_{1}}$, $\Psi _{R}$ and $%
\Lambda _{R}$ remains the same if it is pre- and postmultiplied by $M_{1}$, 
\emph{i.e.}%
\begin{gather}
\Psi _{_{0}}=M_{1}\,\Psi _{_{0}}\,M_{1}\,,\quad \Lambda _{l}=M_{1}\Lambda
_{i}M_{1},\ i=1,\,2,\,3,\,4, \\
\Psi _{1}=M_{1}\,\Psi _{_{1}}\,M_{1}\,,\quad \Psi _{_{R}}=M_{1}\,\Psi
_{R}\,M_{1}\,,\quad \Lambda _{R}=M_{1}\Lambda _{R}M_{1}\,,
\end{gather}%
so $u_{\ast }$ can in turn be replaced by 
\begin{equation}
M_{1}u_{\ast }=-M_{1}[V-g(X_{1},\,X_{2},\,X_{3},\,V,\,\bar{\Pi})]a+M_{1}\,e
\end{equation}%
in (\ref{eq: C_1*u})\thinspace - (\ref{eq: sigma_R u}). Upon division of the
numerator and denominator of each statistic by $\sigma _{1}(\bar{X})$, we
get the expressions (\ref{eq: T_l y_1*(abar)})\thinspace -\thinspace (\ref%
{eq: RH y_1*(abar)}).
\end{proofflexc}

\quad

\begin{proofflexc}
\captionproofflexc{\theoremname}{Th: Invariance-based distributions of exogeneity statistics components with Gaussian errors}
The result follows from well known properties of the normal and chi-square
distributions: if $x\sim N_{n}[\mu ,$\thinspace $I_{n}]$ and $A$ is a fixed
idempotent $n\times n$ matrix of rank $r$, then $x^{\prime }A\,x\sim \chi
^{2}[r\,;\,\mu ^{\prime }A\,\mu ]$\thinspace . Conditional on $\bar{X}$ and $%
V,$ $\Psi _{_{0}}$ is fixed, and 
\begin{equation}
y_{\ast }^{\perp }(\bar{a}\ )=\bar{\mu}_{y\ast }^{\perp }(\bar{a}\
)+M_{1}\varepsilon =M_{1}\{[V-g(X_{1},\,X_{2},\,X_{3},\,V,\,\bar{\Pi})]\bar{a%
}+\varepsilon \}=M_{1}(\mu +\varepsilon )
\end{equation}%
where $\mu =[V-g(X_{1},\,X_{2},\,X_{3},\,V,\,\bar{\Pi})]\bar{a}$ is fixed
and $\varepsilon \sim N_{n}[\mu ,$\thinspace $I_{n}]$. By Lemmas \ref{Th:
Properties of exogeneity statistics components} and \ref{Th: Properties of
weighting matrices in exogeneity statistics}, $T\Psi _{_{0}}$, $T\Lambda
_{1} $, $T\Lambda _{2}$, $T\Lambda _{4}$, $T\Psi _{R}$ and $T\Lambda _{_{R}}$
are symmetric idempotent, and each of these matrices remain invariant
through \text{by pre- and post-multiplication }by $M_{1}$ \text{[}$%
M_{1}\,\Psi _{_{0}}\,M_{1}=\Psi _{_{0}}\,$, etc.]. Thus%
\begin{eqnarray}
S_{T}[y_{\ast }^{\perp }(\bar{a}\ ),\,\Psi _{_{0}}] &=&T\,y_{\ast }^{\perp }(%
\bar{a}\ )^{\prime }\Psi _{_{0}}y_{\ast }^{\perp }(\bar{a}\ )=(\mu
+\varepsilon )^{\prime }M_{1}(T\,\Psi _{_{0}})M_{1}(\mu +\varepsilon ) \\
&=&(\mu +\varepsilon )^{\prime }(T\,\Psi _{_{0}})(\mu +\varepsilon )\sim
\chi ^{2}[\mathrm{rank}(T\,\Psi _{_{0}})\,;\,\mu ^{\prime }(T\,\Psi
_{_{0}})\,\mu ]
\end{eqnarray}%
where%
\begin{equation}
\mathrm{rank}(T\,\Psi _{_{0}})=\mathrm{tr}(T\,\Psi _{_{0}})=\mathrm{tr}%
(T\,C_{1}^{\prime }\hat{\Delta}^{-1}C_{1})=\mathrm{tr}(T\,\hat{\Delta}%
^{-1}C_{1}\,C_{1}^{\prime })=\mathrm{tr}(T\,\hat{\Delta}^{-1}T^{-1}\hat{%
\Delta})=G\,,
\end{equation}%
\begin{equation}
\mu ^{\prime }(T\,\Psi _{_{0}})\,\mu =\mu ^{\prime }M_{1}(T\,\Psi
_{_{0}})\,M_{1}\,\mu =\bar{\mu}_{y\ast }^{\perp }(\bar{a}\ )^{\prime
}(T\,\Psi _{_{0}})\bar{\mu}_{y\ast }^{\perp }(\bar{a}\ )=S_{T}[\bar{\mu}%
_{y\ast }^{\perp }(\bar{a}\ ),\,\Psi _{0}]=\delta (\bar{a},\,\Psi _{_{0}})\,.
\end{equation}%
The proofs for the other quadratic forms are similar, with the following
degrees of freedom vary:%
\begin{gather}
\mathrm{rank}(T\,\Lambda _{1})=\mathrm{tr}\{N_{1}\,\mathrm{\bar{M}}%
[N_{1}Y]\,N_{1}\}=\mathrm{tr}\{N_{1}\,\}-\mathrm{tr}\{\mathrm{\bar{P}}%
[N_{1}Y]\}=\mathrm{tr}\{M_{1}-M\}-\mathrm{tr}\{N_{1}Y(Y^{\prime
}N_{1}Y)^{-1}Y^{\prime }N_{1}\}  \notag \\
=(T-k_{1})-(T-k_{1}-k_{2})-\mathrm{tr}\{(Y^{\prime }N_{1}Y)^{-1}Y^{\prime
}N_{1}Y\}=k_{2}-G\,,
\end{gather}%
\begin{eqnarray}
\mathrm{rank}(T\,\Lambda _{2}) &=&\mathrm{tr}\{T\,M_{1}\left( T^{-1}\mathrm{%
\bar{M}}[M_{1}Y]-\Psi _{_{0}}\right) M_{1}\}=\mathrm{tr}\{M_{1}\mathrm{\bar{M%
}}[M_{1}Y]M_{1}\}-\mathrm{tr}\{T\,\Psi _{_{0}}\}  \notag \\
&=&\mathrm{tr}\{M_{1}\}-\mathrm{tr}\{\mathrm{\bar{P}}[M_{1}Y]\}-\mathrm{tr}%
\{T\,\Psi _{_{0}}\}=T-k_{1}-2G\,,
\end{eqnarray}%
\begin{equation}
\mathrm{rank}(T\,\Lambda _{4})=\mathrm{tr}\{M_{1}\mathrm{\bar{M}}%
[M_{1}Y]M_{1}\}=\mathrm{tr}\{M_{1}\}-\mathrm{tr}\{\mathrm{\bar{P}}%
[M_{1}Y]\}=T-k_{1}-G\,,
\end{equation}%
\begin{equation}
\mathrm{rank}(T\,\Psi _{R})=\mathrm{tr}\{\mathrm{\bar{M}}[\bar{Y}]-\mathrm{%
\bar{M}}[Z]\}=(T-k_{1}-G)-(T-k_{1}-G-k_{2})=k_{2}\,,
\end{equation}%
\begin{equation}
\mathrm{rank}(T\,\Lambda _{R})=\mathrm{tr}(T\,\Lambda _{R})=\mathrm{tr}\{%
\mathrm{\bar{M}}[Z]\}=T-G-k_{1}-k_{2}\,.
\end{equation}%
The independence properties follow from the orthogonalities given in (\ref%
{eq: C_1*Lambda_1 = 0}) and the normality assumption.
\end{proofflexc}

\begin{proofflexc}
\captionproofflexc{\corollaryname}{Th: Doubly noncentral distributions for exogeneity statistics}
These results directly from Theorem \ref{Th: Invariance-based distributions
of exogeneity statistics components with Gaussian errors} and the definition
of the doubly noncentral $F$-distribution.
\end{proofflexc}

\newpage 

\bibliographystyle{agsm}
\bibliography{Doko_Dufour}

\end{document}